\input amstex 
\input amsppt.sty

\hsize 12.2cm
\vsize 19.5cm

\def\nmb#1#2{#2}         % used for renumbering, TeX should ignore. 
\def\cit#1#2{\ifx#1!\cite{#2}\else#2\fi} %for citing references 
\def\totoc{}             %= to table of content, invoked by kms-book.sty 
               % for producing index, invoked by kms-book.sty 
\def\ign#1{}             %=ignore, invisible entry for the index only 

\define\al{\alpha} 
\define\be{\beta} 
\define\ga{\gamma} 
\define\de{\delta} 
\define\ep{\varepsilon} 
\define\ze{\zeta} 
\define\et{\eta} 
\define\th{\theta} 
 
\define\ka{\kappa} 
\define\la{\lambda} 
\define\rh{\rho} 
\define\si{\sigma} 
\define\ta{\tau} 
\define\ph{\varphi} 
\define\ch{\chi} 
\define\ps{\psi} 
\define\om{\omega} 
\define\Ga{\Gamma} 
\define\De{\Delta} 
\define\Th{\Theta} 
\define\La{\Lambda} 
\define\Si{\Sigma}

\define\row#1#2#3{#1_{#2},\ldots,#1_{#3}} 
 
\def\today{\ifcase\month\or 
 January\or February\or March\or April\or May\or June\or 
 July\or August\or September\or October\or November\or December\fi 
 \space\number\day, \number\year} 

%%%%%%%%%%%%%%%%%%%%%%%%%%%%%%%%%%%%%%%%%%%%%%%%%%%%%%%%%%%%%%%%%%%%
%%%%%%%%%%%%%%%  added by Bonnie %%%%%%%%%%%%%%%%%%%%%%%%%%%%%%%%%%%

\def\Z{{\Bbb Z}}

\def\medquad{\hskip0.8em}
\def\mmsquad{\hskip0.7em}

\def\smallquad{\hskip0.4em}
\def\tinyquad{\hskip0.18em}

\def\boxit#1{\vbox{\hrule height 1pt \kern0.1ex
             \hbox{\vrule width 1pt \kern-0.5ex #1
                   \kern-1.1ex \vrule width 1pt} \kern-0ex
             \hrule height 1pt}}

\def\boxiteight#1{\vbox{\hrule height 1pt \kern-0.1ex
             \hbox{\vrule width 1pt \kern-0.5ex #1
                   \kern-1.1ex \vrule width 1pt} \kern-0.1ex
             \hrule height 1pt}}

\def\lowerthis#1{\raise-1ex\hbox{$#1$}}
\def\raisethis#1{\raise1ex\hbox{$#1$}}
\def\raisethisby#1#2{\raise#1\hbox{$#2$}}

\def\spancol{\omit\span\omit\span}

\def\SO{\operatorname{SO}}
\def\SU{\operatorname{SU}}
\def\so{\operatorname{so}}
\def\su{\operatorname{su}}
\def\tr{\operatorname{tr}}

%%%%%%%%%%%%%%%%%%%%%%%%%%%%%%%%%%%%%%%%%%%%%%%%%%%%%%%%%%%%%%%%%%%%

\topmatter 
\title   
Affine orbifolds and rational conformal field theory extensions of  
$W_{1+\infty}$ 
\endtitle 
\leftheadtext{} 
\rightheadtext{} 
\author Victor G\. Kac and Ivan T\. Todorov*  \endauthor 
\affil 
International Erwin Schr\"odinger Institute for Mathematical  
Physics\\
and\\ 
Department of Mathematics, MIT, Cambridge, MA 02139, USA 
\endaffil 
\address 
\endaddress 
%\email \@PAP.UNIVIE.AC.AT \endemail 
\dedicatory \enddedicatory 
\date {December 3, 1996} \enddate 
\thanks  
Supported by the Federal Ministry of Science and Research,
Austria, NSF grants DMS-9103792 and DMS-9622870 and the Bulgarian NFSR under
contract F-404. 
\newline 
*On leave of absence from the Institute for Nuclear Research and  
Nuclear Energy, Bulgarian Academy of Sciences, Tsarigradsko Chauss\'ee  
72, BG-1784, Sofia, Bulgaria  
\endthanks 
%\keywords \endkeywords 
%\subjclass \endsubjclass 
\abstract  
Chiral orbifold models are defined as gauge field theories with a
finite gauge group $\Ga$. We start with a conformal current
algebra $\goth A$ associated with a connected compact Lie group
$G$ and a negative definite integral invariant bilinear form on
its Lie algebra. Any finite group $\Ga$ of inner automorphisms or
$\goth A$ (in particular, any finite subgroup of $G$) gives rise
to a gauge theory with a chiral subalgebra $\goth
A^{\Ga}\subset\goth A$ of {\it local observables} invariant under
$\Ga$. A set of positive energy $\goth A^{\Ga}$ modules is
constructed whose characters span, under some assumptions on
$\Gamma$, a finite dimensional unitary representation of
$SL(2,\Bbb Z)$. We compute their {\it asymptotic dimensions}
(thus singling out the nontrivial orbifold modules) and find
explicit formulae for the modular transformations and hence, for
the {\it fusion rules}.
 
As an application we construct a family of {\it rational conformal  
field theory} (RCFT) extensions of $W_{1+\infty}$ that appear to  
provide a bridge between two approaches to the quantum Hall effect. 
\endabstract 
\endtopmatter 
%\input amspptb.sty 
%\userunningheads 
%\def\leftheadtext{\smc } 
%\def\rightheadtext{\smc } 
%\def\bottremark{\today\hfill} 
 
\document 
 
\heading Contents \endheading 
%\input \jobname.toc 
%\loadtoc 
%\loadindex 
\widestnumber\item{Appendix A}
\roster
\item"{0.}" Introduction
\item"{1.}" {\it Chiral algebras associated with connected compact Lie groups}
\item"{1A.}" Definition of a chiral algebra.  Current algebras
\item"{1B.}" Lattice vertex algebras
\item"{1C.}" Current chiral algebras associated to simple Lie algebras
\item"{2.}" {\it Twisted modules of a current chiral algebra}
\item"{2A.}" Positive energy irreducible $\goth A (G)$ modules
\item"{2B.}" $\Bbb Z_{N}$-twisted current chiral algebra modules
\item"{3.}" {\it Twisted characters and modular transformations}
\item"{3A.}" Kac-Moody and lattice characters
\item"{3B.}" Modular transformations of twisted characters
\item"{3C.}" Small $\tau$ asymptotics of twisted characters of $\goth A (G)$.
\item"{4.}" {\it Affine orbifolds}
\item"{4A.}" Projection on a centralizer's irreducible representation
\item"{4B.}" Asymptotic dimension.  Affine orbifold models for
        non-excep\-tional $\Gamma$.  Action of $Z$
\item"{4C.}" Fusion rules.
\item"{5.}" $U(\ell)$ {\it orbifolds as RCFT extensions of $W_{1+\infty}$}
\item"{6.}" {\it Examples.}
\item"{6A.}" Lattice current algebras for $c=1$
\item"{6B.}" $\SU(2)$ orbifolds
\item"{6C.}" Modular $S$-matrix and fusion rules for an $\SU(3)$ orbifold.
\item"{Appendix A}" Action of the center of a simply connected
        simple Lie group on the coroots and fundamental weights
\item"{Appendix B}" Exceptional elements of a compact Lie group
\item"{References}"
\endroster   
 
\head\totoc 
\nmb0{0}. Introduction \endhead 
 
Given a chiral conformal field theory (CFT) --- i.e., a chiral algebra  
$\goth A$ and a family of positive energy $\goth A$-modules (closed  
under ``fusion'') --- there are two ways of constructing other CFT with  
the same stress-energy tensor $T(z)$ and associated central charge  
$c$. First, one can, in some cases, extend $\goth A$ by adjoining to  
it local primary fields. The stress energy tensor generates an RCFT  
for the minimal models \cite{BPZ} corresponding to central charge $c<1$. For  
$c\geq 1$ one needs in addition a chiral current algebra or a $W$-algebra to  
construct an RCFT (for special rational values of $c$). All RCFT  
extensions of the $(c=1)u(1)$-current algebra have been classified in  
\cite {BMT}; all local extensions of the $\su(2)$ current algebras  
have been described in \cite {MST}. The second path goes in the  
opposite direction: one restricts $\goth A$ to a distinguished  
subalgebra of ``observables'' including $T(z)$; we shall be concerned  
here with the case in which the subalgebra $\goth A^{\Ga}$ consists  
of all elements of $\goth A$ invariant under a finite automorphism  
group $\Ga$. The resulting CFT is called a $\Ga$-{\it orbifold}.  
Examples of orbifolds (first in the context of a ``Gaussian model''  
\cite {G} \cite {H}) have been studied in detail in \cite {DV$^3$}  
where some general properties of arbitrary orbifold models have also  
been pointed out. Non unitary models of $c=1$ have been considered in  
\cite {F}. 
 
The present paper provides a systematic approach to orbifold RCFT.  
Our starting point is a chiral algebra $\goth A=\goth A(G)$  
associated with a connected compact Lie group $G$ whose Lie algebra  
$\goth g$ is equipped with a negative definite integral invariant  
bilinear form. It appears as a tensor product of a lattice chiral  
algebra $\goth A(L)$ and (chiral) affine Kac-Moody algebras  
(corresponding to the simple components $\goth g^j$ of $\goth g$): 
$$ 
\goth A(G)=\goth A(L)\otimes 
(\otimes^s_{j=1}\goth A_{k_j}(\goth g^j)) \tag {0.1} 
$$ 
where $k_j(\in\Bbb Z_{+})$ is the level of the vacuum $\hat\goth
g^j$- module. The lattice $L$ consists of all vectors $\om$ in
the direct sum $\goth g^0$ of $u(1)$-components of the centre of
$\goth g$ such that $e^{2\pi i\om}=1$. To each $\om$ of length
square 2 we can associate a ``charge shift'' operator $E^{\om}$
providing a non-abelian extension of the Lie algebra $\goth g^0$.
Let $G_c$ be the corresponding maximal compact group extension of
$G$. Its significance stems from the fact that each finite order
{\it inner automorphism} of $\goth A(G)$ is given by (the adjoint
action of) an element of $G_c$.

An {\it orbifold chiral algebra} is the fixed point  
set $\goth A^{\Ga}$ of a finite group of automorphisms $\Ga$ of a chiral  
algebra $\goth A$. For any ``non-exceptional'' finite subgroup  
$\Ga \subset G_{c}$ we construct a finite  
family of $\goth A^{\Ga}$-modules $V$, which is complete in the sense  
that their characters transform among themselves under the modular  
group $SL(2,\Bbb Z)$. Each $V$ is labeled by a weight $\La$  
(characterizing an $\goth A(G)$-module), a conjugacy class  
$\bar b\subset \Ga$, and an irreducible representation $\si=\si^b$ of  
the centralizer $\Ga_b$ of an element $b\in\bar b$ in $\Ga$. It  
involves a choice of ``phases'' $\be(b)\in\goth g$, for  
non-exceptional conjugacy classes, satisfying the  
following two conditions: 
$$
\text{(i)~} b=e^{2 \pi i\be(b)} \; , 
\quad 
\text{(ii)~} \be(gbg^{-1}) =Ad_g\be(b) 
\text{ for } b\in\bar b,g\in\Ga \; .
\tag {0.2} 
$$ 
Condition (ii) implies that the centralizer of $b$ should  
stabilize $\be$. Two $\be$'s satisfying \thetag {0.2} differ  
by a co root $m$ which is also stabilized by $\Ga_b$. Any such $m$  
gives rise to a 1-dimensional representation $\si_m$ of $\Ga_b$. The  
change $\be\to\be +m$ can be compensated by a change in the  
representation $\si$: 
$$ 
V_{\La,\bar b,\si}^{\be+m} 
= 
V_{\La,\bar b,\si\otimes\si^{*}_{m}}^{\be} \; .
\tag {0.3} 
$$ 
Thus the family of $\goth A^{\Ga}$-modules is independent of the  
choice of $\be$ (allowing us to skip the superscript $\be$ on $V$). 
 
Knowing the character $\ch_{\La}$ of an $\goth A(G)$-module
$V_{\La}$ \cite {K1} we are able to calculate the $\goth
A^{\Ga}$- characters $\ch_{\La,\bar b,\si}$ of $V_{\La,\bar
b,\si}$.  Similarly, the modular transformation properties of
$\ch_{\La}$ \cite {KP2} determine those of $\ch_{\La,\bar b,\si}$
and hence the orbifold fusion rules. We point out that the group
factors of fusion coefficients $N_{\bar b_{1}\si_{1},\bar
b_{2}\si_{2},\bar b_{3}\si_{3}}$ $(\si_i\in \hat\Ga_{b_i})$ of an
affine orbifold differ from those of the associated Grothendieck
ring (see \cite {Lus} as well as the discussion in Sect.~4 of
\cite {DV$^3$}) due to multipliers $\mu(h|\Si\be_i)$ which define
(for $b_1b_2b_3=1$) 1-dimensional representations of the
intersection $\Ga_{b_1}\cap\Ga_{b_2}(\ni h)$.  This difference
shows up already for (finite) subgroups of $\SU(2)$.  For higher
rank $G$ it may yield a change of charge conjugation, as
displayed in the examples of a 1080 element subgroup of $\SU(3)$
which admits a conjugacy class of involutive elements with a
non-abelian centralizer.
 
We compute (in Sect.~4A) the asymptotic dimensions of orbifold  
characters singling out, in particular, the non-trivial orbifold  
modules. 
 
If $G$ is a simple simply-connected Lie group then the non-trivial  
elements of 
$$ 
Z=Z(G) \cap\Ga \; ,
\tag {0.4} 
$$ 
where $Z(G)$ is the center of $G$, are exceptional --- they cannot be  
written in the form \thetag {0.2} (with $\be$ satisfying (ii)). Each  
element of $Z$ (different from the group unit) is associated with a  
fundamental weight $\La_j$ satisfying $(\La_j|\th)=1$ where  
$\th$ is the highest root. We associate with it (in Sect.~4) a  
permutation of the orbifold modules which maps, in particular,
the (affine) vacuum weight $\tilde\La_0$ into $\tilde\La_j$ and
thus cannot be viewed as an automorphism (``gauge
transformation'') of the (vacuum) chiral
algebra. Knowing the action of $e^{2\pi i\La_j}$ on  
$\{V_{\La\bar b\si}\}$ we can extend our treatment to all  
exceptional elements of a $\Ga\subset \SU(n)$. The treatment of  
Ad-exceptional elements (described in Appendix A), which are  
encountered in other simple Lie groups, remains however, outside the  
scope of the present paper. 
 
Note an essential difference between coset models and orbifold
models. For the construction of a modular invariant family of
characters of coset modules it suffices to take characters of
isotypic components of all (untwisted) modules of the chiral
current algebra with respect to its chiral current subalgebra
\cite {KP0}, \cite {KP2}, \cite {KW}, \cite {K1}. In a sharp
contrast, for an orbifold model one has to take in addition
decompositions into isotypic components of twisted chiral current
algebra modules which become untwisted when restricted to the
orbifold chiral subalgebra.
 
As an application we construct a family of RCFT extensions of  
$W_{1+\infty}$ -one for each value $l(\in\Bbb N)$ of the central  
charge and for each finite subgroup $\Ga$ of $U(l)$. 
It is designed to provide a bridge between two  
current attempts to understand the fractional quantum Hall effect in  
terms of chiral conformal algebras (see \cite {FT} and \cite {CTZ}). 
 
\head \nmb.{1}. 
Chiral algebras associated with connected compact  
Lie groups \endhead 
 
We shall first recall the general notion of a chiral algebra and will  
then introduce a class of such algebras which appear to be of  
paramount importance in the study of RCFT.  
 
\subhead\nmb.{1A}. Definition of a chiral algebra. Current
algebras \endsubhead 

The mathematical concept of a vertex or chiral algebra was introduced  
by R\. Borcherds \cite {Bor} and later developed by a number of  
authors (see, e\.g\. \cite {FLM}, \cite {Go}, \cite {DGM},  
\cite {FZ}, \cite {LZ}, \cite {FKRW}, \cite {KR2}). The version  
adopted here is a specialization of \cite {K2} to $\Bbb Z$-graded  
algebras (restricting from the outset attention to fields of a given  
conformal dimension).  
 
Let $V$ be a $\Bbb Z_{+}$-{\it graded inner product space}  with a  
{\it unique vacuum state}, 
$$ 
V=\oplus_{n=0}^{\infty} V^{(n)} \; ,
\quad 
\dim V^{(0)}=1, \quad\dim V^{(n)}<\infty \; ;
\tag {1.1} 
$$ 
the gradation defines (and can be, conversely, defined by) a  
distinguished hermitian operator $L_0$ called the (chiral) {\it energy  
operator} such that  
$$ 
(L_0-n)V^{(n)}=0 \; .
\tag {1.2} 
$$ 
The unique (up to a phase factor) vector $|0\rangle\in V^{(0)}$ normalized  
by $\langle 0|0\rangle=1$ is called the {\it vacuum}. A {\it
  chiral field} $Y^{(s)}$ {\it of dimension} $s$ is a power series 
$$ 
Y^{(s)}(z)=\sum_{n\in\Bbb Z} 
Y_nz^{-n-s} \; , 
\quad 
s\in \Bbb Z_{+}
\tag {1.3} 
$$ 
with $Y_n(=Y_n^{(s)})\in End V$ satisfying the commutation  
relations (CR) 
$$ 
[Y_n,L_0]=nY_n\Leftrightarrow 
[L_0,Y^{(s)}(z)] 
=\left(z \frac{d}{dz} + s\right) 
Y^{(s)}(z) \; ,
\tag {1.4} 
$$ 
$$ 
Y_n|v_{m}\rangle = 0 
\text{ for } 
v_{m}\in V^{(m)} \; ,
\quad 
n>m \; .
\tag {1.5} 
$$ 
Equation~\thetag {1.5} expresses the postulate that {\it the vacuum is  
the lowest energy state} in $V$. In physical terms $V$ is the  
{\it vacuum space of finite energy states}. 
 
A {\it chiral (vertex) algebra} structure on $V$ is a linear map,  
called the {\it state-field correspondence}, from $V^{(s)}$ to the  
space of fields of dimension $s$:  
$V^{(s)}\ni v_s\to Y(v_s,z)=\sum_n Y_n(v_s)z^{-n-s}$, defined for all  
$s\in \Bbb Z_{+}$ and satisfying the following three axioms: 
 
V1. {\it Vacuum axioms:}  the vacuum vector  
corresponds to the identity operator in $V$ 
$$ 
Y(|0\rangle,z)=1_V \; ;
\tag {1.6} 
$$ 
the field $Y(v_s,z)$ allows to recuperate the vector $v_s$: 
$$ 
%\align 
\lim_{z\to 0} Y(v_s,z)|0\rangle 
 =v_s \; ,
\quad 
\text{i.e.} Y_{-s}(v_s)|0\rangle=v_s 
 \text{ and } 
Y_{n-s}(v_s)|0\rangle=0 \text{ for } n>0 \; .
\tag {1.7} 
%\endalign 
$$ 
 
V2. The {\it translation operator} $L_{-1}:V\to V$ defined by  
$L_{-1}v_s=Y_{-s-1}(v_s)|0\rangle$ satisfies the translation covariance  
condition: 
$$ 
[L_{-1}, Y(v_s,z)]=\frac{d}{dz}Y(v_s,z) \; .
\tag {1.8} 
$$ 
 
V3. The chiral fields are {\it local}:  
$$ 
(z-w)^n[Y(v_s,z), Y(v_{s'},w)]=0 
\text{ for }  
n\geq s+s' \; .
\tag {1.9} 
$$ 
 
Note that the inner product is not logically necessary in this  
generality. It is, however, present in all CFT (being indefinite for  
non-unitary theories) and gives rise to a distinguished  
(anti-involutive) star operation (\cite{DGM}). 
 
We shall be concerned with (orbifolds of) {\it chiral current  
algebras} described below. Let $G$ be a compact Lie group of the form  
$G= G^0\times G^{1}\times \dots\times G^s$ where $G^0=U(1)^r$, and  
$G^j,j=1,\dots,s$, are simple simply-connected groups. (Every compact Lie  
group can be viewed as a product of the above form factored by a  
finite central subgroup). Let $\goth g^j$ denote the Lie algebra of  
$G^j(j=0,\dots,s)$ and let $L=\{\omega \in \goth g^0|\exp 2\pi i
\omega =1\}$.  
We assume that $\goth g$ is equipped with a symmetric integral negative  
definite invariant bilinear form. A bilinear form on $\goth g$ is  
called {\it integral} if the length square of any $a\in i\goth g^j$  
$(j=1,\dots,s)$ such that $\exp 2\pi i \omega = 1$ (resp.\ of any
$\omega \in L$) is an even  
integer (respectively an integer). When restricted to a simple  
$\goth g^j$, the integrality property means that the bilinear form is  
equal to $k_j(v|v')$, where $k_j\in \Bbb N$ will be identified with  
the level of the affine Kac-Moody algebra $\hat\goth g^j$ and  
$$ 
(v|v')=\frac{1}{2g_j^{\spcheck}} 
tr_{\goth g^j} (ad_v ad_{v'}) 
$$ 
($g_j^{\spcheck}$ is the dual Coxeter number of $\goth g^j$;
recall that with such a normalization $(\al|\al)=2$ for long
roots $\al$).
 
In what follows we let also $k_0=1$. 
 
\subhead\nmb.{}{\it Remark 1.1}\endsubhead 
%%%\demo {Remark 1.1} 
Admitting lattice vectors $\al$  
with odd square lengths requires, as it will become clear shortly,  
extending the $\Bbb Z_+$ gradation of the vacuum space \thetag {1.1}  
to a $\frac{1}2 \Bbb Z_+$ gradation. In physical terms it amounts  
to admitting locally anti-commuting (Fermi) fields of half-integer  
conformal dimensions in the chiral algebra. Such fields do not  
describe local observables (in the strict sense of the word) and  
could alternatively be incorporated in the positive energy  
representations of a chiral Bose algebra corresponding to an even  
integral lattice. A way to get rid of Fermi fields is to  
go to a double cover of the group $G$, which makes the lattice $L$  
even. 
%%%\enddemo
 
Given the above data one can construct a chiral algebra  
$$ 
\goth A (G) =\goth A(L) 
\otimes (\otimes_{j=1}^s\goth A_{k_j} 
(\goth g^j)) \; ,
\tag {1.10} 
$$ 
called an {\it affine (or current) chiral algebra} as follows. 
 
For each $\goth g^j (j=0,1,\dots,s)$  consider its affinization  
\cite {K1}: 
$$ 
\hat\goth g^j=\Bbb C[t,t^{-1}]\otimes_\Bbb R\goth g^j 
+\Bbb C K_j \; . 
$$ 
It is a $\Bbb Z$-graded algebra, the energy operator $L_0$ acting on it as  
$-t\frac{d}{dt}$. 
 
Let $V_0(\goth g_j,k_j)$ denote the unique irreducible  
$\hat\goth g^j$-module which admits a non-zero vector $|0\rangle$ such that  
$(\Bbb C[t]\otimes\goth g^j)|0\rangle=0$ and  
$K_j|0\rangle=k_j|0\rangle$. Given an  
element $v\in\goth g^j$ and $n\in\Bbb Z$ we let $v_n$ denote the  
operator on $V_0(\goth g^j, k_j)$ corresponding to $t^n\otimes v$. Let  
$v(z)=\Si_{n\in\Bbb Z}v_n z^{-n-1}$ be the {\it current}  
corresponding to $v$. 
 
Then the chiral algebra structure $\goth A_{k_j}(\goth g^j)$ on the  
vacuum space $V_0(\goth g^j,k_j)$ is defined for each $j=1,\dots,s$  
by the following state-field correspondence: 
$$ 
Y(v_{-i_1-1}^1\dots v_{-i{_n}-1}^n|0\rangle_j,z) 
=:\partial^{i_1}v^1(z)\dots 
\partial^{i_n}v^n(z): 
/ i_1!\dots i_n! 
$$ 
(with appropriately defined normal products, \cite {K2}). 
 
The vacuum space $V$ is given by 
$$ 
V=V(L)\otimes \otimes_{j=1}^s 
V_0(\goth g^j,k_j) \; .
\tag {1.11} 
$$ 
In the next section we describe the first factor in \thetag
{1.11} and the corresponding chiral algebra structure $\goth
A(L)$ (cf.~\cite {K2}, Sect.~5.4).
 
\subhead\nmb.{1B}. Lattice vertex algebras \endsubhead 
 
Let $\Bbb C_\ep[L]$ be the twisted group algebra of the lattice $L$  
with basis $e^\om(\om\in L)$ and multiplication rules 
$$ 
e^{\om} e^{\om '}=\ep(\om,\om ')e^{\om+\om '} \; , 
\quad
\om,\om' \in L \; , 
\tag {1.12} 
$$ 
where $\ep(\om_1,\om_2)$ is a $\pm 1$-valued cocycle: 
$$ 
\align
\ep(\om,0) &= \ep(0,\om)=1 \; ,
\tag {1.13a} \\
\ep(\om,\om')\ep(\om+\om',\om{''}) &= \ep(\om,\om'+\om{''}) 
\ep(\om',\om{''}) \; .
\tag {1.13b} 
\endalign
$$ 
(Equation~\thetag {1.13a} means that $e^0=1$ and equation  
\thetag {1.13b} is equivalent to associativity.) 
 
Let $S=V_0(\goth g^0,1)$. This is the symmetric algebra over the positive  
energy subspace  
$$ 
\hat\goth g^{0(+)}=\oplus_{n<0} 
\Bbb C t^n\goth g^0 \; .
\tag {1.14} 
$$ 
(Here and below we omit the tensor product sign between $t^n$ and  
$\goth g$). The space $V(L)$ is then defined as the tensor product 
$$ 
V(L)=S\otimes \Bbb C_{\ep}[L] \; .
\tag {1.15} 
$$ 
It is an infinite direct sum (over the lattice) of irreducible  
positive energy $\hat\goth g^0$ modules with $k_0=1$: 
$$ 
V(L)=\oplus_{\om\in L}S\otimes e^{\om} \; .
\tag {1.16} 
$$ 
The corresponding ground state vectors are $1\otimes e^{\om}$; in
particular, the $V(L)$ vacuum is $|0\rangle=1\otimes 1$. The
chiral subalgebra $\goth A(S\otimes 1)$ is generated by currents.
The ground state vector $|\om\rangle\equiv 1\otimes e^{\om}$ of
each term in \thetag {1.16} is characterized by being an
eigenvector of $\hat \goth g^{0(-)}=\oplus_{n\geq0}\Bbb C
t^n\goth g^0$:
$$ 
(v_0-(v|\om))|\om\rangle =0  
 =v_n|\om\rangle,n=1,2,\dots 
\tag {1.17} 
$$ 
To display the full chiral algebra $\goth A(L)$ it remains to recall  
the {\it Frenkel-Kac construction} for the charged fields \cite {FK}: 
$$ 
Y(e^{\om},z)=e^{\om}e^{\ph_{+}(z,\om)} 
z^{\om_0}e^{\ph_{-}(z,\om)}\tag {1.18} 
$$ 
where 
$$ 
\ph_{\pm}(z,\om) 
= \pm \sum_{n=1}^\infty 
\om_{\mp n}\frac{z^{\pm n}}{n} \; .
\tag {1.19} 
$$ 
$Y(e^{\om},z)$ is a {\it primary field} with respect to the current  
subalgebra $\goth A(S\otimes 1)$: 
$$ 
[v(z), Y(e^{\om},w)]= 
(v|\om)\de(z-w)Y(e^{\om},w) \; .
\tag {1.20} 
$$ 
Let $p(\om)\in \Bbb Z/2\Bbb Z$ denote the parity of  
$(\om|\om),\om\in L$. The fields $Y(e^{\om},z)$ and $Y(e^{\om'},z)$  
are local if and only if  
$$ 
\ep(\om,\om')= 
(-1)^{p(\om)p(\om')+(\om|\om')} 
\ep(\om',\om) \; .
\tag {1.21} 
$$ 
 
The state-field correspondence for the chiral algebra $\goth A(L)$ is  
defined by  
$$ 
Y(v_{-i_1-1}^1 \dots v_{-i_{n-1}}^n\otimes e^{\om}, z) 
=:\partial^{i_1}v^1(z)\dots \partial^{i_n}v^n(z) 
Y(e^{\om},z):/_{i_1!\dots i_n!}  \; . 
\tag{1.22}
$$ 
 
Note that a 2-cocycle $\ep(\om,\om')$ satisfying \thetag {1.21}  
always exists and the chiral algebra $\goth A(L)$ is independent of  
the choice of this cocycle. 
 
The conformal properties of $Y(e^{\om},z)$ are given by 
$$ 
[T(z), Y(e^{\om},w)]= 
\de(z-w)\frac{\partial}{\partial w}Y(e^{\om},w)+ 
\frac{|\om|^2}{2}Y(e^{\om},w) 
\frac{\partial}{\partial w} \de(z-w), 
|\om|^2\equiv(\om|\om) \; . 
\tag {1.23} 
$$ 
Here $T$ is the {\it stress energy tensor} 
$$ 
T(z)=\sum_{n} L_n z^{-n-2} \ \ 
([v_m,L_n]=m v_{m+n})
\tag {1.24} 
$$ 
expressed in terms of an orthonormal basis $v^i(z)$ of $u(1)$  
currents, $(v^i|v^j)=\de_{ij}$, by the Sugawara formula 
$$ 
T(z)=\frac{1}2 \sum_{i=1}^r:(v^i(z))^2: \; ,
\tag {1.25} 
$$ 
where the normal product can be thought as a limit 
$$ 
\lim_{z_{1,2}\to z} 
\left\{ v^i(z_1)v^i(z_2)-\frac1{z_{12}^2} \right\} 
$$ 
(see comment following Eq.~\thetag {1.36} below). The fusion rules  
for the $\goth A(L)$ vertex operators have the form 
$$ 
\lim_{z\to w} 
\{(z-w)^{-(\om|\om')} Y(e^{\om},z)Y(e^{\om'},w)\} 
= \ep(\om,\om') Y(e^{\om+\om'},w) \; ;
\tag {1.26} 
$$ 
the operator product expansion for oppositely charged fields can be  
written in more detail as 
$$ 
z_{12}^{|\om|^2} 
Y(e^{\om},z_1)Y(e^{-\om},z_2)=: 
\exp \left\{ \int_{z_2}^{z_1} \om(z)dz \right\}:\; ,
\quad
z_{12}=z_1-z_2 \; .
\tag {1.27} 
$$ 
(The normal ordered exponential is defined in such a way that the  
$n$th term of its Taylor expansion is an integral over a single  
quasiprimary field of dimension $n$ --- cf.~\cite {FST}). 
 
\subhead\nmb.{1C.} Current chiral algebras associated to simple
Lie algebras \endsubhead 

The CR between two currents $Y(t^{-1} v^i, z_i)$, $i=1,2$, for two  
arbitrary elements $v^1$ and $v^2$ of $\hat\goth g$ are given by 
$$ 
\left[ Y(t^{-1}v^1,z_1), Y(t^{-1}v^2,z_2) \right] 
Y \left( t^{-1}[v^1,v^2],z_2 \right)
\de(z_{12}) 
- (v^1|v^2)\de'(z_{12}) \; . 
\tag {1.28} 
$$ 
Here and further $z_{12} = z_1-z_2$.
We shall write down for later reference these relations for the  
Chevalley-Cartan basis of a simple component $\goth g^j$ of  
$\goth g$. We shall set 
$$ 
(v_1|v_2)_k=k_j(v_1|v_2) 
\text{ for } v_1,v_2\in\goth g^j \; .
\tag {1.29} 
$$ 
Restricting attention to a simple component we skip the index $j$ on  
$\goth g$. We choose a Cartan subalgebra $\goth h$ in $\goth g$ and a basis  
$\al_i,i=1,\dots,l$, of {\it simple roots} in its dual thus  
introducing a standard partial order in $\goth h^*$, which from now  
on we shall identify with $\goth h$ using the bilinear form $(.\vert.)$. 
 
To each positive root $\al>0$ we associate a certain current representing the  
corresponding coroot $\al^{\spcheck}$: 
$$ 
H^{\al}(z)=\sum_n H_n^{\al}z^{-n-1}, 
H_0^{\al}=\al^{\spcheck}:=\frac{2\al}{|\al|^2}, 
|\al|^2=(\al|\al) \; .
\tag {1.30} 
$$ 
We shall use the positive integer {\it marks} $a_i$ (and $a_i^{\spcheck}$) of  
the Dynkin diagram of $\goth g$ which enter the  
expression for the {\it highest root} 
$$ 
\th=\sum_{i=1}^l a_i\al_i=\sum_{i=1}^l a_i^{\spcheck}\al_i^{\spcheck} 
=\th^{\spcheck}
\tag {1.31} 
$$ 
(see~\cite{K1}, Chap.~4, Tables). Their ratio relates the  
{\it Cartan matrix} $a_{ij}$ of $\goth g$ to the symmetric Gram  
matrix of the coroots, 
$$ 
(\al_i^{\spcheck}|\al_j^{\spcheck}) 
=a_{ij}\frac{a_i}{a_j^{\spcheck}} 
\quad (a_{ij}= \left( \al_i^{\spcheck}|\al_j) \right) 
$$ 
while the sum of check marks of the extended Dynkin diagram gives the  
dual Coxeter number  
$$ 
g^{\spcheck}=1+a_1^{\spcheck}+\dots+a_l^{\spcheck} 
\quad \left( \tr\left( ad_{v_1}ad_{v_2} \right) =
   2g^{\spcheck}(v_1|v_2) \right) \; .
\tag {1.32} 
$$ 
 
The set of indices $(j\in) J$ for which the exponentials  
$e^{2\pi i\La_j}$ of the corresponding fundamental weights $\La_j$  
generate the center $Z(G)$ of the simply connected group $G$ with the  
Lie algebra $\goth g$ is given by  
$$ 
J=\{j=1,\dots, l | \quad a_j=1\} \; .
\tag {1.33} 
$$ 
 
Let $E^{\al}$ be a raising or a lowering operator, depending on the  
sign of $\al$. Then the current CR \thetag {1.28} assume the form: 
$$ 
\align 
[H^{\al}(z_1),E^{\be}(z_2)] 
    & = (\al^{\spcheck}|\be) E^{\be}(z_2)\de(z_{12}) \quad
        (\al,\be  \text{~roots}) \; ,  \tag {1.34} \\ 
[H^{\al}(z_1),H^{\be}(z_2)] & =k(\al^{\spcheck}|\be^{\spcheck})\de'(z_{12}),\\ 
[E^{\al_i}(z_1),E^{-\al_j}(z_2)]& =0\text{ for } i\neq j,\\ 
[E^{\al}(z_1),E^{-\al}(z_2)] &= H^{\al}(z_2)\de(z_{12}) 
-\frac{2 k}{|\al|^2}\de'(z_{12}) \; . 
\endalign 
$$ 
 
The affine chiral algebra $\goth A_k(\goth g)$ contains the  
{\it Sugawara stress energy tensor} (see e.g.~\cite {K2} Sect.~5.7.): 
$$ 
\align 
T(z)& =\frac{1}{2h} 
       \left\{\sum_{\al > 0} \frac{(\al|\al)}{2}: 
       (E^{\al}(z) E^{-\al}(z)+ E^{-\al}(z) E^{\al}(z)):\right. \\   
& \quad \left. {} + \sum_{i=1}^l: H_i(z)H^i(z):\right\} \; , 
\quad 
h=k+g^{\spcheck} \; .
\tag {1.35} 
\endalign 
$$ 
 
Here $H^i$ and $H_i$ correspond to dual bases in the Cartan  
subalgebra: 
$$ 
H^i=\al_i^{\spcheck} \; , \quad 
H_{i}=\La_i \; , \quad 
(\al_i^{\spcheck}|\La_j)=\de_{ij} \; .
\tag{1.36} 
$$ 
The normal product $::$ can be defined by either subtracting the  
singular in $z_{12}$ part of an ordinary product $J_a(z_1)J^a(z_2)$  
or by ordering the frequency parts of the currents (inequivalent  
definitions of the normal product used in \cite {FST} and \cite {K2}  
yield the same expression for the stress energy
tensor). Equations~\thetag{1.34} \thetag{1.35} imply the Virasoro CR 
$$ 
\left[ T(z_1), T(z_2) \right] = \de(z_{12})\partial_2 T(z_2) 
+2 T(z_2) \partial_2 \de(z_{12}) 
+\frac{c}{12} \partial_2^3 \de (z_{12}) 
\left( \partial_2\equiv\frac{\partial}{\partial z_2} \right)
\tag {1.37} 
$$ 
where the Virasoro central charge exceeds the rank $l$ of $\goth
g$.  Denoting by $d (\goth g)$ the dimension of $\goth g$, we have 
$$ 
c=c_k (\goth g)=\frac{k}{h}d(\goth g)\geq l \; .\tag {1.38} 
$$ 
The positive integer $h$ entering \thetag {1.35} and \thetag {1.38}  
(the sum of the level and the dual Coxeter number) is called the  
{\it height}. The last inequality in \thetag {1.38} follows from the  
fact, that $T$ can be split into a sum of two commuting terms, the  
stress tensor $T_H$ of the Cartan subalgebra and a remainder $T_R$: 
$$ 
T=T_H+T_R \; , \quad
T_H(z)= \frac{1}{2k}\sum_{i=1}^l: 
H_i(z)H^i(z):
\tag {1.39} 
$$ 
We find, as a consequence of \thetag {1.34}, \thetag {1.35} and  
\thetag {1.36}  
$$ 
[T_H(z_1), H^i(z_2)]= \partial _2 
(\de(z_{12})H^i(z_2)) = [T(z_1), H^i(z_2)]\tag {1.40a} 
$$ 
and hence 
$$ 
[T_R(z_1), H^i(z_2)]= 0 = [T_R (z_1), T_H(z_2)]\tag {1.40b} 
$$ 
For a level $k> 1$ simply laced (A-D-E) simple Lie algebra the  
RCFT with stress energy tensor $T_R$ correspond to (generalized) G/H  
parafermions --- see \cite {KP0} and \cite {Gep}. For a simply laced level $1  
\quad\hat\goth g$ we have $c_1(\hat\goth g)=l$ and hence $T_R=0$. 
 
Note that the lattice chiral algebra $\goth A(L)$ could also contain  
a level 1 simply laced current subalgebra. In fact, each even  
(integral) lattice $L_r$ has a sublattice  
$W_{r-\nu}\oplus L_{\nu}\subset L_r$ of the same dimension $r$. Here  
$W_{r-\nu}$ is the root lattice of a direct sum of A-D-E (simple) Lie  
algebras, generated by vectors of length square 2, and $L_{\nu}$ is its  
orthogonal complement (with no vector of length square 2), so that  
$L/(W_{r-\nu}\oplus L_{\nu})$ is a finite abelien group, the glue  
group. 
 
The stress energy tensor $T(z)$ of the chiral algebra $\goth A(G)$ is  
defined as the sum of the stress energy tensors of the factors of  
$\goth A(G)$. 
 
\head\totoc\nmb0{2}. Twisted modules of a current chiral algebra \endhead 
 
\subhead\nmb.{2A}. Positive energy irreducible $\goth
A(G)$-modules \endsubhead
 
Let $\goth A(G)=\goth A(L)\otimes (\otimes_{j=1}^s)\goth
A_{k_j}(\goth g^j)$ be a current chiral algebra. Its positive
energy irreducible modules are tensor products of such modules
for each factor.
 
Let $L^*=\{\mu\in\goth g_0 |(\mu|\om)\in \Bbb Z \text{ for all }  
\om\in L\}$ be the dual lattice. It is easy to see that the positive  
energy irreducible modules over $\goth A(L)$ are labeled by the  
elements of the finite abelian group $L^*/L$ as follows. Extend the  
cocycle $\ep(\om_1,\om_2)$ to $L^*$ in such a way that \thetag {1.13}  
holds for $\om,\om'\in L$ an $\om''\in L^*$. We choose a   
vector $\mu$  of a coset of $L^*$ mod $L$, and let  
$$ 
V_{\mu}(L)=\sum_{\om\in\mu+L} S\otimes e^{\om} \; .
\tag {2.1} 
$$ 
Then Eqs.~\thetag {1.18}, \thetag {1.19} and~\thetag {1.22} define  
an irreducible positive energy module over $\goth A(L)$. 
 
As a consequence of the Sugawara formula \thetag {1.25}, the ground  
state energy $\De(\mu)$ of the module $V_{\mu}(L)$ is given by  
$$ 
\De(\mu)=\frac{(\mu|\mu)}{2}, \text{ if }  
\mu \text{ is a minimal length vector in } \mu+L \; .
\tag {2.2} 
$$

Let $\goth g$ be the Lie algebra of a simple connected compact Lie  
group and let $k$ be a non-negative integer. Then the integrable  
positive energy irreducible modules over $\hat\goth g$ of level $k$  
are labeled by the highest weight $\La$ of $\goth g$ in the lowest  
energy subspace (which is a finite-dimensional irreducible  
$\goth g$-module). We denote these modules by $V_\La(\goth g,k)$.  
Recall that $\La$ then satisfies the integrability condition  
(\cite{K1}, Chap.~12): 
$$ 
(\La|\al_i^{\spcheck})\in\Bbb Z_+ 
\text{ for } i=1,\dots, l \; ,
\quad 
(\La|\th)\leq k \; . 
\tag {2.3} 
$$ 
 
Each of the $\hat\goth g$-modules $V_{\La}(\goth g,k)$ extends to a  
$\goth A_k(\goth g)$-module and all positive energy irreducible  
$\goth A_k(\goth g)$-modules are obtained in this way \cite {FZ}. 
 
As a consequence of Eq.~\thetag {1.35}, the ground state energy (=  
conformal dimension) $\De(\La)$ of the module $V_\La(\goth g,k)$ is  
given by: 
$$ 
\De(\La)= 
\frac{(\La+2\rh|\La)}{2h} \; ,  
\text{ where } 
h=k+g^{\spcheck}, 2\rh=\sum_{\al>0}\al \; . \tag {2.4} 
$$ 
 
\subhead\nmb.{2B}. $\Bbb Z_N$-twisted current chiral algebra
modules \endsubhead
 
Let $G^0$ be the connected compact Lie group whose maximal torus is  
$U(1)^r=\Bbb R^r/L$,  
i.e.\ $L$ is the coroot lattice of $G^0$. ($G^0$  
contains the torus $U(1)^{r}$ but can, in general,  
be larger due to the presence of $\om$'s in $L$ of length square 2;  
the semi-simple part of $G^0$ is a product of simply laced  
compact simple Lie groups). Let
$$ 
G_{c}=G^0 \times G^1 
\times \dots \times G^s \; , \tag {2.5} 
$$ 
the corresponding  decomposition of Lie algebras being 
$$ 
\goth g=\goth g^0 \oplus \goth g^1 
\oplus\dots\oplus \goth g^s \; . \tag {2.6} 
$$ 
 
Let $Z^j\subset G^j$ denote the center of  
$G^j, j=1,\dots,s$, and let $Z^0=L^*/L$ 
($Z^0$ is central subgroup of $G^0$). The  
following finite subgroup of $G_{c}$ will play an important role  
in the sequel: 
$$ 
Z(G_c)=Z^0\times Z^1 \times\dots\times Z^s \; . \tag {2.7} 
$$ 
Recall (see \thetag{1.33}) that the center of a simple connected
simply connected compact Lie group consists of 1 and the elements
$$ 
\exp 2\pi i \La_j \; , 
\text{ where } j\in J \; . 
\tag {2.8} 
$$ 
 
Recall that if $Y(v_1,z)=\sum_{n\in\Bbb Z} Y_n(v_1)z^{-n-1}$ is a
field of conformal dimension 1 of a chiral algebra, then
$Y_0(v_1)$ is a derivation of $\goth A$ and $\exp Y_0(v_1)$
converge in any positive energy $\goth A$-module (see
e.g.~\cite {K2}, Sect.~4.9.). Since such derivations of the
chiral algebra $\goth A(G)$ form the Lie algebra $\goth g_{\Bbb
C}$ (the complexification of $\goth g$),   the
group $G_{c}$ acts on $\goth A(G)$ by automorphisms, and
moreover, acts on each positive energy $\goth A(G)$-module $U$ in
a consistent way (i.e.\ $g(au)=g(a)g(u)$ for $g \in G_c$,
$a\in\goth A(G)$, $u\in U)$ preserving the Hilbert metric.
 
It follows from the usual properties of the Casimir operator that the  
stress energy tensor $T(z)$ is a $G_{c}$-invariant observable: 
$$ 
T(z)\in \goth A(G)^{G_{c}} \; .\tag {2.9} 
$$ 
 
Now we recall briefly the notion of a twisted module $U$ over a  
chiral algebra~$\goth A$. Let $b$ be an automorphism of order $N$ of  
$\goth A$; then we get a $\Bbb Z/N\Bbb Z$-grading 
$$ 
\goth A=\oplus_{m\in\Bbb Z/N\Bbb Z} \goth A_m \; , 
$$ 
where $\goth A_m $ is the $\exp 2\pi i m/N $ eigenspace of $b$. A
$b$-twisted $\goth A$-module $U$ is a linear map $a\mapsto
\pi(a)$ from $\goth A$ to the space of fields with values in
End $U$ such that 
the twisted Borcherds identity holds (see e.g.~\cite{KR2}), in
particular all the CR are preserved and 
$$ 
e^{2\pi i L_0} \pi(a)e^{-2\pi i L_0} 
= (-1)^{p(a)} 
e^{\frac{2\pi i m}{N}}\pi(a) 
\text{ for } a\in\goth A_m \; . \tag {2.10} 
$$ 
If $\goth A=\goth A_0$, we get a usual (untwisted)  
$\goth A$-module. 
 
Returning to $\goth A(G)$, fix $\be\in i \goth g$ such that the  
corresponding element $b=\exp 2\pi i\be$ $\in G_{c}$ has finite  
order $N$ and choose a Cartan subalgebra of $\goth g$  
containing $i\be$. 
 
Given a positive energy representation $\pi$ of $\goth A(G)$ in a  
vector space $U$, we construct a $b$-twisted representation  
$\pi_{\be}$ in $U$ as follows. First, due to the decomposition  
\thetag {1.10} of $\goth A(G)$ and the corresponding decomposition  
$U=\otimes_{j=0}^s U^j$, it suffices to construct the $b_j$-twisted  
representation $\pi_{\be^j}$ in $U^j$ for each $j$, where  
$\be=\sum_j\be^j$ is the decomposition \thetag{2.6} and  
$b_j=\exp 2\pi i\be^j$. 
 
Next, for a positive energy representation $\pi$ of  
$\goth A(\goth g)$ we let  
$$ 
\pi_{\be}(E^{\al}(z))= \pi(E^{\al}(z))z^{-(\al|\be)} 
=\sum_{n\in \Bbb Z} E_{n+(\al|\be)}^{\al} 
z^{-n-(\al|\be)-1} \; ,
\tag {2.11} 
$$ 
and extend to the whole ${\goth A} ({\goth g})$ using the twisted
Borcherds identity.

In order to preserve CR we should have 
$$ 
\pi_{\be}(H^{\al}(z))=\pi(H^{\al}(z)) - 
\frac{k(\al^{\spcheck}|\be)}{z} \; . 
\tag {2.12} 
$$ 
Similarly, for a positive energy representation $\pi$ of $\goth A(L)$  
we let 
$$ 
\pi_{\be}(Y(e^{\om},z))=\pi(Y(e^{\om},z)) 
z^{-(\om|\be)} \; , \tag {2.13} 
$$ 
$$ 
\pi_{\be}(\om(z))=\pi(\om(z))- 
\frac{(\om|\be)}{z} \; ,
\tag {2.14} 
$$ 
and extend to ${\goth A} (L)$ using the twisted Borcherds identity.

The constructed $b$-twisted $\goth A(G)$-module will be denoted by  
$U^{(\be)}$. 
 
Going to the stress tensor, which is a sum of a torus part, $T_L$  
\thetag {1.25}, and a contribution of type \thetag {1.35},  
\thetag {1.39} for each simple factor in $G$, we shall see that only  
the Cartan part 
$$ 
T_{\goth h}=T_L+T_H \; ,
\quad 
T_L=\frac{1}2 \sum_{i=1}^r: v_i(z)^z: \; , 
\quad 
T_H=\frac1{2k} \sum_{j=1}^l: H_j H^j:(z)
\tag {2.15} 
$$ 
changes following \thetag {2.12}, \thetag {2.14} while the remainder  
$T_R$ in \thetag {1.39} is left unchanged. 
 
\proclaim{\nmb.{} Proposition 2.1}  If we set  
$$ 
\tilde T_{\goth h} 
=T_{\goth h}-\frac1{2}\be(z) 
+\frac1{2z^2}(\be|\be)_k \; , 
\quad
\tilde T_R = T_R
\tag {2.16a} 
$$ 
implying for the Laurent modes of $\tilde T$ 
$$ 
\tilde L_n 
=L_n - \be_n + \frac1{2} 
(\be|\be)_k 
\de_{n0} \; ,
\tag {2.16b} 
$$ 
where $(\be|\be)_k=k|\be|^2$ for each simple component of  
\thetag {2.6}, then $\tilde T$ and $\tilde J$ satisfy the same CR as $T$  
and $J$ ($J$ standing for any of the $\goth g$-currents,  
$H^{\al},E^{\al}, v^i$) e\.g\. 
$$ 
\left[ \tilde L_n,\tilde J(z) \right] 
=\frac d{dz} \left( z^{n+1} \tilde J(z) \right) \; .
\tag {2.17} 
$$ 
\endproclaim 
 
\demo{Proof} It is straightforward to verify that the commutator of  
$\tilde L_n$ with $\tilde E^{\al}\equiv \pi_{\be}(E_{\al})$  
\thetag {2.11} reproduces \thetag {2.17}. One further uses the fact  
that $\pi_{\be}$ defines a Lie algebra homomorphism on the currents,  
preserving their CR. The constant term in $\tilde L_0$ is obtained by  
computing $[\tilde L_1,\tilde L_{-1}]$. \quad $\square$ \enddemo 
 
\head\totoc\nmb0{3}. Twisted characters and modular  
transformations \endhead 
 
The {\it complete character} of a positive energy $\goth A(G)$-module  
$V$ is defined on the product of the upper half plane $\ta$ and the  
group $G$ as follows: 
$$ 
\ch_V(\ta,z, u)=e^{2\pi i(k,u)} 
tr{_V}
\left( q^{L_0-\frac{c_k}{24}} e^{2\pi i z} \right) 
\; .
\tag {3.1a} 
$$ 
Here 
$$ 
q=e^{2\pi i\ta} (|q|<1),z\in i\goth g,  
(k,u)=u_0+\sum_{j=1}^s 
k_ju_j \; ,
\tag {3.1b} 
$$ 
$u_j$ are auxiliary (complex) parameters, $L_0$ is the chiral energy  
operator \thetag {1.2}, \thetag {1.4} (the zero mode of the stress  
energy tensor \thetag {1.24}), $c_k$  is the Virasoro central charge  
(cf.~\thetag {1.38}):  
$$ 
c_k=r+\sum_{j=1}^s c_{k_j}(\goth g^j) \; .
\tag {3.1c} 
$$ 
 
If $V$ is irreducible then $\ch_V$ splits  
into a product of Kac-Moody and lattice characters; we reproduce  
their expressions and transformation properties separately. This will  
allow us to write down the general orbifold characters. 
 
\subhead\nmb.{3A}. Kac-Moody and lattice characters \endsubhead 

Let  
now $G$ be a connected simply connected compact Lie group with a  
simple Lie algebra $\goth g$. We shall use the following  
notation: $M^*$ is the weight lattice dual to the  
co root lattice $M$; the set of level $k$ dominant weights is  
\cite {K1} 
$$ 
P_+^k= 
\left\{ \La \in M^*|(\La|\al_i^{\spcheck})\geq0, 
i=1,\dots,l; 
(\La|\th^{\spcheck})\leq k \right\} \; ;
\tag {3.2} 
$$ 
$Q\subset M^*$ is the root lattice; the quotient $M^*/kM$ is a
finite abelian group of order $|M^*/k M|=k^l |M^*/M|$. The values
of $|M^*/ M|$ may be found e\.g\.  in \cite {KW} (in the simply
laced case $|M^*/ M|$ is the determinant of the Cartan
matrix). The Kac-Moody character $\ch_{\La}(\ta,z,u)\equiv
\ch_{V_{\La}(\goth g,k)} (\ta,z,u)$ can be expressed in terms of
classical $\Th$ functions of weight $l/2$ and certain almost
holomorphic modular forms $c^{\Lambda}_{\lambda}$, the {\it
string functions}, of opposite weight (\cite {K1}, Eq.~(12.7.12)):
$$ 
\ch_{\La}(\ta,z,u)=\sum\Sb\la\in M^*/kM \\ \La-\la\in Q\endSb 
c_{\la}^{\La}(\ta)\Th_{\la k}^M (\ta,z,u) \; ,
\tag {3.3a} 
$$ 
$$ 
\Th_{\la k}^M (\ta,z,u)=e^{2\pi i ku} 
\sum_{\ga\in M+\frac\la{k}} 
q^{\frac{k}{2}(\ga|\ga)}  
e^{2\pi i k(\ga|z)} \; .
\tag {3.3b} 
$$ 
We assume here that $iz$ is an element of $\goth g$ and  
choose a Cartan subalgebra containing~$iz$.  
 
The modular transformation  
law for $\Th$ is given by (see Theorem 13.5 of \cite {K1}): 
$$ 
\align 
S& =\pmatrix 
0 & -1 \\ 1&0 \endpmatrix: 
\Th_{\la k}^M(\ta,z,u) 
\to \Th_{\la k}^M 
\left( -\frac 1{\ta},\frac {z}{\ta},u-\frac {(z|z)}{2\ta} \right) \\ 
& = 
(-i\ta)^{l/2} 
|M^*/kM|^{-1/2} 
\sum_{\la'\in M^*/kM} 
e^{-2\pi i \frac{(\la|\la')}{k}} 
\Th_{\la' k}^M (\ta,z,u) \; ,
\tag {3.4} 
\endalign 
$$ 
$$ 
T=\pmatrix 1&1\\0&1\endpmatrix: 
\Th_{\la k}^M(\ta,z,u) 
\to \Th_{\la k}^M(\ta+1,z,u) 
= 
e^{i\pi\frac{(\la|\la)}{k}} 
\Th_{\la k}^M (\ta,z,u) \; .
\tag {3.5} 
$$ 
The characters $\ch_{\La}$ span a finite  
dimensional representation of $SL(2,\Bbb Z)$ as well (see \cite {KP} or  
Theorem 13.8 of \cite {K1}): 
$$ 
\ch_{\La} 
\left( -\frac{1}{\ta}, \frac{z}{\ta}, u-\frac{(z|z)}{2\ta} \right) 
= 
\sum_{\La'\in P_+^k} 
S_{\La\La'} \ch_{\La'} (\ta,z,u) \; ; 
\tag {3.6} 
$$ 
here the $S_{\La\La'}$ are given by the Kac-Peterson formula: 
$$ 
S_{\La\La'}=i^{|\De_+|} 
|M^*/kM|^{-1/2} 
\sum_{w\in W(\goth g)} 
\ep(w)\exp  \left\{ -2\pi i \frac{(\La+\rh|w(\La'+\rh))}{h}\right\} \; ,
\tag {3.7} 
$$ 
where $|\De_+|$ is the number of positive roots, $W(\goth g)$ is the Weyl  
group of $\goth g, \ep(w)=\pm$ according to the parity of $w$,  
$2\rh$ and $h$ are defined in \thetag {2.4}, 
$$ 
\ch_{\La}(\ta+1,z,u)=e^{2\pi im_{\La}} 
\ch_{\La}(\ta,z,u) \; , \tag {3.8} 
$$ 
$$ 
m_{\La}=\De(\La)-\frac{c_k(\goth g)}{24}\tag {3.9} 
$$ 
where $\De(\La)$ is the conformal dimension  
\thetag {2.4}, $c_k(\goth g)$ is the Virasoro central charge  
\thetag {1.38}. 
 
In the special case of $\goth g=\su(2)$ we have  
$$ 
S_{\la\la'}=\sqrt{2/h}\sin \pi  
\frac{(\la+1)(\la'+1)}{h} \; , 
\quad
h=k+2 \; ,
\tag {3.10a}  
$$ 
$$ 
m_{\la}=\frac{\la(\la+2)}{4h} - \frac{c}{24} \; , 
\quad 
c=c_k(\su(2)) = \frac{3k}{h} \; .
\tag {3.10b} 
$$ 
 
Note that for a simply laced affine algebra at level 1 (so that  
$c=l$) there is only one non-zero string function, 
which is a negative power of the Dedekind $\et$-function:  
$c_{\La}^{\La}(\ta)|_{k=1} =(\et(\ta))^{-l}$. Recall the  
transformation properties of the $\et$-function: 
$$ 
\et(-\frac {1}{\ta}) = (-i\ta)^{1/2}\et(\ta) \; , \quad  
\et(\ta+1)=e^{\pi i/12} \et(\ta) \; . 
\tag {3.11} 
$$ 
 
The matrix $S$ simplifies in this case as it coincides with that for  
the lattice characters (see \thetag {3.14} below). 
 
It is clear from the construction that the lattice character  
$\ch_{\mu}$ of the module $V_{\mu}(L)$ is given by  
$$ 
\ch_{\mu}(\ta,z,u)=(\et(\ta))^{-r} 
\Th_{\mu 1}^L(\ta,z,u) \; . \tag {3.12} 
$$ 
Here, as before, $z$ is an element of  
$\goth g^0$ and we choose a Cartan subalgebra containing  
$z$. (The expression \thetag {3.12} has, of course, the same form  
as the level 1 simply laced Kac-Moody character; it coincides with  
\thetag {3.3}, \thetag {3.11} for $L=M,r=l$). The modular  
transformation law for $\ch_{\mu}$ can be read off \thetag {3.4},  
\thetag {3.5} and \thetag {3.11} (the expression for $S$ in the  
counterpart of \thetag {3.6} being simpler than \thetag {3.7}): 
$$ 
\ch_{\mu}
\left( -\frac 1{\ta}, \frac{z}{\ta}, u-\frac{1}{2\ta}|z|^2 \right) 
=\sum_{\mu'\in L^*/L} 
S_{\mu\mu'}\ch_{\mu}(\ta,z,u)
\tag {3.13} 
$$ 
where 
$$ 
S_{\mu\mu'}=|L^*/L|^{-1/2} 
e^{-2\pi i (\mu|\mu')} \; ; \tag{3.14}
$$ 
$$ 
\ch_{\mu}(\ta+1,z,u) = e^{2\pi i m_{\mu}} \ch_{\mu}(\ta,z,u) \; , 
\quad
m_{\mu}=\De(\mu)-\frac{r}{24} \; , 
\quad
\De(\mu)=\frac{1}{2} |\mu|^2 \; . 
\tag{3.15} 
$$ 
As mentioned above, an irreducible positive energy  
$\goth A(G)$-module $V$ is the tensor product of the  
$\goth A(L)$-module $V_{\mu}(L)$ and $\goth A(\goth g^j)$-modules  
$V_{\La^j}(\goth g^j)$. Hence positive energy irreducible  
$\goth A(G)$-modules are parameterized by the set  
$$ 
P_+^k=(L^*/L)\times P_+^{k_1}\times\dots\times P_+^{k_s} \; . 
$$ 
We let $\mu=\La^0$, call $\La=\sum_{j=0}^s\La^j$ the {\it highest  
weight} of $V$, and write $V=V_{\La}$. The character of $V_{\La}$,  
$\La\in P_+^k$, is the product 
$$ 
\ch_{\La}(\ta,z,u)= 
\prod_{j=0}^s\ch_{\La^j}(\ta,z^j,u^j) \; .
\tag {3.16} 
$$ 
 
\subhead\nmb.{3B}. Modular transformations of twisted characters \endsubhead 
 
Recall that the affine chiral algebra $\goth A(G)$ is defined by the  
data consisting of a compact group $G$ and an invariant bilinear form  
on its Lie algebra $\goth g$. This invariant bilinear form looks as follows: 
$$ 
(x|y)_k\equiv \sum_{j=0}^r k_j(x^j|y^j) \; . 
\tag {3.17a} 
$$ 
We will also use the normalized invariant bilinear form 
$$ 
(x|y)=\sum_{j=0}^r (x^j|y^j) \; . 
\tag {3.17b} 
$$ 
 
Let now $\be\in i\goth g$ be such that $b=\exp 2\pi i\be\in G$ has  
finite order and choose a Cartan subalgebra of $\goth g$ containing  
$i\be$. It follows from \thetag {2.16b} and \thetag {3.1} that the  
value of the character of the $b$-twisted $\goth A(G)$-module  
$V_{\La}^{(\be)}$ at $e^{2\pi i\al}\in G$ is given by the following  
formula: 
$$ 
\align 
\ch_{\La}^{\al,\be}(\ta)& \equiv tr_{V_{\La}} 
q^{L_0-\be+\frac{1}{2}(\be|\be)_k-\frac{c_k}{24}} 
e^{2\pi i\al}\\ 
& = e^{\pi i(\al|\be)_k} 
\ch_{\La}
\left( \ta,\al-\ta\be, -\frac{1}{2}(\al-\ta\be|\be) \right) \; .
\tag {3.18} 
\endalign 
$$ 
 
Each factor in \thetag {3.18} can be written in a similar form for the  
Kac-Moody and the lattice case (assuming that $\al$ and $\be$ lie in  
the same Cartan subalgebra): 
$$ 
\align
\ch_{\La}^{\al,\be}(\ta) &= 
    \sum\Sb\la\in M^*/kM \\ \La-\la\in Q\endSb 
    c_{\la}^{\La}(\ta)\Th_{\la k}^{\al,\be} (\ta)
    \tag {3.19} \\
\ch_{\mu}^{\al,\be}(\ta) &=
    [\et(\ta)]^{-r} \Th_{\mu 1}^{\al,\be}(\ta)
\tag {3.20} 
\endalign
$$ 
where in both cases 
$$ 
\Th_{\la k}^{\al,\be} (\ta)= 
e^{i\pi k
( \al|\be)} \Th_{\la k}^{M}
\left( \ta,\al-\be \ta, \frac{1}{2}(\be\ta-\al|\be) \right) = 
\sum_{\ga\in M+\frac{\la}{k}} 
q^{\frac{k}{2}|\ga-\be|^2} 
e^{2 \pi i k(\ga|\al)}\tag {3.21} 
$$ 
(We can read off the lattice $\Th$-function from \thetag {3.21}  
setting $M=Q=L$, $\la=\mu$, $k=1$). 
 
The modular transformation law for twisted characters is deduced from  
the known transformation properties of Kac-Moody and lattice  
characters \thetag {3.6--3.9} and \thetag {3.13--3.15} using the following  
lemma (cf.~\cite {KP2} and \cite {K1}). 
 
\proclaim{\nmb.{}Lemma 3.1} Let the finite set of functions  
$\{F_i(\ta,z,u),i\in I\}$ be closed under modular transformations: 
\endproclaim 
$$ 
F_i \left(
      \frac{a\ta+b}{c\ta+d},  
      \frac{z}{c\ta+d},  
      u-\frac{c(z|z)}{2(c\ta+d)}
      \right) 
= 
\sum_{j\in I} A_{ij} F_j (\ta,z,u) \; ,
\quad
A_{ij} \in {\Bbb C} \; ,
\tag {3.22} 
$$ 
{\it for } $\pmatrix a & b \\ c& d \endpmatrix$$\in SL(2,\Bbb Z)$.  
{\it Define} 
$$ 
\align
  F_i^{\al,\be}(\ta) &= 
      F_i
      \left( \ta,\al-\ta\be, - \frac{1}{2}(\al-\ta\be|\be)
      \right) 
      \; .
      \tag {3.23} \\
\intertext{\it Then} 
F_i^{\al,\be} \left( \frac{a\ta+b}{c\ta+d} \right) &= 
    \sum_{j\in I} A_{ij} F_j^{d\al-b\be, a\be-c\al} (\ta) \; .
    \tag {3.24} 
\endalign
$$ 
 
\demo{Proof} If we set 
$$ 
\al-\be\frac{a\ta+b}{c\ta+d} 
= \frac{\tilde z}{c\ta+d} \; , 
\text{ with } 
\tilde z=d\al-b \be-(a\be-c\al)\ta 
$$ 
then 
$$ 
F_i^{\al,\be} 
\left( \frac{a\ta+b}{c\ta+d} \right) 
= F_i \left( \frac{a\ta+b}{c\ta+d},
             \frac{\tilde z}{c\ta+d}, 
             \tilde u-
             \frac{c(\tilde z|\tilde z)}{2(c\ta+d)}
      \right) 
$$ 
where $\tilde u=\frac{1}{2}(z|c\al-a\be)$. The law \thetag {3.24}  
then follows from \thetag {3.22}. \quad $\square$ 
 
It is now straightforward to apply Lemma 3.1 to \thetag {3.18} to  
find the transformation formula of twisted $\goth A(G)$-characters  
$\ch_{\La}^{\al,\be}$ using the transformation formula for complete  
characters from the previous section. Introduce the following 
notation: 
$$ 
\align
S_{\La,\La'} &= \prod_{j=0}^s S_{\La^j,\La^{\prime j}} \; ,
\tag {3.25a} \\
m_{\La} &= \sum_{j=0}^s m_{\La^j} \; ,
\tag {3.25b} 
\endalign
$$ 
where the $S_{\La^j,\La^{\prime j}}$ are given by \thetag {3.7} and  
\thetag {3.14} and the $m_{\La^j}$ are given by \thetag {3.9} and  
\thetag {3.15}. Then we have 
$$ 
\align
\ch_{\La}^{\al,\be} 
    \left( -\frac{1}{\ta} \right) 
    &= e^{2\pi i(\al|\be)_k} 
    \sum_{\La'} S_{\La\La'} 
    \ch_{\La'}^{\be,-\al}(\ta) \; , 
    \tag {3.26} \\
\ch_{\La}^{\al,\be}(\ta+1) 
    &= e^{2\pi i(m_{\La}+\frac{1}{2}(\be|\be)_k)} 
    \ch_{\La}^{\al-\be,\be}(\ta) \; . 
    \tag {3.27} 
\endalign
$$ 
\enddemo 
 
\subhead\nmb.{3C}. Small $\tau$ asymptotics of twisted characters of  
$\goth A(G)$ \endsubhead 
 
The small $\ta$ asymptotics will be used in the sequel for singling  
out non-trivial orbifold modules. Since the parameter  
$\be=\frac{2\pi}{i}\ta$ (which has a positive real part) can be  
interpreted as inverse temperature, the small $\ta$ asymptotics can  
be interpreted as the high temperature behavior. 
 
\proclaim{\nmb.{}Lemma 3.2} 
(a) The $q$-expansions of  
$\Th_{\la k}^{\al,\be}(\ta)$, $q^{c_k/24} c_{\la}^{\La}(\ta)$ and  
$q^{c_k/24}\ch_{\La}^{\al,\be}(\ta)$ involve only non-negative  
powers of $q$.  
\roster
\item"{\it (b)}" The $q$-expansion of $\Th_{\la k}^{\al,\be}(\ta)$
  has a non-zero constant term iff $\la-k\be\in kM$. This
  constant term equals $e^{2\pi i(\al|\be)_k}$.
 
\item"{\it (c)}" The $q$-expansion of $q^{c_k/24}c_{\la}^{\La}(\ta)$
  has a non-zero constant term iff $\La=k\La_j$ with $j\in J$
  (see \thetag {1.33}) or $\La = 0$, and $\la-\La\in k M$. This
  constant term equals 1. (Recall that $\La_j$ are fundamental
  weights).
 
\item"{\it (d)}" The $q$-expansion of $q^{c_k/24}\ch^{\al,\be}(\ta)$
  has a non-zero constant term iff $\La=k\La_j$ with $j\in J$ or
  $\La=0$, and $\La-k\be$ $\in k M$. This constant term equal
  $e^{2\pi i(\al|\be)_k}$.  
\endroster
\endproclaim
 
\demo{Proof} (a) and (b) are clear. (c) is proved in \cite {KW}.
(d) follows from (b) and (c) by making use of \thetag {3.19}.
\quad $\square$ \enddemo
 
The modular inversion $S$ relates low temperature to high temperature  
behavior and is a key to computing small $\ta$ asymptotics.  
 
By Lemma 3.2(a) and (d)  each term in the expansion of  
$e^{-\frac{\pi i c_k}{12\ta}} \ch_{\La}^{\al,\be}$ $(-\frac{1}{\ta})$  
vanishes exponentially for $\ta\downarrow 0$ unless $\La=k\La_j$ with  
$j\in J$ or $\La=0$, and $\La-k\be\in kM$, hence, by Lemma 3.2(d): 
$$ 
\align 
&\qquad \lim_{\ta\downarrow 0} 
\; e^{-\frac{\pi i c_k}{12\ta}} 
\ch_{\La}^{\al,\be} 
\left( -\frac{1}{\ta} \right) \\ 
&\qquad\qquad  = 
  \cases  
    e^{2\pi i(\al|\be)_k} 
         & \text{for~$\La=k \La_j$,
           $j\in J$, 
           or~$\La=0$,  
           and~$\La-k\be\in kM$,} \\ 
    0    & \text{otherwise}.  
\endcases
\tag {3.28} 
\endalign 
$$ 
 
Similarly, we have: 
$$ 
\lim_{\ta\downarrow 0} 
\; e^{-\frac{\pi i c_k}{12\ta}} 
\ch_{\mu}^{\al,\be}
\left( -\frac{1}{\ta} \right) 
= 
\cases
e^{2\pi i(\al|\be)} & \text{for~$\be-\mu\in L$,} \\ 
0                   & \text{otherwise}. 
\endcases
\tag {3.29} 
$$ 
 
Substituting $\ta$ by $-\frac{1}{\ta}$ in \thetag {3.26}:  
$$ 
\ch_{\La}^{\al,\be}(\ta)= 
e^{2\pi i(\al|\be)_k} 
\sum_{\La'} S_{\La\La'}  
\ch_{\La'} ^{\be,-\al} 
\left( -\frac{1}{\ta} \right) \; , 
$$ 
we find, using \thetag {3.28} and \thetag {3.29}, that  
$e^{-\frac{\pi i c_k}{12 \ta}}$  
$\ch_{\La}^{\al.\be}(\ta)$  
$\sim |L^*/L|^{-1/2}$  
$e^{2\pi i(\La^0|\al)}$\break
$\prod_{j=1}^s$  
$S_{\La^j,\ga_j}$,  
as $\ta\downarrow 0$,  
where $\ga_j=k_j\La_i$ with  
$i\in J^j$ or ${\ga_j=0}$,  
if $\al+\sum_{j=1}^s\ga_j$  
$\in L\oplus(\oplus_{j=1}^s M^j)$,  
and tends to 0 otherwise. Recalling that \cite {KW} 
$$ 
S_{\La,k\La_j} = S_{\La,0} 
 e^{-2\pi i(\La|\La_j)} 
\text{ if } j\in J \; ,
\tag {3.30} 
$$ 
we arrive at the following result. 
 
\proclaim{\nmb.{} Proposition 3.1} The high temperature
asymptotics of twisted ${\goth A} (G)$ characters is given by 
$$ 
\lim_{\ta\downarrow 0} 
\; e^{-\frac{\pi i c_k}{12\ta}} 
\ch_{\La}^{\al,\be}(\ta) 
= 
\cases
S_{\La,0} e^{2\pi i(\La|\be)_k} 
    & \text{if } \exp 2\pi i  \al\in Z(G) \\ 
0   & \text{otherwise}. 
\endcases 
$$ 
\endproclaim 
 
Here $Z(G_c)$ is the finite central subgroup of $G_{c}$ defined by  
\thetag {2.7} and we use~\thetag {2.8}. 
 
\goodbreak
\head\totoc\nmb0{4}. Affine orbifolds \endhead 
 
\subhead\nmb.{4A}. Projection on a centralizer's irreducible  
representation.  Asymptotic dimension \endsubhead 
 
Let as before $\be\in i\goth g$ be such that $b=\exp 2\pi i\be \in G_{c}$ has  
finite order. Given a positive energy $\goth A(G)$-module $U$, we  
have the $b$-twisted module $U^{(\be)}$ constructed in Sect.~2B{}.  
Consider the chiral subalgebra $\goth A(G)^{b}$ of fixed elements of  
$\goth A(G)$ with respect to $Ad_b$. When restricted to  
$\goth A(G)^{b}$, $U^{(\be)}$ becomes an untwisted  
$\goth A(G)^{b}$-module. This simple, but important observation  
allows one to construct in many cases all untwisted modules of a  
chiral algebra (see e.g.~\cite {KR2}). 
 
We shall use in the sequel the following orthogonality relations of  
irreducible characters of a finite group $\Ga$ : 
$$ 
\align
\frac{1}{|\Ga|} \sum_{h\in \Ga} \si^{*}(h)\si'(hg)
    &= \frac{\sigma (g)}{\sigma (1)} \de_{\si,\si'} \; , 
    \quad \si, \si' \in\hat\Ga \; , 
    \tag {4.1} \\
\frac{1} {|\Ga_{g}|} \sum_{\sigma \in \hat{\Ga}} \si^{*}(g)\si(h)
    &= \de_{\bar g,\bar h} \; , 
    \quad g,h\in\Ga \; . 
    \tag {4.2} 
\endalign
$$ 

Here and further $\hat\Ga$ denotes the set of all irreducible  
characters (= representations) of $\Ga$, $\si^*$ stands for the  
complex conjugate character, $\Ga_g$ stands for the centralizer of  
$g\in\Ga$. We shall also denote by $\bar g$ the conjugacy class of  
$g$ in $\Ga$. Recall that $|\Ga|=|\Ga_g| |\bar g|$. 
 
Let $\Ga$ be a finite subgroup of the compact group $G_{c}$. We
shall consider $\Ga$ as the {\it gauge group} of our CFT and
define the chiral subalgebra $\goth A^{\Ga}$ of gauge invariant
observables as the set of $Ad_{\Ga}$-invariant elements of $\goth
A(G)$. This is called an {\it orbifold chiral algebra}. One can
ensure that $\goth A^{\Ga}$ only contains local Bose fields (even
when $\goth A(L)$ involves fermionic vertex operators) replacing
$L$ by $L_{\text{even}}$ (the maximal even sublattice of $L$) and
$\Gamma$ by its extension by $L/L_{\text{even}}$. It will be the
objective of this section to construct a set of positive energy
representations of $\goth A^{\Ga}$ which again give rise to an
RCFT. That will be demonstrated in the next section by displaying
the $SL(2,\Bbb Z)$ properties of their characters. (This is, in
general, not the case if the subgroup $\Ga$ of $G$ is infinite.)
The $\goth A^{\Ga}$-modules in question are obtained by splitting
the twisted $\goth A(G)$-modules into $\goth A^{\Ga}$-invariant
parts.
 
%%%\subhead\nmb.{}{\it Remark 4.1} \endsubhead 
\remark {Remark 4.1}
It is clear that $\goth
A^{\Ga}=\goth A^{\Ga Z(G_c)}$ where $\Ga Z(G_{c})$ is the finite
subgroup of $G_{c}$ generated by $\Gamma$ and $Z(G_{c})$. Hence
the orbifold model does not change if we enlarge $\Ga$ by the
central group $Z(G_{c})$ and in principle we may assume that
$\Ga$ contains $Z(G_{c})$ (but we shall not do that).
\endremark
  
Pick $b\in\Ga$ and write it in the form $b=\exp 2\pi i\be$, where  
$i\be\in\goth g$. Let $\Ga_\be$ be the stabilizer of $\be$ in $\Ga$  
with respect  to the adjoint action of $\Ga$ on $\goth g$. Then the  
twisted $\goth A(G)$-module $U^{(\be)}$ becomes untwisted with  
respect to the chiral subalgebra $\goth A(G)^{\Ga_{\be}}$ of fixed  
elements with respect to $\Ga_{\be}$. It follows from the  
construction that the group $\Ga_{\be}$ acts on $U^{(\be)}$.  
 
Let $\si$ be an irreducible character of the group $\Ga_{\be}$.
It follows from \thetag {4.1} that the projector on the
$\si$-isotypic component of a representation of $\Ga_{\be}$ is
given by
$$ 
P_{\si}= 
\frac{\si(1)}{|\Ga_{\be}|} 
\sum_{h\in\Ga_\be} 
\si^*(h)h \; .
\tag {4.3} 
$$ 
The subspace $P_\si$ $U^{(\be)}$ is irreducible with respect to the  
pair $(\Ga_{\be}, \goth A(G)^{\Ga_{\be}})$. This can be proved in the same  
way as Theorem 1.1 from \cite {KR2}. It follows that the  
$\goth A(G)^{\Ga_{\be}}$-module $P_{\si} U^{(\be)}$ is isomorphic to the  
sum of $\si(1)$ copies of an irreducible module which we denote by  
$U_{\si}^{(\be)}$. 
 
Since the affine orbifold $\goth A(G)^{\Ga}$ is contained in  
$\goth A(G)^{\Ga_{\be}}$, we obtain a $\goth A(G)^{\Ga}$-module  
$U_\si^{(\be)}$ by restriction. Take now $U=V_{\La}$. It follows from  
\thetag {3.18} and \thetag {4.3} that the character  
$\ch_{\La,\si}^{\be}=tr 
q^{L_0-\be+\frac{1}{2}(\be|\be)_k-\frac{c_k}{24}}$ of the  
$\goth A(G)^{\Ga}$-module $V_{\La,\si}^{(\be)}$ is given by 
$$ 
\ch_{\La,\si}^{\be}(\ta)= 
\frac{1}{|\Ga_{\be}|} 
\sum\Sb h\in \Ga_{\be} \\ h= e^{2\pi i\al},[\al,\be]=0\endSb 
\si^*(h)\ch_{\La}^{\al,\be}(\ta) \; . 
\tag {4.4} 
$$ 
Applying the orthogonality relation \thetag {4.2}, we can invert  
\thetag {4.4}: 
$$ 
\ch_{\La}^{\alpha,\be}(\ta)= 
\sum_{h\in \hat\Ga_{\be}} 
\si(h)\ch_{\La,\si}^{\be}(\ta) 
\text{ for }  
h=e^{2\pi i\al} \; . 
\tag {4.5} 
$$ 
 
Let $Z=\Ga\cap Z(G_{c})$ denote the {\it small center} of the subgroup  
$\Ga$ of $G_{c}$. 
 
\proclaim{\nmb.{} Theorem 4.1} The orbifold character  
$\ch_{\La,\si}^{\be}(\ta)$ is nontrivial iff $\La$ and $\si$
agree on~$Z$: 
$$ 
\La|_Z = \si|_Z. \tag {4.6} 
$$ 
Provided that \thetag {4.6} holds, one has: 
$$ 
\lim_{\ta\downarrow 0} 
e^{-\frac {\pi ic}{12\ta}} 
\ch_{\La,\si}^{\be}(\ta) 
= S_{\La,0}\si(1)|Z 
|/|\Ga_{\be}| \; .
\tag {4.7} 
$$ 
\endproclaim 
 
\demo{Proof} It is clear from the construction that  
$V_{\La,\si}^{(\be)}=0$ if \thetag {4.6} fails. Furthermore, by  
Proposition 3.1 and \thetag {4.4} we have: 
$$ 
\lim_{\ta\downarrow 0} 
e^{-\frac {\pi ic}{12\ta}} 
\ch_{\La,\si}^{\be}(\ta) 
= \frac{S_{\La,0}}{|\Ga_{\be}|}  
\sum\Sb h\in Z \\ h= e^{2\pi i\al}\endSb 
\si^*(h) e^{2\pi i(\La|\al)_k} \; .
$$ 
 
It follows from the orthogonality \thetag {4.1} of characters of the  
group $Z$ that this is zero unless \thetag {4.6} holds, in which case  
it is given by the right-hand side of \thetag {4.7}. The latter is  
positive since $S_{\La,0}$ is a positive real number (see discussion  
below). 
\qed\enddemo 
 
An important characteristic of a chiral algebra module $V$ is its  
{\it asymptotic dimension} \cite {KP2} \cite {KW} and Sect.~13.13 of  
\cite {K1}. It is defined as the coefficient $a(V)$ of the leading  
term of the small $\ta$ (or high temperature) expansion of the  
specialized character $\ch_V$: 
$$ 
\ch_V(\ta)=tr_V 
q^{(L_0-\frac{c}{24})} 
\approx a(V) 
e^{\frac{\pi i c}{12\ta}}  \; .
\tag {4.8} 
$$ 
For example Theorem 4.1 states that the asymptotic dimension of
the orbifold module $V^{(\beta)}_{\Lambda ,\sigma}$ is given by
the right hand side of \thetag {4.7} provided that condition
\thetag{4.6} holds.  The positive reals $a(V)$ have multifold
interpretations.  If $\goth A(V_1)\subset\goth A(V_2)$ are two
chiral algebras (with $V_1\subset V_2$) then $a(V_2)/a(V_1)$
gives the {\it index of embedding} of the associated von Neumann
algebras (see \cite {R}, \cite {LR} and \cite {RST} and
references therein). If $V_{\La k}$ is an affine algebra module
and $V_{0 k}$ the corresponding vacuum module of height $h$ then
$a_h(\La)/a_h(0)$ is the ``quantum dimension'' of $V_{\La}$ \cite
{V}. In the case at hand the knowledge of $a(V)$ appears as an
efficient tool for singling out non-trivial orbifold modules,
and, as we shall see, for handling the splitting of reducible
modules into irreducible components.
 
An $\goth A(G)$-module $V_{\La}$ appears as an outer product of  
representation of the chiral algebras $\goth A(L)$ and  
$\goth A_{k_j}(\goth g^j)$. (We use the term {\it outer tensor product} to  
be distinguished from the tensor product of representations of a  
group $G$ that is again regarded as a representation of $G$ rather  
than as a representation of the direct product $G\times G$.) 
 
The asymptotic dimension of an outer product of representations  
obviously equals the product of asymptotic dimensions of factors.  
Hence the asymptotic dimension $a(V_{\La})$ of a $\goth A(G)$-module  
$V_{\La}$ is equal to the product of $a(V_{\mu}(L))$ and  
$a(V_{\La^j}(\goth g_j))$, $j=1,\dots,s$. The asymptotic dimension of  
lattice modules is independent of $\mu$: 
$$ 
a(V_{\mu}(L))=S_{\mu,0}=|L^*/L|^{-\frac1{2}} \; .
\tag {4.9} 
$$ 
The asymptotic dimension of Kac-Moody modules is given by (see  
\cite {KP},\cite {KW} or \cite {K1}(13.8.10)): 
$$ 
a(V_{\Lambda}(\goth g)) = S_{\Lambda ,0} = |M^{*}/hM|^{-\frac{1}{2}}
\prod_{\al>0} 2 \sin 
\frac{\pi(\La+\rh|\al)}{h} \; .
\tag {4.10} 
$$ 
This number is positive since $(\La|\al)\leq k$ and  
$(\rh|\al)\leq(\rh|\th)=g^{\spcheck}-1$, so that  
$(\La+\rh|\al)<h=k+g^{\spcheck}$. 
 
\subhead\nmb.{4B.} Affine orbifold models for non-exceptional $\Ga$.  
Action of $Z$. Modular transformations \endsubhead 
 
In order to construct a modular invariant family of $\Ga$-orbifold  
modules we need to impose some restrictions on the subgroup $\Ga$ of  
$G_{c}$. Let $Z$ be the small center of~$\Ga$.  
 
\proclaim{\nmb.{}Definition 4.1}  An element $b\in\Ga$  is called  
{\it non-exceptional} if there exists $\be(b)\in i \goth g$ such that
$ b=\exp 2\pi i\be (b) \text{ and } \Ga_b=\Ga_{\be}$.
The subgroup $\Ga$ of the compact group $G_{c}$ is called a {\it
  non-exceptional subgroup} if for any $g\in\Ga$ there exists
$\ze\in Z$ such that $\ze g$ is a non-exceptional element of
$\Ga$.  \endproclaim
 
An element $g\in G_{c}$ is called {\it Ad-exceptional} element of
$G_{c}$ if it cannot be written in the form $g=b\ze$, where $b$
is a non-exceptional element of $G_{c}$ and $\ze\in Z(G)$.
Obviously, a subgroup $\Ga$ of $G_{c}$ containing $Z(G_{c})$
(recall that, due to Remark 4.1, we may assume that $\Ga\supset
Z(G_{c})$) which does not contain Ad-exceptional elements of
$G_{c}$ is a non-exceptional subgroup of $G_{c}$.  We shall
describe Ad-exceptional elements of a compact group $G$ in
Appendix B. Here we only note that $U(n)$ contains no exceptional
elements and $\SU(n)$ contains no Ad-exceptional elements. Any
connected simple compact Lie group other than $\SU(n)$ does
contain Ad-exceptional elements.

>From now on let $\Gamma$ be a non-exceptional finite subgroup of
the compact Lie group~$G_{c}$.

It follows from the definition that for each $g \in \Gamma$ there
exists a $\zeta \in Z$ such that $b = \zeta^{-1}g$ is
non-exceptional.  Moreover for each $g$ of a conjugacy class
$\bar g$ we can choose the same $\zeta \in Z$ and a map $\beta :
\bar{b} \rightarrow i \goth g$ satisfying
$$ 
b = e^{2\pi i \beta (b)} \; ,
\quad 
\beta (hbh^{-1}) = Ad_{h} \beta (b)
\quad 
\text{for all } b \in \bar{b}, \; 
h \in \Gamma \; . 
\tag{4.11}
$$
Note that a choice of $\beta (b)$ such that $\Gamma_{b} =
\Gamma_{\beta (b)}$, determines uniquely the map $\beta$
satisfying \thetag{4.11}.

A quadruple $(\Lambda ,b,\beta ,\sigma )$ where $\Lambda \in
P^{k}_{+}$, $b$ is a non-exceptional element of $\Gamma$, $\beta$
is a map satisfying \thetag{4.11} and $\sigma \in
\hat{\Gamma}_{\beta}$ is called an {\it admissible quadruple} if
the compatibility condition condition \thetag{4.6} holds.  Due to
Theorem 4.1 the $\goth A^{\Gamma}$-module $V^{(\beta
  (b))}_{\Lambda ,\sigma}$ is nontrivial for any admissible
quadruple $(\Lambda ,b, \beta ,\sigma)$; we shall denote it by
$V^{\beta}_{\Lambda ,b,\sigma}$.  We have for any $g \in \Gamma$
the identity
$$
V^{Ad_{g}\beta}_{\Lambda ,gbg^{-1},\sigma^{g}} =
V^{\beta}_{\Lambda ,b,\sigma} \; ,
\tag{4.12a}
$$
where $\sigma^{g} \in \hat{\Gamma}_{gbg^{-1}}$ is defined by
$$
\sigma^{g}(h) = \sigma(g^{-1}hg) \; . 
\tag{4.12b}
$$
We thus obtain the first equivalence of admissible quadruples:
$$ 
(\Lambda ,b,\beta ,\sigma) \sim (\Lambda ,gbg^{-1},Ad_{g}\beta
,\sigma^{g}) \; . 
\tag{4.13}
$$
Recalling that \thetag{4.11} defines a map $\beta: \bar{b}
\rightarrow i\goth g$ and dropping the superscript $g$ on
$\sigma$ we may denote the character of the module \thetag{4.12}
by $\chi^{\beta}_{\Lambda ,\bar{b} ,\sigma}$.  Furthermore, if
$\beta (b)$ is replaced by $\beta (b)+m$ where
$$ 
e^{2\pi im} = 1 \; , 
\quad
[\beta (b),m] = 0 \; , 
\quad 
\Gamma_{\beta (b)+m} = \Gamma_{b} \; , 
\tag{4.14}
$$
then
$$ 
V^{\beta+m}_{\Lambda ,b,\sigma \otimes \sigma_{m}} =
V^{\beta}_{\Lambda , b,\sigma} \; , 
\tag{4.15a}
$$
where $\sigma_{m}$ is a $1$-dimensional representation of
$\Gamma_{b}$ defined by
$$ 
\sigma_{m}(h) = e^{2\pi i(m|\alpha)_{k}}\ \text{for}\ h =
e^{2\pi i\alpha} \in \Gamma_{b} \; . 
\tag{4.15b}
$$
Here and further we are using the following simple fact.

\proclaim{\nmb.{}Lemma 4.1} Let $G$ be a connected compact Lie
group with Lie algebra $\goth g$ and let $\lambda \in i \goth g$
be a weight, i.e. 
$$ 
e^{2\pi i(\lambda | m)} =1 \text{ if }  e^{2\pi im} = 1
\text{ and } [\lambda ,m] = 0 \; . 
\tag{4.16a}
$$
Then $\lambda$ defines a $1$-dimensional representation
$\sigma_{\lambda}$ of its stabilizer $G_{\lambda}$ by the formula
$$ 
\sigma_{\lambda}(g) = e^{2\pi i (\lambda |\gamma)}
\text{ for }
g = e^{2\pi i\gamma} \in G_{\lambda} , \; 
\gamma \in i\goth g_{\lambda} \; . 
\tag{4.16b}
$$
\endproclaim

\demo{Proof} Since the group $G_{\lambda}$ is connected, it is
generated by elements $g$ of the form \thetag{4.16b}.  The map
$\sigma_{\lambda}$ is independent of the choice of $\gamma$
representing $g$ due to \thetag{4.16a}.  If $g_{j} = e^{2\pi
i\gamma_{j}} \in G_{\lambda}$ where $\gamma_{j} \in i\goth
g_{\lambda}$, $j = 1,2$, then the Cambell-Hausdorff formula
implies $\sigma_{\lambda}(g_{1}g_{2}) = \exp \{2\pi
i[ ( \lambda|\gamma_{1}+\gamma_{2})+(\lambda |\gamma)]\}$ where
$\gamma$ is a linear combination of commutators
$[\gamma_{1},\gamma_{2}],\ldots,
[[\gamma_{i_{1}},\gamma_{i_{2}}],\ldots ,]$, 
for $i_{1},i_{2},\ldots \in \{ 1,2\}$.  But $(\lambda
|[\gamma_{1},\gamma_{2}]) = ([\lambda ,\gamma_{1}]|\gamma_{2}) =
0$ and the same holds for multifold commutators of $\gamma_{j}$.
Thus \thetag{4.16b} does indeed define a $1$-dimensional
representation of $G_{\lambda}$. \quad $\square$
\enddemo

The isomorphism \thetag{4.15} gives a second equivalence relation
for admissible quadruples:
$$ 
(\Lambda, b,\beta(b), \sigma) \sim (\Lambda ,b,\beta(b)+m
,\sigma \otimes \sigma_{m}) 
\tag{4.17}
$$ 
provided that $m \in
i\goth g$ satisfies \thetag{4.14}.  In deriving the equality of
the corresponding characters we use the identity
$$
e^{-2\pi i(m|\alpha)_{k}} \chi^{\alpha ,\beta+m}_{\Lambda} (\tau)
= \chi^{\alpha ,\beta}_{\Lambda} (\tau) \; . 
\tag{4.18}
$$

The least obvious equivalence relation appears when two
non-exceptional elements of $\Gamma$ are obtained from each other
by multiplication with an element $\zeta \in Z$.

Every element of $Z$ can be written in the form 
$$
\zeta = (\zeta^{(0)}_{j_{0}},\ldots, \zeta^{(s)}_{j_{s}}) \in
Z_{0} \times \cdots \times Z_{s} \; , 
\quad 
\zeta^{(\nu)}_{j} = e^{2\pi i
  \Lambda^{(\nu)}_{j}} \text{ or } 1 \; .
$$
Here
$\{\Lambda^{(0)}_{j}\}$ generate the finite abelian group
$L^{*}/L$; for each simple component $\goth g$ the fundamental
weight $\Lambda_{j}$ belongs to the set $j$ of indices with
$a_{j} = 1$, see \thetag{1.33}.  If both $b$ and $\zeta_{j}b$ are
non-exceptional we can write
$$
\align
k\beta(\zeta_{j}b) &= k\beta(b) + k\Lambda_{j}+m 
\tag{4.19a} \\
Ad_{\Gamma_{b}}(k\Lambda_{j}+m) &= k\Lambda_{j}+m \; ,
\quad 
e^{2\pi im} = 1 \; . 
\tag{4.19b}
\endalign
$$

We proceed to define the action $\zeta_{j}$ on $\sigma$ and
$\Lambda$.  According to Lemma 4.1 the phase factor
$$ 
\sigma_{j}(b^{\prime}) = e^{2\pi i(k\Lambda_{j} +
  m|\beta^{\prime})} 
\text{ for } b^{\prime} = e^{2\pi i\beta^{\prime}} \; ,
\quad 
Ad_{\Gamma_{b}}\beta^{\prime} = \beta^{\prime} 
\tag{4.20}
$$ 
gives rise to a $1$-dimensional
representation $\sigma_{j}$ of $\Gamma_{b}$.  The transformation
$\Lambda \rightarrow \zeta_{j}(\Lambda)$ of a lattice weight
$\Lambda \in L^{*}$ is given by $\zeta_{j}(\Lambda) = (\Lambda +
\Lambda_{j}) \mod L$.  If $\goth g$ is a simple rank $\ell$ Lie
algebra and $\Lambda \in P^{k}_{+}$, then $\zeta_{j} (\Lambda)$
is defined by
$$
\zeta_{j}(\Lambda) = k \Lambda_{j} + w_{j}\Lambda 
\tag{4.21a}
$$
where $w_{j}$ is the unique element of the Weyl group $W$ of
$\goth g$ that permutes the set $\{ - \theta,\alpha_{1},\ldots,
\alpha_{\ell}\}$ and satisfies
$$
-w_{j} \theta = \alpha_{j} \; . 
\tag{4.21b}
$$

\proclaim{\nmb.{}Theorem 4.2}  The pair of non-exceptional quadruples
$$
\split 
x &= (\Lambda, \bar{b}, \beta, \sigma) \; ,
\quad 
\left( \Lambda =
    \sum^{s}_{\nu = 0} \Lambda^{\nu} \right) \text{~and} \\ 
&\qquad
\zeta(x) = 
\left( \sum_{\nu}(w_{j_{\nu}}\Lambda^{\nu} +
k_{\nu}\Lambda_{j_{\nu}}) ,
\overline{\zeta b}, \beta + \sum_{\nu} 
\left( \Lambda_{j_{\nu}} + \frac{m_{\nu}}{k_{\nu}} \right) ,
\sigma \otimes \left( \otimes_{\nu} \sigma_{j_{\nu}} \right) \right) 
\endsplit 
\tag{4.22}
$$
gives rise to the same orbifold module leaving the same
corresponding character invariant.  \endproclaim
 
The action of the center on non-exceptional quadruples for which
$b$ and $\zeta b$ belong to the same conjugacy class $\bar{b}$
has no fixed points for level $k = 1$ in the simply laced case, 
but may have a fixed
point for higher levels.  For $G = \SU(2)$ this happens for even
$k$ and $\Lambda = \frac{1}{2} k$.  An example of this type is
provided in Sect.~6 (see Example~6.4).  The corresponding twisted
orbifold module turns out to be reducible in this case.
Understanding its splitting into irreducible components requires
more work and will be postponed to a subsequent publication.
Here we shall {\it restrict} our {\it attention to the case when}
$Z$ {\it acts on the admissible quadruples without fixed points}
(thus including all level $1$ orbifolds, all $\SU(p)$ orbifolds
(with $p$ prime) for levels not divisible by $p$, as well as all
$\Gamma \subset G$ orbifolds with a trivial small center).

We denote by ${\Cal X}$ the set of equivalence classes of all
admissible quadruples with equivalence relations \thetag{4.13},
\thetag{4.17} and \thetag{4.22}.

One may use the following description of ${\Cal X}$.  Consider
the action of $Z \times \Gamma$ on $\Gamma$ for which $Z$ acts by
multiplication and $\Gamma$ by conjugation.  Choose a subset $B
\subset \Gamma$ consisting of non-exceptional representatives of
orbits of this action, and for each $b \in B$ choose $\beta(b)
\in i\goth g$ satisfying \thetag{4.11}.  We call such $B$ an
admissible subset of $\Gamma$.  Then ${\Cal X}$ may be identified
with the set of admissible quadruples $(\Lambda , b, \beta (b),
\sigma)$, where $\Lambda \in P^{k}_{+} , b \in B, \sigma \in
\hat{\Gamma}_{b},$ with the equivalence relation that occurs only
if
$$
\zeta b = gbg^{-1}\ \text{for some}\ \zeta \in Z\ \text{and} \ g
\in \Gamma  \; .
\tag{4.23a}
$$
Then we let (cf.~\thetag{4.22}):
$$
(\Lambda , b, \beta(b),\sigma) \sim 
\left( \sum_{\nu}
(w_{j_{\nu}}\Lambda^{\nu} + k\Lambda_{j_{\nu}}), b,\beta(b) ,
\sigma \otimes \sigma^{*}_{\sum_{\nu}m_{\nu}} \otimes
\sigma_{(1-Ad_{g^{-1}})\beta(b)} 
\right)
\; .
\tag{4.23b}
$$
We can state now our main result.

\proclaim{\nmb.{}Theorem 4.3} (a) Under the modular inversion $S$
the characters $\chi_{x}(x \in {\Cal X})$ transform among themselves:
$$ 
\chi^{\beta}_{\Lambda ,\bar{b},\sigma}
\left( -\frac{1}{\tau} \right) =
\sum_{\bar{g} = \overline{\zeta^{\prime}b^{\prime}} \subset
  \Gamma} \sum_{ \Sb b=e^{2\pi i\beta} \in \bar{b}\\ 
  b^{\prime}=e^{2\pi i\beta^{\prime}}\in b^{\prime}\\ 
  [\beta,\beta^{\prime}]=0 \endSb} \sum_{\sigma^{\prime} \in
  \hat{\Gamma}_{b}} \sum_{\Lambda^{\prime}} S_{\Lambda
  \Lambda^{\prime}} S^{\beta}_{\bar{b}\sigma ,
  \bar{b}^{\prime}\sigma^{\prime}}
  \chi_{\Lambda^{\prime}, \bar{b}^{\prime}, \sigma^{\prime}} (\tau)
\tag{4.24a}
$$
where $S_{\Lambda\Lambda^{\prime}}$ is the affine Kac-Moody
$S$-matrix \thetag{3.25a}, and the ``group theoretic'' factor
looks as follows:
$$
S^{\beta}_{\bar{b}\sigma,\bar{b}^{\prime}\sigma^{\prime}} =
\frac{1}{|\Gamma|} \sum_{\Sb b \in \bar{b},b^{\prime}\in
\overline{b^{\prime}}\\ bb^{\prime}=b^{\prime}b \endSb}
\sigma^{\prime}(b)\sigma(b^{\prime})e^{-2\pi i (\beta (b)|\beta
(b^{\prime}))_{k}} \; . 
\tag{4.25}
$$
For levels and groups $\Gamma \subset G$ for which the small
center $Z$ acts without fixed points each equivalence class of
quadruples in ${\Cal X}$ is encountered $|Z|$ times and we can write
$$
\chi^{\beta}_{\Lambda,\bar{b},\sigma} 
\left( -\frac{1}{\tau} \right) =
\sum_{(\Lambda^{\prime},b^{\prime},\beta^{\prime},\sigma^{\prime})\in
{\Cal X}} |Z|S_{\Lambda \Lambda^{\prime}}
S^{\beta}_{\bar{b}\sigma,\bar{b}^{\prime}\sigma^{\prime}}
\chi^{\beta^{\prime}}_{\Lambda^{\prime},\bar{b}^{\prime},\sigma^{\prime}}
(\tau) \; . 
\tag{4.24b}
$$
\roster
\item"{\it (b)}" If the lattice $L$ is even then the characters
  $\chi_{x}$ are eigenfunctions of the modular translation $T$:
$$
\chi^{\beta}_{\Lambda,\bar{b},\sigma} (\tau +1) = \exp 
\left\{  2\pi i
\left( m_{\Lambda} + \frac{1}{2} 
  \left(
    \beta(b)|\beta(b^{\prime}) \right)_{k}
  \right) \right\}
\frac{\sigma^{*}(b)}{\sigma(1)} \chi^{\beta}_{\Lambda
  ,\bar{b},\sigma} (\tau) \; . 
\tag{4.26}
$$
They are eigenfunctions of $T^2$ also for odd lattices.

\item"{\it (c)}" The inverse matrix $S^{-1}$ is complex conjugate
  to $S$.  The matrix $S$ in \thetag{4.24b} is manifestly
  symmetric and hence also unitary.

\item"{\it (d)}" The matrix elements of $S$ and $T$ remain
  unchanged under the equivalence relations
  \thetag{4.13}, \thetag{4.17}, \thetag{4.22}, \thetag{4.23}.

\item"{\it (e)}" The charge conjugation operator $C = S^{2}$
  gives rise to the following involutive permutation of the set
  ${\Cal X}$:
$$
C: (\Lambda ,b,\beta(b),\sigma) \longmapsto
(\Lambda^{c},b^{-1},\beta(b^{-1}),\sigma^{c}) 
\tag{4.27a}
$$
where $\Lambda^{c} = -\Lambda$ in the lattice case, $\Lambda^{c}$
is the highest weight of the contragredient to $\Lambda$
representation of $\goth g$ in the affine case, and
$$
\sigma^{c}(h) = \sigma^{*}(h)e^{2\pi
  i(\beta(b)+\beta(b^{-1})|\alpha)_{k}} \text{ for }  h =
e^{2\pi i \alpha} \in \Gamma_{b} \; . 
\tag{4.27b}
$$
\endroster
\endproclaim

\demo{Proof of Theorem 4.2}  We shall content ourselves with
verifying the equality of characters for admissible quadruples
\thetag{4.22}.  The crux of the argument is the proof of the relation
$$
\chi^{\alpha, \beta + \Lambda_{j}+m}_{k \Lambda_{j}
  +w_{j}\Lambda} (\tau) = e^{2\pi i(\Lambda_{j}+m|\alpha)_{k}}
\chi^{\alpha, \beta}_{\Lambda} (\tau) 
\tag{4.28}
$$
(for an appropriate choice of $m \in M$) in the case of a (rank
$\ell$) simple Lie algebra~$\goth g$.  To prove it we use the
Weyl-Kac formula for the affine characters (\cite{K1} Chap.~10).
We first extend the coroot and weight spaces of $\goth g$ by
introducing the central element
$$
K = \sum^{\ell}_{\nu =0}
a^{\spcheck}_{\nu}\alpha^{\spcheck}_{\nu} \,
\left( \leftrightarrow \alpha^{\spcheck}_{0} = K -
  \theta^{\spcheck} \right)
\tag{4.29}
$$
and the gradation operator $d(\leftrightarrow -L_{0})$ (see
Chap.~7 of \cite{K1}).  The bilinear form $(.|.)$ is extended to
the resulting $\ell + 2$ dimensional space by
$$
(K|K) =(d|d) = 0 = (K|\alpha_{i}) = (d|\alpha_{i}) \; ,
\quad
i=1,\ldots,\ell 
\; ; \quad 
(K|d) = 1 \; . 
\tag{4.30}
$$
The Weyl-Kac formula then gives:
$$
\chi^{\alpha ,\beta}_{kd+\Lambda} (\tau) = 
\frac{\sum_{\tilde{w}} \varepsilon(\tilde{w})e^{2\pi i
    \left( \tau \left( \frac{|\beta|^{2}}{2} K-\beta-d \right)
      +\alpha|\tilde{w} \left( k d+\Lambda + \tilde{\rho}\right)
    \right) } }
{\sum_{\tilde{w}}\varepsilon(\tilde{w})e^{2\pi i
    \left( \tau \left( \frac{|\beta|^{2}}{2} K-\beta-d \right)
      +\alpha|\tilde{w}\tilde{\rho} \right)}} 
\tag{4.31}
$$ 
where the sum is over the affine Weyl group $W(\hat{\goth
  g})$, $\tilde{\rho}$ is defined by
$$
\tilde{\rho} = g^{\spcheck}d + \rho \; , 
\quad
\rho = \sum^{\ell}_{i=1}
\Lambda_{i} \; , 
\tag{4.32}
$$
and $\varepsilon (\tilde{w}) = \pm 1$ according to the parity of
$\tilde{w}$.  We define the element $\tilde{w}_{j}$ of the extended
affine Weyl group $\hat{W}$ as follows (cf.~Sect.~1 of \cite{FKW}
and Appendix~B below):
$$
\align
\quad
\tilde{w}_{j} &= t_{j}w_{j} \; , 
\quad 
t_{j}d = d+ \Lambda_{j} - \frac{|\Lambda_{j}|^{2}}{2} K \; , 
\quad
t_{j}v = v-(v|\Lambda_{j})K (v \in \goth h) \; , \\
%%\quad
& w_{j}d = d \; , 
\quad 
\tilde{w}_{j}K = K  \; , 
\tag{4.33}
\endalign
$$
(where $w_{j} \in W(\goth g)$ is defined on $\goth h$ as above).

We shall use the following three properties of $\tilde{w}_{j}$:
\roster
 \item"{(i)}" it preserves the extended Killing form;

 \item"{(ii)}" it leaves $\tilde{\rho}$ invariant;

 \item"{(iii)}" it normalizes $W(\hat{\goth g})$. 
\endroster

They allow us to write down the exponent in the numerator of \thetag{4.31} as 
$$ 
\align
  & \left( \tilde{w}_{j}
      \left\{ \tau \left( \frac{|\beta|^{2}}{2} K - \beta - d \right) 
      + \alpha \right\} 
      |w \left\{ \tilde{w}_{j} \left( k d + \Lambda \right) +
      \tilde{\rho} \right\} \right)  
  = \\ 
& \;\;  \tau \frac{k}{2} |w_{j}\beta+\Lambda_{j}|^{2} - k
    \left( w_{j} \alpha |\Lambda_{j} + (w_{j}\alpha - \tau
       \left( d+\Lambda_{j}+w_{j}\beta 
       \right) | w
       \left\{ \tilde{\rho} + \tilde{w}_{j}
           (k d+\Lambda) 
       \right\}
    \right)
\, .  
\endalign
$$
It follows that
$$
\chi^{w_{j}\alpha,w_{j}\beta+\Lambda_{j}}_{\tilde{w}_{j}(k
  d+\Lambda)} (\tau) = e^{2\pi i k(w_{j}\alpha |\Lambda_{j})}
\chi^{\alpha ,\beta}_{kd+\Lambda}(\tau) \; . 
\tag{4.34}
$$
Observing on the other hand the invariance relation
$$
\chi^{w^{-1}_{j}\alpha ,w^{-1}_{j}\beta}_{kd+\Lambda}(\tau) =
\chi^{\alpha,\beta}_{kd+\Lambda} (\tau)
$$
and the fact that $\tilde{w}_{j}(kd+\Lambda)$ can be substituted
by $\zeta_{j}(kd+\Lambda)$ in the expression \thetag{3.18} for
the character, we complete the proof of \thetag{4.28}.  It
remains to insert the result into \thetag{4.4} in order to
conclude that
$$
\chi^{\beta+\Lambda_{j}+m}_{w_{j}\Lambda
  +\Lambda_{j},\bar{\zeta}_{j}\bar{b}, \sigma \otimes \sigma_{j}}
(\tau) = \chi^{\beta}_{\Lambda,\bar{b},\sigma} (\tau)  \; ,
\tag{4.35}
$$
thus proving Theorem 4.2. \quad $\square$
\enddemo

\demo{Proof of Theorem 4.3} We use the assumption that $\Gamma$ is a non-exceptional subgroup of $G$ in order to express $h$ in the formula \thetag{4.4} for the orbifold character by a non-exceptional element $b^{\prime-1}$:
$$
h = \zeta b^{\prime -1} = e^{ 2\pi i(\alpha_{\zeta} +
  \beta^{\prime}_{-}) } \tag{4.36a}
$$
where
$$
\zeta = e^{2\pi i\alpha_\zeta} \in Z \; , 
\quad
[\alpha_{\zeta},\beta(b)]=0 \; , 
\quad 
b^{\prime -1} = e^{2\pi
  i\beta^{\prime}_{-}} \; . 
\tag{4.36b}
$$
This allows to rewrite \thetag{4.4} in the form
$$
\chi^{\beta}_{\Lambda,\bar{b},\sigma}(\tau) = \frac{1}{|\Gamma|}
\sum_{b \in \bar{b}} \sum_{\Sb
h =\zeta b^{\prime -1}\in\Gamma_{b}\\
[\beta (b),\beta'_{-}]=0
\endSb} \sigma (b')\chi^{\beta'_{-},\beta}_{\Lambda} (\tau)
\tag{4.37}
$$
where we have used the relation
$$
\chi^{\alpha_{\zeta}+\beta'_{-},\beta}_{\Lambda} (\tau) =
e^{2\pi i
(\Lambda|\alpha_{\zeta})_{k}}X^{\beta'_{-},\beta}_{\Lambda}
(\tau) 
\tag{4.38a}
$$
for $e^{2\pi i(m|\alpha_{\zeta})} =1$ whenever $m \in M$,
$[\alpha_{\zeta},m] = 0$, implying
$$
\sigma^{*}(h)
\chi^{\alpha_{\zeta}+\beta^{\prime}_{-},\beta}_{\Lambda} (\tau) =
\sigma(b^{\prime})\chi^{\beta^{\prime}_{-},\beta}_{\Lambda}(\tau)
\tag{4.38b}
$$
for $\sigma|_{Z} = \Lambda|_{Z}$ (we have also used
$\sigma^{*}(b^{\prime -1}) = \sigma(b')$).  Inserting the modular
inversion law \thetag{3.26} into \thetag{4.37} we find
$$
\chi^{\beta}_{\Lambda,\bar{b},\sigma} \left( -\frac{1}{\tau} \right) =
\frac{1}{|\Gamma|} \sum_{b\in\bar{b}} \sum_{\Sb 
h=\zeta b^{\prime -1}\\
b'\in\Gamma_{b\beta,-\beta'_{-}}
\endSb} \sigma(b')e^{2\pi i(\beta|\beta'_{-})_{k}}
\sum_{\Lambda'} S_{\Lambda\Lambda'} \chi^{\beta
  ,-\beta{'}_{-}}_{\Lambda'} (\tau) \; , \tag{4.39a}
$$
where, in view of \thetag{4.5}, we can write
$$
\chi^{\beta,-\beta'_{-}}_{\Lambda} (\tau) =
\sum_{\sigma^{-\beta'_{-}}\in\hat{\Gamma}_{b}}
\sigma^{-\beta'_{-}}(b)
\chi^{-\beta'_{-}}_{\Lambda',\bar{b}',\sigma^{-\beta'_{-}}}
\; .
\tag{4.39b}
$$ 
Finally, we would like to substitute the upper index of $\chi$ by
the phase $\beta{'}$ of $b'$ which differs from
$-\beta'_{-}$ by a coroot:
$$
b' = e^{2\pi i \beta{'}} \Rightarrow e^{-2\pi
  i(\beta'+\beta'_{-})} = 1 
\quad
([\beta'+\beta'_{-},\beta] = 0) \; . 
\tag{4.40}
$$
Applying \thetag{4.15} we obtain
$$ 
\split
&\chi^{\beta}_{\Lambda,\bar{b},\sigma}
\left( -\frac{1}{\tau} \right) \\
& \qquad =
\frac{1}{|\Gamma|} 
\! \sum_{\Sb \bar{g}\subset\Gamma \\   
          \bar{g}=\overline{\zeta' b'} \endSb} 
\! \sum_{\Sb b=e^{2\pi i\beta} \in \bar{b} \\ 
          b' = e^{2\pi i\beta'}\in \bar{b}'\\ 
          bb'=b' b \endSb}
\! \sum_{\sigma'\in\hat{\Gamma}_{b}} \sum_{\Lambda'}
\sigma(b')\sigma'(b)e^{-2\pi i(\beta|\beta')_{k}}
S_{\Lambda \Lambda'}
\chi^{\beta'}_{\Lambda',\bar{b}',\sigma'}(\tau)
\endsplit
\tag{4.41}
$$ 
where
$$
\sigma'(b) = \sigma^{-\beta'_{-}} (b)e^{2\pi
  i(\beta'+\beta'_{-}|\beta)_{k}} \; . 
\tag{4.42}
$$ 
If the small center $Z$ acts on admissible quadruples for
which $\zeta b \in \bar{b}$ without fixed points, then each term
in the sum is encountered exactly $|Z|$ times and we end up with
\thetag{4.24b}, \thetag{4.25}.

The $T$-transformation law \thetag{4.26} follows from Eq.~\thetag{3.27}: 
$$ 
\split 
\chi^{\beta}_{\Lambda,\bar{b},\sigma}(\tau+1) &=
e^{2\pi i\{ m_{\Lambda} + \frac{1}{2} (\beta|\beta)_{k} \} }
\sum_{\Sb h\in \Gamma_{b} \\ 
          h=e^{2\pi i\alpha}\\ 
          [\alpha ,\beta] =0\endSb} 
\sigma^{*}(h)\chi^{\alpha -\beta,\beta}_{\Lambda}
(\tau) \\ 
& = e^{2\pi i\{
  m_{\Lambda}+\frac{1}{2}(\beta|\beta)_{k}-(\sigma|\beta)\}}
\chi^{\beta}_{\Lambda,\bar{b},\sigma}(\tau) \; . 
\endsplit
\tag{4.43}
$$ 
Here we have used the fact that $b$ is in the center
of $\Gamma_{b}$ and $\sigma(\in \hat{\Gamma}_{b})$ is
irreducible, so that
$$
\sigma^{*}(h) = \sigma^{*}(hb^{-1})
\frac{\sigma^{*}(b)}{\sigma(1)} \tag{4.44a}
$$
where the last factor is a complex number of absolute value $1$
which can be written as
$$
\frac{\sigma^{*}(b)}{\sigma(1)} =: e^{-2\pi
i(\sigma|\beta)} \; . 
\tag{4.44b}
$$
(Equation~\thetag{4.44b} thus defines a linear functional
$(\sigma|\beta)$ in $\beta$ whose exponential agrees with the
value of $\Lambda$ on $Z$.)

Using once more Lemma 3.1 for the inverse transformation
$\pmatrix 0 & 1 \\ -1 & 0 \endpmatrix$ to \thetag{3.4} we derive
$S^{-1} = S^{*}$, where $*$ stands for complex conjugate.  The
symmetry of $S$ is manifest from the expressions for $S_{\Lambda
  \Lambda'}$ and
$S^{\beta}_{\bar{b}\sigma,\bar{b}'\sigma'}$.

To prove the invariance of $S$-matrix elements with
non-exceptional entries under the equivalence relation
\thetag{4.23} we use an extension of \thetag{3.30}:
$$
S_{\zeta_{j}(\Lambda),\Lambda'} = e^{-2\pi i
  (\Lambda_{j}|\Lambda')} S_{\Lambda\Lambda'} \; , 
\tag{4.45}
$$
(cf.~\cite{KW}) and the fact that $\sigma'$ and $\Lambda'$
coincide on the central element $\Lambda_{j}$.  To verify that
$T$ is also invariant under $\zeta_{j}$ one uses
$(\sigma|\Lambda_{j}+m) = (\Lambda|\Lambda_{j}+m)$ and
$(w_{j}\Lambda|\Lambda_{j}) = (\Lambda|w_{j}^{-1}\Lambda_{j}) =
(\Lambda|\Lambda_{j}+m')\ (m' \in M)$ to prove that the
phase
$$
\phi(k d+\Lambda ,\beta ,\sigma) = \frac{1}{2k} (kd + \Lambda +
2\tilde{\rho} |k d+\Lambda)-(\sigma|\beta) +
\frac{k}{2}|\beta|^{2}
$$
changes by an integer:
$$
\split
\Delta \phi &= \frac{k}{2} (|\Lambda_{j}|^{2} - |\Lambda_{j} +
m|^{2}) + (\Lambda|m'-m)+(\Lambda-\sigma|\Lambda_{j}+m)\\
&=(\Lambda|m'-m)-k\{(\Lambda_{j}|m)+\frac{1}{2}|m|^{2}\} \in
\Bbb Z (\Rightarrow e^{2\pi i \Delta \phi} = 1) \; .
\endsplit
$$

We finally proceed to prove \thetag{4.27}.  To this end we
compute $C = S^{2}$ by applying Lemma 3.1 to the central element
of $SL(2,\Bbb Z)$.
$$
\pmatrix 0 & -1 \\
1 & 0\endpmatrix^{2} = \pmatrix -1 & 0 \\ 0 & -1 \endpmatrix  \; .
$$
This gives
$$ 
\split (C\chi)^{\beta}_{\Lambda, \bar{b}, \sigma} (\tau) & =
\sum_{\Lambda'} C_{\Lambda\Lambda'} \frac{1}{|\Gamma_{b}|}
\sum_{b'\in\Gamma_{b}} \sigma^{*}(b')
\chi^{-\beta(b'),-\beta(b)}_{\Lambda'} (\tau) \\ 
&= \sum_{\Lambda'} C_{\Lambda\Lambda'}
  \sum_{\bar{b}'\sigma{'}} C^{\beta\ 
  \beta'}_{\bar{b}\sigma,\bar{b}'\sigma{'}}
  \chi^{\beta'}_{\Lambda',\bar{b}',\sigma'} (\tau)
  \; ,
\endsplit
$$
where $C_{\Lambda\Lambda'}= \delta_{\Lambda^{*}\Lambda'}$
is known from the modular properties of affine Kac-Moody
characters (\cite{K1} Chap.~13), while the second factor is computed to be
$$
C^{\beta\ \beta'}_{\bar{b}\sigma, \bar{b}'\sigma'} =
\delta_{ \overline{b^{-1}},\overline{{b}'} }
\delta_{\sigma^{c},\sigma'}
\delta_{\beta,-\beta'} \; .
\qquad %%\lowerthis \square
\raise-3ex\hbox{$\square$} 
\tag{4.46}
$$
\enddemo 

We note that the equivalence class $v$ of the {\it vacuum
  admissible quadruple}, i.e.~that corresponding to the vacuum
$\goth A(G)^{\Gamma}$-module, is selfconjugate:
$$
v := \text{class of } (0,1,0,1) = Cv.
$$
Note also the following formula for any $x = (\Lambda,
b,\beta(b),\sigma) \in {\Cal X}:$
$$
S_{x,v} = S_{\Lambda ,0} \frac{|\bar{b}|}{|\Gamma|} \sigma(1) \; .
$$

%%%\subhead\nmb.{}Remark 4.2\endsubhead  
\remark {Remark 4.2}
It follows from Lemma 3.2d
that the eigenvalues of $L_{0}$ are strictly positive in all
$\goth A(G)$-modules $V_{x}$, $x \in {\Cal X}$, except for the
vacuum module $V_{v}$.  The $0$-th eigenspace of $L_{0}$ in
$V_{v}$ is $\Bbb C |0\rangle$.
\endremark

%%%\subhead\nmb.{}Remark 4.3\endsubhead 
\remark {Remark 4.3}
The $\goth
A(G)^{\Gamma}$-modules $V_{x}$ and $V_{C_{x}}$ $(x \in {\Cal
  X})$ are contragredient.
\endremark

\subhead\nmb.{4C.} Fusion rules \endsubhead

We can summarize the most important features of the outcome of
the previous section as follows.

Starting with a compact Lie group $G= (\Bbb R^{r}/L) \times
G'$, where $G'$ is simply connected, and a negative
definite integral invariant bilinear form on its Lie algebra
which is even on the lattice $L$,  we have constructed for every
non-exceptional finite subgroup $\Gamma$ of $G$ a collection of
$\goth A(G)^{\Gamma}$-modules parametrized by a finite set ${\Cal
X}$.  This set is equipped with an involutive permutation $C$
(corresponding to taking a contragredient module) and a
distinguished element $v$ (corresponding to the vacuum module)
such that $Cv = v$.  We have also matrices $S = (S_{xy})_{x,y \in
{\Cal X}}$ and $T = (T_{xy})_{x, y \in {\Cal X}}$ satisfying the following
three properties, provided that the small center $Z$ acts on
${\Cal X}$ without fixed points:
\roster
\item"{(a)}" $S$ is symmetric and $T$ is diagonal,

\item"{(b)}" the map $\pmatrix 0 & -1 \\ 1 & 0 \endpmatrix
  \Rightarrow S, \pmatrix 1 & 1 \\ 0 & 1 \endpmatrix \Rightarrow
  T, \pmatrix -1 & 0 \\ 0 & -1 \endpmatrix \Rightarrow C$ give a
  unitary representation of the group $SL_{2}(\Bbb Z)$.

\item"{(c)}" $S_{xv} > 0$ for all $x \in {\Cal X}$.
\endroster

Following Verlinde~\cite{V}, introduce the {\it fusion algebra}
$A({\Cal X})= \oplus_{x \in {\Cal X}} \Bbb C x$ by the formula:
$$
xy = \sum_{z \in {\Cal X}} N_{xyz} Cz \; , 
\tag{4.47a}
$$
where the {\it fusion coefficients} $N_{xyz}$ are defined by
$$
N_{xyz} = \sum_{a \in {\Cal X}} S_{ax} S_{ay} S_{az}/S_{av} \; .
\tag{4.47b}
$$

It follows from the above properties of $S$ that the fusion
algebra $A({\Cal X})$ is a finite-dimensional  commutative
associative semisimple algebra with identity element $v$ and
involutive automorphism $C$.  All homorophisms of the algebra
$A({\Cal X})$ to $\Bbb C$ are labeled by elements $y \in X$ and given by
$$
ch_{y}(x) = S_{xy}/S_{vy} \,
(x \in {\Cal X}) 
\; .
\tag{4.48}
$$
The positive real number $ch_{v}(x)$ is the relative (= quantum) dimension.

The basic observation of [V] is that the fusion algebras arising
in a RCFT have the following fundamental property:
\roster
\item"{(d)}" $N_{xyz} \in \Bbb Z_{+}$.
\endroster

Denote by $X^{af}$ the set $P^{k}_{+}$ labeling all positive
energy irreducible representations of the chiral algebra $\goth
A(G)$ with vacuum element $v=0$, conjugation $C\Lambda
=\Lambda^{c}$, $S$-matrix $S^{af} = (S_{\Lambda\Lambda'})$ and
$T$-matrix $T^{af} = e^{2\pi i
  m_{\Lambda}}\delta_{\Lambda\Lambda'}$.  It follows
from~\cite{KP2} that the properties (a)-(c) hold, and it is a very
difficult theorem established by the efforts of many people that
(d) holds as well.  Denote by
$N_{\Lambda\Lambda'\Lambda''}(\in \Bbb Z_{+})$ the fusion
coefficients.

Similarly, let ${\Cal X}^{gr}$ denote the set of all pairs
$(\bar{g},\sigma)$, where $\bar{g}$ is a conjugacy class of
$\Gamma$ and $\sigma$ is an irreducible character of
$\Gamma_{g}$.  Let $v = (1,1)$ be the vacuum element and let
$C(\bar{g},\sigma) = (\overline{g^{-1}},\sigma^{c})$ where
$\sigma^{c}$ is defined by \thetag{4.27b}.  Let
$S^{gr}_{\bar{b}\sigma,\overline{g'}\sigma'}$ be the matrix
defined by the right-hand side of \thetag{4.25} and let
(cf.~\thetag{4.26}):
$$
T^{gr}_{\bar{g}\sigma,\overline{g'}\sigma'} = e^{2\pi
  i(\beta|\beta)_{k}} \frac{\sigma^{*}(b)}{\sigma(1)}
\; . \tag{4.49}
$$
It follows from the remarks of the previous section that the
properties (a), (b) and (c) hold.  It can be demonstrated by an
appropriate, example of an $\SU(2)$ subgroup of level $1$ (see
Example 6.5), that property (d) does not hold in general.

Lusztig \cite{Lus} studied the ``limiting'' case of our ${\Cal
  X}^{gr}$ when in \thetag{4.26}, \thetag{4.27b} and
\thetag{4.49} one sets all $\beta(b)$ equal zero and $b = g$.  In
this case (d) holds due to his interpretation of the fusion
algebra as the Grothendieck ring of the category of
$\Gamma$-equivariant vector bundles.

Whenever the center of $G$ is trivial like in the case of $E_{8}$
the fusion rules factorize: $N_{xx' x''} =
N_{\Lambda\Lambda'\Lambda''}
N_{\overline{g''}\sigma,\overline{g'}\sigma',\overline{g''}\sigma''}$.
In particular, for a level $1$ orbifold like $\goth
A_{1}(E_{8})^{\Gamma}$ they coincide with the group theoretic
fusion rules which we proceed to compute.

The following cubic sum rule tells us that the fusion coefficient
$$
N_{\bar{g}_{1}\sigma_{1},\bar{g}_{2}\sigma_{2},\bar{g}_{3}\sigma_{3}}
= \sum_{\bar{h}\sigma}
\frac{S_{\bar{g}_{1}\sigma_{1},\bar{h}\sigma}
  S_{\bar{g}_{2}\sigma_{2},\bar{h}\sigma}
  S_{\bar{g}_{3}\sigma_{3},\bar{h}\sigma}}{S_{11,\bar{h}\sigma}}
\tag{4.50}
$$
vanishes unless there are triples $g_{j} \in \bar{g}_{j}$, $j =
1,2,3$ such that $g_{1}g_{2}g_{3} = 1$. 

\proclaim{\nmb.{} Proposition 4.4} (\cite{Gor} Theorem 2.12) Let
$\bar{g}_{i}$, $i=1,2,3$, be three conjugacy classes in a finite
group $\Gamma$. The number $n_{123}$ of triples $g_{i} \in
\bar{g}_{i}$ such that $g_{1}g_{2}g_{3} = 1$ is given by
$$
n_{123} =
\frac{|\bar{g}_{1}||\bar{g}_{2}||\bar{g}_{3}|}{|\Gamma|}
\sum_{\sigma \in \hat{\Gamma}} \frac{1}{\sigma(1)}
\sigma(g_{1})\sigma(g_{2})\sigma(g_{3})
\; .
$$
\endproclaim

In deriving the fusion rules we follow~\cite{DV$^{3}$}, but compute
explicitly the phase factors.

\proclaim{\nmb.{}Theorem 4.5} The fusion rules \thetag{4.50} can
be expressed in either of the two forms:
$$
\alignat2
\quad{} &&
N_{\bar{b}_{1}\sigma_{1},\bar{b}_{2}\sigma_{2},
  \bar{b}_{3}\sigma_{3}}
&= \frac{1}{|\Gamma|} \sum_{h\in \Gamma} \sum_{\Sb b_{i} \in
  \overline{b_{i}} \cap \Gamma_{h} \\ b_{1}b_{2}b_{3} = 1\endSb}
\sigma_{1}(h)\sigma_{2}(h)\sigma_{3}(h)\mu(h|\Sigma \beta_{i})
\tag{4.51a} \\
\quad{} &&
N_{\bar{b}_{1}\sigma_{1},\bar{b}_{2}\sigma_{2},
  \bar{b}_{3}\sigma_{3}} 
&= \sum_{O_{12}} \frac{1}{|\Gamma_{b_{1},b_{2}}|} 
  \sum_{h\in\Gamma_{b_{1},b_{2}}} \sigma_{1}(h)\sigma_{2}(h)
     \sigma_{3}(h)\mu(h|\Sigma \beta_{i}) \; . 
\tag{4.51b}
\endalignat
$$
Here the multiplier $\mu$ is given by
$$
\mu(h|\Sigma \beta_{i}) = e^{2\pi i(\alpha |\sum \beta_{i})_{k}}
\; ,
\quad
\beta_{i} = \beta(b_{i}) \; ,
\quad 
h = e^{2\pi i\alpha} \; . 
\tag{4.52}
$$
The outer sum in \thetag{4.51b} is over different orbits $O_{12}$
of pairs $(b_{1},b_{2})$ under the adjoint action of $\Gamma$;
the number $|O_{12}|$ of such orbits is determined from the
relation
$$
|O_{12}| |\Gamma_{12}| = |\Gamma| \; . 
$$
\endproclaim

The {\it proof} uses the form 
$$
S_{\bar{g}_{j}\sigma_{j},\bar{h}\sigma} = \frac{1}{|\Gamma_{h}|}
\sum_{b_{j} \in \bar{b}_{j} \cap \Gamma_{h}}
\sigma_{j}(h)\sigma(b_{j})e^{-2\pi i(\alpha|\beta_{j})_{k}}
\tag{4.53}
$$
of \thetag{4.25} for the three factors in the numerator of
\thetag{4.47} and reduces to a straightforward application of
Proposition 4.4 (noting the conjugation invariance of $\mu$).
(For $x_{3} = v$ (the vacuum module) $\bar{b}_{3} = \bar{1}$,  
$\sigma_{3} = 1$ $(\beta_{3} = 0)$ we reproduce as a special case
the charge conjugation matrix \thetag{4.46}:
$C_{\bar{g}_{1}\sigma_{1},\bar{g}_{2}\sigma_{2}} =
N_{\bar{g}_{1}\sigma_{1},\bar{g}_{2}\sigma_{2},\bar{1}1}$.)
 
The multiplier \thetag{4.52} does not depend on the choice of the
phase $\alpha$ of $h$ provided it belongs to the stabilizer
$\frak g_{b_{1},b_{2}}$ of the pair $(b_{1},b_{2})$ in $\frak g$;
$\mu$ thus defines a representation of $\Gamma_{b_{1},b_{2}}$
according to Lemma 4.1 applied to $G = G_{b_{1}},\lambda =
\beta_{2}$.

\head\totoc\nmb0{5}. $U(l)$ orbifolds as RCFT extensions of $W_{1+\infty}$  
\endhead 
 
What is now called $W_{1+\infty}$ first appeared as the (unique
nontrivial) central extension $\hat \Cal D$ of the Lie algebra
$\Cal D$ of differential operators on the circle \cite {KP1}. Its
representation theory (including the classification of
quasi-finite positive energy representations) was developed in
\cite {KR1,2} and \cite {FKRW}. It has also attracted the
attention of physicists, in particular, the most degenerate
`minimal series' of unitary representations of $W_{1+\infty}$ of
\cite {FKRW} are being applied in the study of quantum Hall
fluids \cite {CTZ}. (More reference to both physical applications
and related mathematical developments are cited in the above
papers and in the bibliography to \cite {AFMO}.)  The vacuum
$\hat \Cal D$-module (corresponding for unitary representation to
a positive integer central charge $c=l$) was shown \cite {FKRW}
to carry a canonical chiral (vertex) algebra structure.  The
resulting chiral algebra $W^{(l)}_{1+\infty}$ was described in
\cite {BGT} in terms of a series of quasi primary fields of
dimension $\nu+1, \nu=0,1,\dots$:
$$ 
V^{\nu}(z)= \! \sum V_n^{\nu}z^{-n-\nu-1} \, ,
\;
[L_m,V^{\nu}(z)]=z^m \left( z\frac{d}{dz}+ (m+1)(\nu+1) \right)  
V^{\nu}(z) \, , 
\;
m=0 \, , \, \pm 1, 
\tag {5.1a} 
$$ 
satisfying local CR such that  
$$ 
[V_m^{\mu},V_n^{\nu}] =(\nu m-\mu n) 
V_{m+n}^{\mu+\nu-1} 
+\dots+ 
c\frac{(\nu !)^4}{(2\nu)!} 
\left( 
\aligned 
& m +\nu \\ 
& m-\nu-1 
\endaligned 
\right) 
\de_{m,-n} 
\de_{\mu\nu} \; , 
\quad
c=l \; . 
\tag {5.1b} 
$$ 
The (quasi finite) irreducible positive energy modules $V_{\vec
  r}$ of $W^{(l)}_{1+\infty}$ are characterized by $l$ {\it
  exponents} (see \cite {KR1,2}) $\vec r=(r_1,\dots, r_l)$ that
take real values for unitary representations. Each $V_{\vec r}$
has a cyclic minimal energy vector $|\vec r\rangle$ such that
$$ 
V_n^{\nu}|\vec r\rangle=0  \text{ for } 
n=1,2,\dots, 
\{ V_0^{\nu}-v_{\nu}(\vec r)\} 
|\vec r\rangle =0\tag {5.2} 
$$ 
where 
$$ 
\split
v_0(\vec r) &= 
\sum_{i=1}^l r_i \; , \\
v_{\nu}(\vec r) &= \frac{(\nu-1)!\nu!}{(2\nu)!} 
\sum_{j=0}^{\nu-1} 
\binom \nu {j} 
\binom \nu {j+1} 
\sum_{i=1}^l 
(r_i-j)\dots (r_i+\nu-j-1)r_i
\; .
\endsplit
\tag {5.3} 
$$ 
In particular $V^1(z)=T(z)$ so that the ground state energy  
eigenvalue is $v_1(\vec r)=\frac{1}{2}\vec r^2$ =  
$\sum_{i=1}^lr_i^2$: 
$$ 
\left( L_0-\frac {1}{2}\vec r^2 \right) 
|\vec r\rangle =0 
\;\;
(L_0=V_0^1) \, , 
\;\;
v_2(\vec r)=\frac{1}{3} \sum_{i=1}^l r_i^3 \; , 
\;\;
v_3(\vec r)=\frac {1}{4} 
\sum_i (r_i^4+\frac {1}{5}r_i^2) \, .
\tag {5.4} 
$$
The vacuum $\hat \Cal D$-module contains for $c=l\in \Bbb N$ a
unique singular vector of degree $l+1$ such that the quotient by
the submodule generated from this singular vector is irreducible
\cite {KR1}. This irreducible quotient (together with its chiral
algebra structure) is isomorphic to a (level $l$) $W(u_l)$ vacuum
module --- see~\cite {FKRW}, Sect.~5. As a result, any
irreducible representation of $W^{(l)}_{1+\infty}$ has a canonical
structure of an irreducible representation of $W(u_l)$ of level
$l$, and all irreducible representations of $W(u_{l})$ with
central charge $l$ arise in this way.
 
Any $V_{\vec r}$ splits into a tensor product of a $W(\su(l))$
module of central charge $l-1$ and a $W^{(1)}_{1+\infty}$-module.
To see this we rescale the $u(1)$ current and split the stress
energy tensor as in \thetag {1.39}:
$$ 
J(z)=\frac{1}{\sqrt{l}} V^0(z) \; ,
\quad 
T(z)=T_J(z)+T_{su}(z) \; , 
\quad
T_J=\frac{1}{2}:J^2: 
\left( =\frac{1}{2l}:(V^0)^2: \right) \; ,
\tag {5.5a} 
$$ 
$$ 
[T_{su}(z),J(w)]=0 \; .
\tag {5.5b} 
$$ 
The minimal eigenvalue of the energy of the second term, $L_0^{su}$ is  
then given by the difference 
$$ 
\frac{1}{2}\vec r^2-\frac{1}{2l} 
\left( \sum_{i=1}^r r_i \right)^2 = 
\frac{1}{2l} \sum_{i<j} 
(r_i-r_j)^2 =: \om_l(\vec r)
\tag {5.6} 
$$ 
A $W^{(l)}_{1+\infty}$-module $V_{\vec r}$ is {\it degenerate} if
some of the differences $r_i-r_j$ are integer. It is {\it
  maximally degenerate} if all $r_i-r_j$ are integer (such
representations are termed minimal~\cite{CTZ}); the
representation of the second $(su)$-factor is indeed then a limit
of the Zamolodchikov-Fateev-Lukyanov $W_l(p)$-models of central
charge $c=(l-1) \left\{ 1-\frac{l(l+1)}{p(p+1)} \right\}$ as
observed in~\cite{CTZ}.  Since every $V_{\vec r}$ can be viewed
as a tensor product of maximally degenerate (including $c=1$)
modules we shall turn our attention to the case of integer
$r_i-r_j$. Assume that $r_i-r_j\in \Bbb Z$, we then arrange the
$r_i$'s in a decreasing order and denote the set of such $\vec
r$'s as $P^+$:
$$ 
P^+=\{\vec r\in \Bbb R^l| 
r_1\geq r_2\geq\dots\geq r_l, 
r_i-r_j\in\Bbb Z\} \; .
\tag {5.7} 
$$ 
If we interpret the ordered set $\La=(\la_1,\dots,\la_{l-1})$ of  
differences 
$$ 
\la_i=r_i-r_{i+1} \; , 
\quad
i=1,2,\dots,l-1 \tag {5.8} 
$$ 
as defining a highest weight of $\SU(l)$, then for the fundamental  
weights $\La_1=(1,0,\dots,0),\dots,\La_{l-1}=(0,\dots,0,1)$ the  
ground state energy eigenvalues \thetag {5.6} coincide with the level  
1 eigenvalues of the $\hat{su}_l$ current algebra $\goth A_1(\su(l))$: 
$$ 
\om_l(\vec r(i))= 
\frac{(\La_i+2\rh|\La_i)}{2(l+1)} 
\text{ for } 
r_j^{(i)}-r_{j+1}^{(i)}=\de_{ij}\tag {5.9} 
$$ 
(which can be verified by a direct computation). It is natural to  
expect that the $W(su_l)$ representations of such weights obey  
fusion rules given by the tensor product expansion formulae for  
$\SU(l)$ (see Conjecture~6.1 of \cite {FKRW}). 
 
It follows that a CFT with chiral algebra $W(su (l))$ and a  
highest weight module $V_{\vec r}$ with $r_i-r_j$ non-zero integers  
has an infinite number of sectors and hence is not a rational CFT.  
(We are using here the basic property of any quantum field theory to  
be closed under fusion.) This `irrationality' can also be seen from  
an analysis of the characters of these representations (computed in  
\cite {FKRW}). The orbifold construction of the previous sections  
allows to define a large class of RCFT extensions of $W_{1+\infty}$  
with the same stress energy tensor. 
 
In fact the embedding of the vacuum module of
$W^{(l)}_{1+\infty}$ into the Fock space $\Cal F_l$ of $l$ free
complex fermion fields, used from the outset in \cite {FKRW} and
\cite {KR2}, does provide one such (chiral superalgebra)
extension. So does its even (bosonic) part which coincides with
the level 1 current algebra of the rank $l$ (even) orthogonal
group $\goth A_1(\so(2l))$. (Indeed, if we separate the real and
imaginary part of the free fermions writing them as
$$ 
\ps_j=\frac{1}{\sqrt 2} 
(\ph_{2j-1}-i\ph_{2j}) \; ,
\quad 
j=1,\dots,l \; ,
\quad  
\text{then } 
J_{jk}(z)= i\ph_{j}(z)\ph_k(z) \,
(j<k)
\tag {5.10} 
$$ 
satisfy the commutation relations of level 1 $\so(2l)$ currents. The  
complex structure selects a Cartan subalgebra that includes $V^0$: 
$$ 
H^j(z)=:\ps_j^*(z)\ps_j(z): 
= J_{2 j-1, 2j}(z) \; ,
\quad 
V^0(z)=\sum_{j=1}^l 
H^j(z) \; .
\tag {5.11} 
$$ 
Then we can define $W^{(l)}_{1+\infty}$ as the $U(l)$-invariant
subalgebra of $\goth A_1(\so(2l))$ $(u(l))$ and $\so(2l)$ sharing
the same Cartan subalgebra). A more general RCFT extension of
$W^{(l)}_{1 + \infty}$ is provided by the chiral algebra
associated with the compact group $U(l)$, equipped
with a lattice structure $Q$ (see Sect.~1). Here $Q$ is an
$l$-dimensional even integral lattice whose sublattice of vectors
of length square 2 includes the (rank $l-1$) $\su(l)$ lattice.
(The root lattices of rank $l$ semi-simple Lie algebras-
$\so(2l)$, $\su(l+1)$, $\su(l)\oplus \su(2)$) --- appear then as
special cases. Note that the $\su(l)$ Cartan currents are
orthogonal to $V^0$ \thetag {5.11} (or $J$ \thetag {5.5}); they
are
$$ 
H^{\al_i}(z)=H^i(z)- H^{i+1}(z) \; , 
\quad
i=1,\dots, l-1 \; , 
\tag {5.12} 
$$ 
$\al_1,\dots,\al_{l-1}$ being the simple roots of $\su(l)$.  Any
of the extensions $\goth A(Q)$ of $W^{(l)}_{1+\infty}$ where $Q$
is a (rank $l$) lattice with the above properties admits a finite
set of positive energy CFT representations whose characters span
a (finite dimensional) representation of $SL(2,\Bbb Z)$. All
these extensions involve, in particular, $l$ commuting $u(1)$
currents and can be thus related to the approach of Fr\"ohlich,
Thiran et al.\ to the quantum Hall effect (see \cite {FT} and
references therein). A large family of intermediate observable
algebras is provided by $\Ga$ orbifolds of $\goth A(Q)$ where
$\Ga$ is any finite subgroup of $U(l)$. If $\Ga$ is not contained
in any proper Lie subgroup of $\SU(l)$ then $\goth A^{\Ga}$ only
involves a single $u(1)$ current --- the one belonging to
$W^{(l)}_{1+\infty}$. Such $\goth A^{\Ga}$ could be viewed as
RCFT extensions of minimal $W^{(l)}_{1+\infty}$ models (exploited
in \cite {CTZ}).
 
We proceed to state the precise results for the Fock space $\Cal F_l$  
of $l$ free (complex) fermions and its orbifolds. 
 
\proclaim{\nmb.{} Theorem 5.1 \cite {FRKW}} The fermion Fock space  
$\Cal F_l$ viewed as a representation of the pair $(U(l),  
W^{l}_{1+\infty})$ splits into an infinite direct sum of tensor products 
$$ 
\Cal F_l=\oplus_{\vec r\in P_+} 
F(\vec r)\otimes L(\vec r) \; , 
\tag {5.13} 
$$ 
where $P_+=\{\vec r=(r_1,\dots, r_l) 
\in \Bbb Z^ l | r_1\geq \cdots \geq r_l\}$, $F(\vec r)$ is the finite  
dimensional irreducible $U(l)$-module of highest weight $\vec r$,  
$L(\vec r)$ is a unitary $W^{(l)}_{1+\infty}$ positive energy module with  
exponents $\vec r$ and specialized character 
$$ 
\ch_{\vec r}(\ta)=tr_{L(\vec r)} q^{L_{0}-\frac {l}{24}} 
= 
q^{\frac{1}{2}\vec r^{2}} 
\et^{-l}(\ta) 
\prod_{1\leq i<j\leq l} 
(1-q^{r_i-r_j+j-i}) \; .
\tag {5.14} 
$$                                     
\endproclaim 
 
The following result is a specialization of Theorem 4.3 applied to  
the chiral algebra $\goth A(\Bbb Z^l)^{\Ga}$ where $\Bbb Z^l$ is the  
integral lattice with the standard bilinear form, and $\Ga$ is a  
finite subgroup of $U(l)$. Recall that $\goth A(\Bbb Z^{l})$ has a unique  
irreducible representation, hence we may skip the index $\Lambda$. 
 
\proclaim{\nmb.{} Theorem 5.2} Let $\Ga$ be a finite subgroup of  
$U(l)$. Write each $b\in\Ga$ in the form $b=\exp 2\pi i\be$ where  
$i\be\in u(l)$ is fixed by $Ad_{\Ga_b}$. Let $\{\be_i (\vec r)\}$  
denote the set of eigenvalues of $\be$ in $F(\vec r)$. Given an  
irreducible character $\si$ of $\Ga_b$, let  
$$ 
m_{\vec r,\si,\be}(q)= 
q^{\frac{1}{2}(\be|\be)} 
\sum_i m_{\vec r,\si,\be_i(\vec r)} 
q^{-\be_i(\vec r)} \; , 
$$ 
where $m_{\vec r,\si,\be_i(\vec r)}$ is the multiplicity of $\si$  
in the $\be_i(\vec r)$-eigenspace of $\be$ in $F(\vec r)$. Then the  
$\goth A(\Bbb Z^{l})^{\Ga}$-characters can be written in the following  
form: 
$$ 
\ch_{\bar b,\si}^{\be} (\ta)= 
\sum_{\vec r\in P^+} 
m_{\vec r,\si,\be} 
(q) \ch_{\vec r}(\ta) \; .
\tag {5.15} 
$$ 
All these characters are modular functions and their $\Bbb C$-span is  
invariant under the transformation $\ta\mapsto -\frac{1}{\ta}$. 
\endproclaim 
 
In particular, for $\bar b=\bar 1$, we have  $\be=0$ and all  
$m_{\vec r,\si\be}$ $(q)\in\Bbb Z_+$ and we find the characters of  
untwisted orbifold modules, which, unlike $\ch_{\vec r}$ are modular  
functions of~$\ta$. This special case of Theorem 5.2 provides a  
family of solutions to the following problem: find non negative  
integers $n(\vec r)$ such that 
$$ 
\sum_{\vec r\in P^+} n(\vec r) \ch_{\vec r}(\ta) 
$$ 
is a modular function of $\ta$. Each pair  $\Ga\subset U(n)$ ($\Ga$  
finite subgroup), $\si\in\hat\Ga$ gives a solution to this problem with  
$n(\vec r)=n_{\Ga}^{\si}(\vec r)$ being the multiplicity of $\si$  
in $F(\vec r)$ viewed as a $\Ga$-module. 
 
\demo{Proof of Theorem 5.2} In view of \thetag{4.4} and
\thetag{3.18} we can write
$$ 
\ch_{\bar b,\si}^{\be}(\ta) 
=\frac{1}{|\Ga_{b}|}
\sum\Sb a\in\Ga_b \\ a=\exp 2\pi i \al\\ [\alpha ,\beta] = 0\endSb
\si^*(a) \ch(\ta,\al-\be\ta,\frac{1}{2}(\be|\be\ta))
\; ,
\tag {5.16}
$$ 
where, due to \thetag {5.13} 
$$ 
\ch(\ta,z,u)= e^{2\pi i u} 
\sum_{\vec r\in P^{+}} 
\ch_{\vec r}(\ta) 
tr_{F(\vec r)} 
e^{2\pi i z} \; .
\tag {5.17} 
$$ 
Hence we have: 
$$ 
\ch_{\bar b,\si}^{\be} (\tau) = 
\sum_{\vec r\in P^+} 
\ch_{\vec r}(\ta) 
q^{\frac{1}{2}(\be|\be)} 
\sum_{a\in\Ga_b} 
\si^*(a)tr_{F(\vec r)} 
(a q^{-\be})
\; .
\tag {5.18} 
$$ 
Since $\Ga_b$ fixes $\be$, each eigenspace of $\be$ in $F(\vec r)$  
is $\Ga_b$-invariant. The contribution of the  
$\be_i(\vec r)$-eigenspace to the inner sum of \thetag {5.18} is  
clearly equal $m_{\vec r,\si,\be_i(\vec r)}q^{-\be_i(\vec r)}$. This  
proves \thetag {5.15}. \quad $\square$
\enddemo 
 
%%%\subhead{\nmb.{} Remark 5.1}\endsubhead 
\remark {Remark 5.1}
Theorem 5.2 can be  
generalized to any simply laced simple Lie algebra $\goth g$ of   
rank $l$ and $\La\in P_+^1$. Namely, formula \thetag {5.15} holds for  
any non-exceptional element $b$, where the sum is taken over  
$\la\in(\La+Q) \cap  P^1_+$, and (see \cite {K1}, Exercise 12.17): 
$$ 
\ch_{\la}(\ta)=q^{\frac{1}{2}(\la|\la)} 
\et^{-l}(\ta) 
\prod_{\al>0} 
\left( 1-q^{(\la+\rh|\al)} \right) \; . 
$$ 
We have: 
$$ 
\ch_{\La,\bar b,\si}^{\be}(\ta)= 
\sum_{\la\in(\La+Q)\cap P_+} 
m_{\la,\si,\be} 
(q)\ch_{\la}(\ta) \; .
\tag {5.19} 
$$ 
The character $\ch_{\La,\si,b}^{\be}(\ta)$ is a modular function  
and their $\Bbb C$-span is $SL_2$ $(\Bbb Z)$-invariant provided that  
$\Ga$ is a non-exceptional finite subgroup of our simple Lie group. 
\endremark
 
%%%\subhead{\nmb.{} Remark 5.2}\endsubhead 
\remark {Remark 5.2}
Taking $\Ga=\{1\}$ in  
Remark 5.1 we arrive at the following curious identity by comparing  
two expressions for $\Ga$-orbifold characters for each weight $\La$  
and real number $m$:  
$$ 
|\{\la\in\La+Q |(\la|\la)=m \}|= 
\sum\Sb \la\in\La+Q \\ (\la|\la)=m \endSb 
\prod_{\al>0} 
\frac{(\la+\rh|\al)}{(\rh|\al)} \; . 
$$ 
\endremark

\head\totoc\nmb0{6.} Examples.  \endhead

\subhead\nmb.{6A Lattice current algebras for $c = 1$} \endsubhead

The simplest $(c=1)$ case of a lattice current algebra is worth
singling out for at least two reasons:  (1) the basic
$\theta$-functions encountered here also appear in the $\SU(2)$
affine orbifold model; (2) the lattice part of a $U(l)$ orbifold
encountered in a $W_{1+\infty}$ theory is of this ($U(1)-$)type.

A $1$-dimensional lattice $L = \Bbb Z \omega$ is characterized by
a single natural number $m = |\omega |^{2}$; we shall denote
$\goth A (L;|\omega|^{2} = m)$ by $\goth A (m)$.  Note that $m$
is twice the dimension of the basic charged fields $Y(e^{\pm
\omega}, z)$, while $v(z) = m^{-\frac{1}{2}}\omega(z)$ is the
corresponding $u(1)$ current (see Sect.~1B).  The dual lattice is
$L^{*} = \Bbb Z \omega^{*}$ where $(\omega^{*}|\omega) = 1
\Rightarrow |\omega^{*}|^{2} = \frac{1}{m}$.  The factor group
$L^{*}/L$ is the cyclic group of order $m$; there are, correspondingly, $m$
untwisted modules whose weights will be labeled by minimal length
representatives
$$
\mu \omega^{*} \in L^{*}/L \; ,
\quad
\frac{m-1}{2} \leq \mu \leq
\frac{m}{2} \; ,
\quad 
\mu \in \Bbb Z \; . 
\tag{6.1}
$$
The specialized character of the positive energy 
$\goth A (m)$-module $V_{\mu}$ (of ground state 
$|\mu \omega^{*}\rangle$) is
given by (see \cite{DFSZ} \cite{PT})
$$
K_{\mu}(\tau ,m) = \frac{1}{\eta (\tau)} \Theta^{L}_{\mu 1}
(\tau ,0,0) = \frac{1}{\eta (\tau)} \sum_{n \in \Bbb Z}
q^{\frac{m}{2}(n+\frac{\mu}{m})^{2}} \; . 
\tag{6.2}
$$
This set spans a representation of $SL(2,\Bbb Z)$ in the case of
a bosonic algebra ($m$ even) and requires supplementing it with
{\it Ramond sector} ($\Bbb Z_{2}$ twisted) modules corresponding
to half-odd integer $\mu$'s in the interval \thetag{6.1} for $m$
odd and splitting each integer $\mu$ character into two
(corresponding to summing over even and odd $n$'s in \thetag{6.2}).

For $m = 2s$ even the modular transformation law for $K_{\mu}$ is
given, according to \thetag{3.15}, by
$$
K_{\mu} (\tau +1, 2s) = e^{i\pi (\frac{\mu^{2}}{2s} -
\frac{1}{12})} K_{\mu} (\tau ,2s) \tag{6.3}
$$
$$
K_{\mu} \left( - \frac{1}{\tau},2s \right) 
= \frac{1}{\sqrt{2s}}
\sum^{s}_{\nu=1-s} e^{-i \pi \frac{\mu \nu}{s}}
K_{\nu}(\tau,2s). \tag{6.4}
$$  

%%%\subhead\nmb.{Example 6.1}\endsubhead 
\remark {Example 6.1}
{\it A 
$\Bbb Z_{N}$-orbifold of $\goth A (m)$ is given by the
chiral algebra $\goth A (N^{2}m)$ (and its positive energy
modules).}  If indeed we introduce the inner automorphism
$$
\goth A(m) \ni A \rightarrow UAU^{-1} \; ,
\quad  
U = e^{2\pi i \omega^{*}_{0}/N} 
\tag{6.5a}
$$
$$
Ue^{\omega}U^{-1} = e^{2\pi i/N}e^{\omega} 
\quad 
(UJ(z)U^{-1} = J(z)) 
\tag{6.5b}
$$
($J$ being the $u(1)$ current $J(z) = Y(t^{-1}
\frac{\omega}{\sqrt{m}}, z),$ $\omega^{*}_{0} =
\frac{1}{\sqrt{m}} J_{0}$, cf.~Sect.~1B), then the vertex
operators $Y(e^{\pm N\omega},z)$ generate the gauge invariant
subalgebra
$$
\goth A (m)^{\Bbb Z_{N}} = \goth A (N^{2}m) \; . 
\tag{6.6}
$$
The $\Bbb Z_{2}$-orbifold of $\goth A (m)$ with $m$ odd has an
even gauge invariant subalgebra $\goth A (4m)$. The
representation theory of $\goth A (m)$, $m = 2\rho +1, \rho \in
\Bbb Z_{+}$, can be deduced from this remark.
\endremark

%%%\subhead\nmb.{Example 6.2}\endsubhead 
\remark {Example 6.2}
{\it Modular properties of
characters of } $\goth A (m = 2\rho +1)$ {\it derived from those
for }  $\goth A (4m)$.  The characters $K_{\mu}(\tau ,m)$, $m$
odd, $\mu = \frac{1}{2} \Bbb Z \mod m$ are expressed in terms of
$K_{\nu}(\tau ,4m)$ as follows:
$$
K_{\mu}(\tau ,m) = K_{2\mu}(\tau ,4m) + K_{2\mu +2m}(\tau, 4m)
\; . 
\tag{6.7}
$$
The periodicity relation
$$
K_{\nu +m}(\tau ,m) = K_{\nu}(\tau ,m) 
\tag{6.8}
$$
allows to replace (if necessary) the indices in the right hand
side of \thetag{6.7} by equivalent ones in the canonical interval
\thetag{6.1}.  The resulting $SL_2 (\Bbb Z)$ transformation
properties of $K_{\mu}(\tau ,m)$ then read
$$
K_{\mu}(\tau +1,m) = e^{i\pi 
  \left( \frac{\mu^{2}}{m}-\frac{1}{12} \right)}
\left\{ K_{2\mu}(\tau ,4m)+(-1)^{2\mu +m} K_{2\mu +2m}(\tau, 4m)
\right\}
\tag{6.9a}
$$
$$
\split
K_{\mu} \left( -\frac{1}{\tau} ,m \right) 
    &= \frac{1}{\sqrt{m}} 
       \sum^{m}_{\Sb \nu=1-m \\ (\nu \in \Bbb Z )\endSb} 
       e^{-2\pi i \mu \nu/m}
       K_{2\nu}(\tau ,4m) 
       \\
&= \frac{1}{\sqrt{m}} \sum_{\nu \! \mod m}
   e^{-\frac{2\pi i\mu \nu}{m}} K_{\nu}(\tau ,m) \; . 
\endsplit
\tag{6.9b}
$$
Thus, for $m$ odd, only the entire set of $4m$ characters
$K_{\nu}(\tau ,4m)$ is closed under $SL_2 (\Bbb Z)$. The original
set $\{ K_{\mu}(\tau ,m), \mu \in \Bbb Z /m\Bbb Z\}$,
corresponding to the Neveu-Schwarz sector of the supersymmetric
theory, is however invariant under the subgroup of the modular
group generated by $T^{2}(\tau \rightarrow \tau +2)$ and $S$.  It
is remarkable that the diagonal partition function (in which we
restore the dependence on the $u(1)$ variable $z$),
$$
Z(\tau ,z) = \sum_{\mu \mod m} \chi_{\mu m} (\tau ,z)
\bar{\chi}_{\mu m}(\tau ,z) \tag{6.10}
$$
where
$$
\chi_{\mu m}(\tau ,z) = \sum_{n} q^{\frac{1}{2m}(mn+\mu)^{2}}
e^{2\pi i z \frac{mn+\mu}{m}} \tag{6.11}
$$
is related to the Laughlin plateaus of the quantum Hall effect
(corresponding to filling factor $\nu = \frac{1}{m}$, charge
$\frac{n}{m}$ and fractional spin $J = \frac{n^{2}}{2m}, n \in
\Bbb Z$ --- see \cite{CZ}).  (The characters used in \cite{CZ}
differ from \thetag{6.11} by a non-analytic factor, $\exp \{
-\frac{\pi}{m} \frac{(Im z)^{2}}{Im \tau}\}$ corresponding to a
modified Hamiltonian and ensuring invariance under $z \rightarrow
z + \tau$.)
\endremark

%%%\subhead\nmb.{Example 6.3}\endsubhead 
\remark {Example 6.3}
{\it Charge conjugation
orbifolds.}  The involutive lattice conjugation
$$
C_{L}: \;  
e^{\omega} \rightarrow e^{-\omega} \; ,
\quad  
J \rightarrow -J 
\tag{6.12}
$$  
provides, for $m \neq 2$, an example of an outer automorphism of
the chiral algebra $\goth A (m)$.  Our construction of orbifold
modules does not apply, strictly speaking, to this case.
Nevertheless, it is easy to construct a modular invariant set of
$C_{L}$-orbifold characters.  We shall write them down for the
bosonic ($m = 2s$, $s \in \Bbb N$) case.
\endremark

The $C_{L}$-orbifold chiral algebra $\goth A (2s)^{C_{L}}$ is
generated by a single primary field $\phi = \phi (z,\omega)$ with
respect to its $\goth A (S \otimes 1)^{C_{L}}$ subalgebra, the
real part of the vertex operator $Y(e^{\omega},z)$:
$$
\phi (z,\omega) 
= \frac{1}{\sqrt{2}}
\left\{ Y(e^{\omega},z) + Y(e^{-\omega},z) \right\} \; .
$$
Here $\goth A(S\otimes 1)$ is the $u(1)$ chiral current
subalgebra corresponding to the subspace $S \otimes 1$
\thetag{1.16}.  The operator product expansion of two $\phi$'s
involves the stress energy tensor $T$ and the Virasoro primary
field $:J^{4}(z):$ that generates $\goth A (S \otimes
1)^{C_{L}}$.  The chiral algebra splits into a $C_{L}$-even and a
$C_{L}$-odd parts.  The vacuum module character splits,
accordingly, into two pieces:
$$
K_{0}(\tau ,2s) = K^{+}_{0} (\tau ,2s) + K^{-}_{0}(\tau ,2s)
\tag{6.13a}
$$
where
$$
K^{\pm}_{0}(\tau ,2s) = \frac{1}{2} 
\left\{ K_{0} (\tau ,2s) \pm
    (K_{0}(\tau ,8) - K_{4}(\tau ,8))
\right\} 
\; . 
\tag{6.13b}
$$
The difference of $\Bbb Z_{2}$ twisted level $1\ A^{(1)}_{1}$
characters (that appears in parentheses) can be written in the
form
$$
K_{0}(\tau ,8) - K_{4}(\tau ,8) = \frac{1}{\eta (\tau)} \sum_{n}
(-q)^{n^{2}}
\; . 
\tag{6.13c}
$$
Each pair of representations of weights $\pm \mu \omega^{*}$ of
$\goth A (2s) (|\omega^{*}|^{2} = \frac{1}{2s})$ for $1 \leq \mu
\leq s-1$ gives rise to a single representation of the gauge
invariant subalgebra $\goth A (2s)^{C_{L}}$.  The characters
$K^{\pm}_{0}$ \thetag{6.13}, being expressed in terms of
$K_{\mu}$, have known modular transformation properties; in
particular,
$$
\split
K^{\pm}_{0}(-\frac{1}{\tau} ,2s) &= \frac{1}{2\sqrt{2s}} 
\Bigl\{
  K^{+}_{0} (\tau ,2s) + K^{-}_{0}(\tau ,2s) + K_{s}(\tau ,2s) 
  \\
&\qquad + 2 \sum^{s}_{\mu =1} K_{\mu}(\tau ,2s) \Bigr\} \pm
  \frac{1}{\sqrt{2}}(K_{1}(\tau ,8) + K_{3}(\tau ,8))
\; . 
\endsplit
\tag{6.14}
$$
Analyzing this relation together with the unitarity requirement
for the $S$-matrix one concludes that there are altogether $s+7$
inequivalent representations of $\goth A(2s)^{C_{L}}$
(see~\cite{DV$^3$}) corresponding to $s+3$ untwisted and $4$
twisted orbifold modules.  The $\mu = s$ $\goth A(2s)$-module
splits, in particular, into two $\goth A(2s)^{C_{L}}$-modules
with the same specialized character
$$
\frac{1}{2} K_{s}(\tau ,2s) = \frac{1}{\eta} \sum^{\infty}_{n=0}
q^{s(n+\frac{1}{2})^{2}} 
\tag{6.15}
$$
Similarly, there are two pairs of twisted representations with
characters $K_{i}(\tau ,8)$, $i = 1,3$, each $K_{i}$ appearing
twice (with a coefficient $\pm \frac{1}{2\sqrt{2}}$) in
\thetag{6.14}.

For $s = 1$ the model reduces to a $\Bbb Z_{2}$ affine orbifold.
For $s=2,3,4$ and $6$ it has been identified with known models in
[DV$\ ^3$].  We conjecture that these $C_{L}$-orbifolds can be
shown to exist for all values of $s$ using the vertex operator
construction of Sect.~1A.

\subhead\nmb.{6B. $\SU(2)$ orbifolds}\endsubhead

The finite subgroups of $\SU(2)$ being thoroughly studied,
\footnote{For a modern treatment based on the McKay
correspondence --- see~\cite{Kos}.} 
the $\goth A_{k}(\su(2))$
orbifold characters and their modular properties can be worked
out quite explicitly.  Noting that the Cartan subalgebra of
$\su(2)$ is $1$-dimensional we can express its elements $\alpha
,\beta ,\gamma, \lambda$ by (rational) numbers identifying each
of them with the coefficient to $\Lambda^{\spcheck}_{1} =
\frac{1}{2} \sigma_{3}$ ($\sigma_{j}$ are the Pauli matrices --- see
\thetag{6.23}); then
$$
\split
|\gamma-\beta|^{2} &= \frac{1}{2} 
    (2n + \frac{\lambda}{k} - \beta)^{2} \; ,
    \quad 
    n \in \Bbb Z \; , 
    \quad
    \lambda = 1-k ,\ldots,0,1,\ldots,k
    \; , \\
(\gamma |\alpha) &= \left( n+\frac{\lambda}{2k} \right) \alpha 
    \; , \quad 
    \alpha, \beta \in \Bbb Q 
    \; .
\endsplit 
\tag{6.16}
$$
The character \thetag{4.36}, \thetag{3.18}, \thetag{3.3} can be
written in the form
$$
\chi^{\beta}_{\Lambda,\bar{b},\sigma}(\tau) = \sum^{k}_{\Sb
\lambda =1-k \\ \Lambda-\lambda \in 2 \Bbb Z\endSb}
c^{\Lambda}_{\lambda}(\tau) \Theta^{\beta}_{\lambda
,k,\sigma}(\tau) \; , 
\tag{6.17}
$$
where
$$
\Theta^{\beta}_{\lambda,k,\sigma}(\tau) = \sum_{n \in {\Bbb Z}}
q^{\frac{k}{4}(2n+\frac{\lambda}{k}-\beta)^{2}}
\sigma_{2kn+\lambda} \; , 
\tag{6.18}
$$
$$
\sigma_{2kn+\lambda} = \frac{1}{|\Gamma_{b}|} \sum_{\Sb h \in
\Gamma_{b} \\ \tr h = 2 \cos \pi \alpha\endSb}
\sigma^{*}(h)e^{i\pi(2kn+\lambda)\alpha} \; . 
\tag{6.19}
$$
For $b \neq 1$ and non-exceptional, $\Gamma_{b}$ is abelian and
$h$ can be assumed diagonal.

We have treated in Sects.~2, 3 and 6A the case of a $\Bbb Z_{N}$
orbifold (as an automorphism group of $\goth A(\SU(2)), \Bbb
Z_{N}$ appears as a subgroup of $\SO(3)$; $\Gamma$ in this case
should be identified with its double cover $\Bbb Z_{2N} \subset
\SU(2)$).  Each $\Bbb Z_{N}$ automorphism group leaves a $u(1)$
(Cartan) current invariant.  The remaining {\it non-abelian
subgroups of} $\SO(3)$ can be described as groups on two
generators, $s$ and $t$, obeying three relations:
$$
s^{n_{1}} = t^{n_{2}} = (st)^{n_{3}} = 1 \; , 
\quad 
\frac{1}{n_{1}} +
    \frac{1}{n_{2}} + \frac{1}{n_{3}} = 1 + \frac{2}{|\text{Ad } 
    \Gamma|} > 1 \tag{6.20}
$$
($n_{1},n_{2},n_{3}$ are natural numbers and we denote the group
unit by $1$).  The double cover $\Gamma(\subset \SU(2))$ of
$\text{Ad } \Gamma$ is again generated by two elements $s$ and
$t$ but the group unit in the first relation \thetag{6.20} is
replaced by the non-trivial central element  $\varepsilon$ of
$\SU(2)$:
$$
s,t \in \Gamma \Rightarrow s^{n_{1}} = t^{n_{2}} = (st)^{n_{3}}
= \varepsilon \; , 
\quad
\varepsilon^{2} = 1 (|\Gamma | = 2|\text{Ad } \Gamma|) \; . 
\tag{6.21}
$$

%%%\subhead\nmb.{Example 6.4}\endsubhead 
\remark {Example 6.4}
{\it The} $\Bbb H_{8}
\subset \SU(2)$ {\it orbifold.}  The abstract group of quaternion
units has $8$ elements, $\{ 1,\varepsilon,q_{i},\varepsilon
q_{i}, i=1,2,3\}$; they obey multiplication rules $q^{2}_{i} =
\varepsilon$, $q_{1}q_{2} = q_{3}$ which fit \thetag{6.21} with
$n_{1} = n_{2} = n_{3} = 2$.  It corresponds (according to McKay)
to the affine Dynkin diagram $D^{(1)}_{4}$ (see~\cite{K1}
Chap.~4, Table~Aff 1).  The dimensions of its non-trivial
representations coincide with the coefficients $a_{j}$ in the
expansion of the highest root $\theta$ of $D_{4}$ in terms of
simple roots:
$$
\theta = \alpha_{1} + 2\alpha_{2} + \alpha_{3} +
\alpha_{4} \; . 
\tag{6.22}
$$
We shall denote the (equivalence classes of) irreducible
representations (IR) of $I\!\!H_{8}$ by the simple roots
$\alpha_{\nu}$ of $D^{(1)}_{4}$ ($\alpha_{0}$ corresponding to
the trivial representation).  Then $\alpha_{2}$ maps $I\!\!H_{8}$
into a subgroup of $\SU(2)$:
$$
\alpha_{2}(q_{j}) = \frac{1}{i} \sigma_{j} \, ,
\;\;
j = 1,2,3 
\;\;
\left( \sigma_{1} = 
       \pmatrix 0 & 1 \\ 1 & 0 \endpmatrix \, ,
       \;\;
       \sigma_{3} = 
           \pmatrix 1 & 0 \\ 0 & -1 \endpmatrix \, , 
       \;\;
       \sigma_{2} = i\sigma_{1}\sigma_{3}
\right) \, . 
\tag{6.23}
$$
We reproduce in Table~1, for reader's convenience, the character table for
$\Gamma = I\!\!H_{8}$ also indicating the centralizer
$\Gamma_{g}$ of an element in each conjugacy class (CC).
\endremark

\midinsert
\topcaption{Table 1} 
        $\Gamma = I\!\!H_{8}$: characters and centralizers.
\endcaption
%%$$\matrix
\vskip-1ex
$$
\text{\boxit{$\matrix
 \format
    \medquad\c\medquad    &&
    \vrule height5ex depth3ex \medquad\c\medquad \\
\lowerthis{IR} \raisethisby{0.5ex}{\Bigg \backslash} \raisethis{cc} 
        & 1 & \varepsilon & \{q_{1},\varepsilon q_{1}\} 
        & \{ q_{2},\varepsilon q_{2}\} & \{ q_{3},\varepsilon q_{3}\} \\
\noalign{\hrule height 1pt}
\alpha_{0} & 1 & 1 & 1 & 1 & 1\\
\noalign{\vskip-0.5ex}
\alpha_{1} & 1 & 1 & 1 & -1 & -1 \\
\noalign{\vskip-0.5ex}
\alpha_{2} & 2 & -2 & 0 & 0 & 0 \\
\noalign{\vskip-0.5ex}
\alpha_{3} & 1 & 1 & -1 & 1 & -1 \\
\noalign{\vskip-0.5ex}
\alpha_{4} & 1 & 1 & -1 & -1 & 1 \\
\noalign{\vskip-0.5ex}
\Gamma_{g} & \Gamma & \Gamma & \Bbb Z_{4} & \Bbb Z_{4} & \Bbb Z_{4}
\endmatrix$}}
$$
\endinsert

Using Table~1 and symmetrizing with respect to $2kn + \lambda$ we
compute the sum \thetag{6.19} for the {\it untwisted characters}
(i.e., for $\Gamma_{g} = \Gamma$, $\beta = 0$):

$$
\aligned
(\alpha_{0})_{2kn+\lambda} & = \frac{1}{8}
    [1+(-1)^{\lambda}][1+3(-1)^{kn}i^{\lambda}] \; , \\
(\alpha_{j})_{2kn+\lambda} &= \frac{1}{8}
    [1+(-1)^{\lambda}][1-(-1)^{kn}i^{\lambda}] \; ,
    \quad j =1,3,4  \; , \\
(\alpha_{2})_{2kn+\lambda} &=
\frac{1}{4}[1-(-1)^{\lambda}] \; . 
\endaligned 
\tag{6.24}
$$
Inserting these expressions in \thetag{6.17}, \thetag{6.18} we
recover for $k=1$ the characters \thetag{6.13} of the
$C_{L}$-orbifold for $s=4$:
$$
\aligned
k=1: \; \chi_{0,1,\alpha_{0}}(\tau) 
     &= \frac{1}{4\eta(\tau)}
        \sum_{n} [1+3(-1)^{n}]q^{n^{2}} 
        = K^{+}_{0}(\tau,8) \; , \\
\chi_{0,1,\alpha_{j}}(\tau) 
     &= \frac{1}{2\eta(\tau)}\sum_{n}
     q^{(2n+1)^{2}} = \frac{1}{2} K_{4}(\tau ,8) \; ,
     \quad 
     j=1,3,4 \; ,
     \quad 
     m=0,1 \; , \\
\chi_{1,1,\alpha_{2}}(\tau) &= \frac{(-1)^{m}}{2\eta(\tau)}
     \sum_{n} q^{\frac{1}{4}(2n+1)^{2}} 
     = K_{2}(\tau ,8)
     \left( = \frac{1}{2}K_{1}(\tau,2) \right)
     \; , 
\endaligned 
\tag{6.25a}
$$
where
$$
K^{+}_{0}(\tau,8) = K_{0}(\tau,8)-\frac{1}{2}K_{4}(\tau, 8)
\; . 
\tag{6.25b}
$$

The characters of the $\Bbb Z_{2}$-{\it twisted orbifolds} are
also computed from \thetag{6.17}, \thetag{6.18} for $\beta =
\frac{1}{2}$ and $\sigma(q^{\mu}_{j}) = i^{\sigma \mu}
(q^{\mu}_{j}, \ \mu \in \Bbb Z /4\Bbb Z$ is the general form of
an element of the centralizer $\Bbb Z_{4}$ of
$q_{j}$). Equation~\thetag{6.19} then gives
$$
\sigma_{2kn+\lambda} = \frac{1}{4} \sum_{\mu \! \mod 4}
i^{2kn+\lambda -\sigma)\mu} = \frac{1+(-1)^{\lambda-\sigma}}{4}
[1+(-1)^{kn}i^{\lambda-\sigma}] \; , 
\tag{6.26}
$$
reproducing, for $k=1$ the $C_{L}$-twisted characters of $\goth A (8)$:
$$
\aligned
\chi_{0,\bar{q}_{j},0}(\tau) &=\sum_{n}
    q^{\frac{1}{4}(4n-\frac{1}{2})^{2}} = K_{1}(\tau ,8) =
    \chi_{1,\bar{q}_{j},1}(\tau) \; , 
    \quad 
    j=1,2,3 \; , \\
\chi_{0,\bar{q}_{j},2}(\tau) &=
    K_{3}(\tau,8)=\chi_{1,\bar{q}_{j},-1}
    \; . 
\endaligned 
\tag{6.27}
$$   
(We label throughout the irreducible representations of $\Bbb
Z_{4}$ --- and their characters --- by the exponents $\sigma = 0,\pm
1,2$.)

The number of inequivalent orbifold modules of a level $1$
current algebra (for a simple $\frak g$) is
$$
N(\Gamma \subset G;k=1) = \frac{1}{|Z|} \sum_{\bar{g} \subset
\Gamma} |\hat{\Gamma}_{g}|
\; . 
\tag{6.28}
$$
In the case at hand it is $\frac{1}{2}(5+5+3\times 4) =11$ thus
coinciding with the number $s+7$ of $C_{L}$-orbifold modules for
$s=4$.

Equations~\thetag{6.24} and \thetag{6.26} also allow to compute
orbifold characters for higher levels; in particular, for
$k=2,g=1$, we obtain (expressing the string functions
$c^{\Lambda}_{\lambda}$ in terms of the {\it branching
coefficients} $b^{\Lambda}_{\lambda} =
\eta^{l}c^{\Lambda}_{\lambda}$, for a rank $l$ $\frak g$ ---
see~\cite{K1} Sect.~12.12):
$$
\aligned
\chi_{\Lambda,1,\alpha_{0}}(\tau) &= \frac{1}{\eta (\tau)}
\left\{ b^{\Lambda}_{0} (\tau) \sum_{n} q^{2n^{2}} - \frac{1}{2}
    b^{\Lambda}_{2} (\tau) \sum_{n} q^{\frac{1}{2} (2n+1)^{2}}
    \right\} \\
&= b^{\Lambda}_{0}(\tau) K_{0} (\tau,4) - \frac{1}{2}
    b^{\Lambda}_{2} (\tau) K_{2}(\tau ,4) \; ,
    \quad 
    \Lambda =0,2 \; , 
\endaligned
\tag{6.29a}
$$
$$
\chi_{\Lambda,1,\alpha_{j}}(\tau) = \frac{1}{2}
b^{\Lambda}_{2}(\tau)K_{2}(\tau ,4) \;, 
\quad 
j=1,3,4 \; ,
\quad 
\Lambda =0,2 \; ,
\tag{6.29b}
$$
$$
\chi_{1,1,\alpha_{2}}(\tau) = b^{1}_{1}(\tau)K_{1}(\tau,4) \
(\text{since}\ b^{\Lambda}_{\lambda} =
b^{\Lambda}_{-\lambda}) 
\; . 
\tag{6.29c}
$$
Similarly, using \thetag{6.26}, we can evaulate the twisted
characters.  For those permuted by the action of the centre we
find
$$
\aligned
&\chi_{0,\bar{q}_{j},0}(\tau) = b^{0}_{0}(\tau) K_{1}(\tau,4) =
    \chi_{2,\bar{q}_{j},2}(\tau) \; , \\
&\chi_{2,\bar{q}_{j},0}(\tau) = b^{2}_{0}(\tau)K_{1}(\tau ,4) =
    \chi_{0,\bar{q}_{j},2}(\tau) \; ,
    \quad 
    j=1,2,3 \; . 
\endaligned 
\tag{6.30a}
$$
The remaining twisted characters are split by the action of the
centre, and we only obtain their sums:
$$
\aligned
\chi^{+}_{1,\bar{q}_{j},1}(\tau) +
\chi^{-}_{1,\bar{q}_{j},1}(\tau) 
    &= b^{1}_{1}(\tau) K_{0}(\tau, 4) \; , \\
\chi^{+}_{1,\bar{q}_{j},-1}(\tau)+\chi^{-}_{1,\bar{q}_{j},-1}(\tau)
    &= b^{1}_{1}(\tau) K_{2} (\tau ,4) \; . 
\endaligned 
\tag{6.30b}
$$
Here the branching coefficients can be expressed in terms of the
Virasoro characters $\chi_{\Delta}(\tau ,c)$ of the Ising model
(corresponding to $c=\frac{1}{2}, \Delta = 0, \frac{1}{16},
\frac{1}{2}$):
$$
\aligned
b^{0}_{0}(\tau) &= b^{2}_{2} (\tau) = \chi_{0}
    \left( \tau, \frac{1}{2} \right) \; , \\
b^{2}_{0}(\tau) &= b^{0}_{2}(\tau) =
    \chi_{\frac{1}{2}}(\tau ,\frac{1}{2}) \; , \\
b^{1}_{1}(\tau)&=b^{1}_{-1}(\tau) 
    = \chi_{\frac{1}{16}}
    \left( \tau ,\frac{1}{2} \right) \; . 
\endaligned 
\tag{6.30c}
$$
It follows from \thetag{6.29} and \thetag{6.30} that there are
$2\times 4+1=9$ untwisted and $3\times 6 =18$ twisted level $2$
orbifold modules or altogether $27$  $\goth
A_{2}(\su(2))^{I\!\!H_{8}}$-representations.

%%%\subhead\nmb.{Example 6.5}\endsubhead 
\remark {Example 6.5}
{\it Group theoretic} $S${\it -matrix and fusion
rules for} $I\!\!H_{8} \subset \SU(2)$ {\it and for} $I\!\!H_{8} \subset
\SU(2) \subset E_{8}$.  The simply connected compact group $E_{8}$
is singled out (among the Lie groups with simple simply laced Lie
algebras) for having a trivial centre.  The corresponding current
algebra has a single level $1$ representation, the vacuum $\goth
A_{1}(E_{8})$ module; the modular $S$-matrix is then the identity
operator (multiplication by $1$).  Hence, if $\Gamma$ is a
(non-exceptional) finite subgroup of $E_{8}$ then the $\Gamma
\subset E_{8}$ group theoretic $S$-matrix coincides with the
$\goth A_{1}(E_{8})^{\Gamma}$ orbifold $S$-matrix.  The
possiblity to embed the pair $I\!\!H_{8} \subset \SU(2)$ in
$E_{8}$ thus provides an additional justification for the study
of the group theoretic $S$-matrix per-se.
\endremark

We observe that the $S$-matrix elements depend on both the Lie
group $G$ containing the pair $I\!\!H_{8} \subset \SU(2)$ and on
the {\it level of embedding} of $\SU(2)$ in $G$ which is defined
as follows.  Let the bases in $\su(2)$ and $\frak g$ be chosen in
such a way that the Cartan generator $H$ of $\su(2)$ is expressed
as a linear combination of the Cartan generators $H^{i}$ with
non-negative integer coefficients $m_{i}: H=\sum^{l}_{i=1}
m_{i}H^{i}$.  Then the integers $m_{i}$ satisfy the quadratic
relation
$$
\frac{1}{2} \sum^{l}_{i,j=1} m_{i}a_{ij}m_{j} =
\sum^{l}_{j=1}m_{j} =: N \; ,
$$
where, for a simply laced $\goth g , (a_{ij})$ is its Cartan
matrix.  The positive integer $N$ is the level of embedding of
$\su(2)$ in $\goth g$.

For a level $1$ embedding the $S$-matrix elements involving at
least one non-exceptional entry are independent of $G$.  In the
case of $\Bbb H_{8}$ the phase factor in \thetag{4.25} for a
non-exceptional $b$ and an arbitrary $g$ is only non-trivial if
both $b$ and $g$ belong to the same conjugacy class $q_{j}$.  We
shall then set
$$
\beta (\varepsilon^{m}q_{3})=\frac{(-1)^{m-1}}{4} \sigma_{3}
\Rightarrow \exp\{-2\pi
ik(\beta(\varepsilon^{m}q_{j})|\beta(\varepsilon^{n}q_{j}))\} =
\exp \left\{(-1)^{m+n}\frac{k\pi}{4i} \right\} 
\tag{6.31}
$$
Omitting the upper index $\beta$ on $S$ (for this fixed choice) we obtain
$$
4S_{\varepsilon^{m}\alpha_{\mu},\bar{q}_{j}\sigma}=
(-1)^{m\sigma}\alpha_{\mu} (q_{j})
\tag{6.32a}
$$
$$
2S_{\bar{q}_{j}\sigma,\bar{q}_{j}\sigma'} =
\frac{i^{\sigma+\sigma'}}{2}
\left\{ 
  e^{-i\frac{k\pi}{4}}+(-1)^{\sigma+\sigma'}e^{i\frac{k\pi}{4}}
\right\}
=\cos \left\{ \left(
    \sigma+\sigma'-\frac{k}{2}
    \right) \frac{\pi}{2}
    \right\}
\; . 
\tag{6.32b}
$$
(In computing the sum in the $2$ elements $b=\pm q_{j}$ of the
conjugacy class $\bar{q}_{j}$ in the expression \thetag{4.25} for
$S$ it is important to change at the same time $\sigma$ according
to \thetag{4.12}.  This yields \thetag{6.32b}.)

The only $G$ dependence appears if the central element
$\varepsilon$ of $\Gamma$ is present in both entries:
$$
8S_{\varepsilon^{m}\alpha_{\mu},\varepsilon^{n}\alpha_{\nu}} =
p^{mnk}_{\varepsilon}(-1)^{n\delta\mu_{2} + m\delta\nu_{2}}
2^{\delta\mu_{2} + \delta\nu_{2}} \; ,
\quad
p_{\varepsilon}:= e^{-2\pi i|\beta(\varepsilon)|^{2}} \; ,
\tag{6.33a}
$$
where $\beta(\varepsilon) = 0$ if $G = \SU(2)$, or, more
generally, if it is an exceptional element of $\Gamma \subset G$,
while
$$
p_{\varepsilon} = -1\ \text{if} \
\Gamma_{\beta(\varepsilon)}=\Gamma_{\varepsilon} \tag{6.33b}
$$
(in a level $1$ embedding).  It turns out that the fusion rules
involving a pair of $q_{j}$ and an $\varepsilon$ are integer iff
$\varepsilon$ is a regular element of $\Gamma \subset G$ (i.e.,
if \thetag{6.33b} takes place).  Indeed we have
$$
N_{\bar{q}_{j}\sigma_{1},\bar{q}_{j}\sigma_{2},1\alpha_{\mu}} =
\frac{1+(-1)^{\sigma_{1}+\sigma_{2}+\delta_{\mu 2}}}{4}
\alpha_{\mu}(1) + \frac{\alpha_{\mu}(q_{j})}{2} \cos \pi
\frac{\sigma_{1}-\sigma_{2}}{2} \; ,
$$
which is a $k$ independent non-negative integer, but 
$$
N_{\bar{q}_{j}\sigma_{1},\bar{q}_{j}\sigma_{2},\varepsilon\alpha_{\mu}}
=
\frac{1+p^{k}_{\varepsilon}(-1)^{\sigma_{1}+\sigma_{2}+\delta_{\mu_{2}}}}{4}
\alpha_{\mu}(1) + \frac{\alpha_{\mu}(q_{j})}{2} \cos \left(
\frac{k-\sigma_{1}-\sigma_{2}}{2} \pi\right)
$$
which is only integer for odd $k$ if $p_{\varepsilon}=-1$.

%%%\subhead{\nmb.{}Remark 6.1}\endsubhead  
\remark {Remark 6.1}
Equation~\thetag{6.33b} always
takes place for a level 1 embedding\break 
$\SU(2) \subset E_{8}$.  In
spite of the fact that $\varepsilon$ is an involution
$(\varepsilon^{2} = 1)$ and every involution in $E_{8}$ is
exceptional (as a consequence of the description of finite order
automorphims of a simple Lie algebra presented in Appendix B)
$\varepsilon$ is not exceptional in $\Gamma \subset \SU(2) \subset
E_{8}$ whenever $\SU(2)$ is generated by a pair of opposite roots
of $E_{8}$ --- which is always the case (up to conjugation) for a
level $1$ embedding.  
In other words $(E_8)_{\beta (\varepsilon)}$ is strickly smaller
than $(E_8) \varepsilon$ but $\SU (2) \cap (E_8)_{\beta (\varepsilon)} =
\SU (2) \cap (E_8)_\varepsilon$.
By contrast, for the maximal embedding
$\SU(2) \subset E_{8}$ given by
$$
E = \sum^{8}_{i=1} E^{\alpha_{i}} \; ,
\quad 
H = 2 \rho \; ,
$$
$\varepsilon$ is exceptional in $\Gamma
\subset E_{8}$.  However, the level of this embedding,
$$
N = 2 (\rho | \rho) = g^{\spcheck} \dim E_8 / 6 = 1240
\; ,
$$
is divisible by $4$, hence the group theoretic fusion rules (with
$\beta (\varepsilon) = 0=1-p_{\varepsilon}$) coincide with those
of the Grothendieck ring proven to be non-negative integers in
\cite{Lus1}.  
We have an exceptional subgroup $\Gamma \subset \SO (3) \subset
E_8$ in this case.  The image of any $4$-th order element of
$\tilde{\Gamma} \subset \SU(2)$ is an involution whose centralizer
in $\SO (3)$ is disconnected (see the discussion at the end of
Appendix~B).  It is likely that at least in the case when orders
of all elelments of $\Gamma$ divide $N$ the corresponding twisted
orbitold modules do exist and the resultuing modular $S$-matrix
coincides with the one for the Grothendieck ring.
To compute the fusion rules for the $\goth
A_{k}(\SU(2))^{\Gamma}$ orbifold we shall use the
(non-factorizable) $|{\Cal X}| \times |{\Cal X}|$ 
$S$-matrix of the full theory.
\endremark

For the {\it level 1 orbifold} ordering the states as
$(\Lambda,\bar{b},\sigma) = (0,1,\alpha_{\nu}),\nu =0,1,3,4$,
$(1,\bar{1},\alpha_{2})$, $(0,\bar{q}_{j},0)(\simeq
(1,\bar{q}_{j},1)),(1,\bar{q}_{j},-1)(\simeq (0,\bar{q}_{j},2)),
j=1,2,3$, we can write the $11 \times 11$ $S$-matrix as
$$
2\sqrt{2} S= \bmatrix 
\frac{1}{2} & \frac{1}{2} & \frac{1}{2} & \frac{1}{2} & 1 & 1 &1 &1 &1 &1 & 1\\
\frac{1}{2} & \frac{1}{2} & \frac{1}{2} & \frac{1}{2} & 1 &1 &1
    &-1 &-1 & -1 &-1 \\
\frac{1}{2} & \frac{1}{2} & \frac{1}{2} & \frac{1}{2} 
    &1 &-1 &-1 &1 &1 &-1 &-1 \\
\frac{1}{2} & \frac{1}{2} & \frac{1}{2} & \frac{1}{2} 
    &1 &-1 &-1&-1&-1&1 &1\\
1&1&1&1&-2&0&0&0&0&0&0\\
1&1&-1&-1&0&\sqrt{2}&-\sqrt{2}&0&0&0&0\\
1&1&-1&-1&0&-\sqrt{2}&\sqrt{2}&0&0&0&0\\
1&-1&1&-1&0&0&0&\sqrt{2}&-\sqrt{2}&0&0\\
1&-1&1&-1&0&0&0&-\sqrt{2}&\sqrt{2}&0&0\\
1&-1&-1&1&0&0&0&0&0&\sqrt{2}&-\sqrt{2}\\
1&-1&-1&1&0&0&0&0&0&-\sqrt{2}&\sqrt{2}\\
\endbmatrix 
\; . 
$$
The resulting fusion rules differ, in general, from the group
theoretic ones even for admissible entries.  We have, for
instance, 
$$
\align
N_{0\bar{q}_{j}0,\Lambda\bar{q}_{j}-\Lambda,1 1 \alpha_{2}}
   &= 1 \text{ for } \Lambda =0,1 \; ,
\text{ while } N_{\bar{q}_j 0}, \bar{q}_j - \Lambda,
    \bar{1} \alpha_2 = \frac{1 + (-1)^{1 - \Lambda}}{2} \; ,
\\
N_{0\bar{q}_{j}0,1\bar{q}_{j}-1,0\bar{1}\alpha_{\mu}} &=
    \frac{1-\alpha_{\mu}(q_{j})}{2}
\text{ for }  \mu \neq 2
\; .  \\
& \qquad \text{ while } N_{\bar{q}_j 0}, \bar{q}_j - 1, 1 \alpha \mu = 0
    \text{ for } \mu \neq 2
\; . 
\endalign
$$

%%%\subhead{\nmb.{} Example 6.6} \endsubhead 
\remark {Example 6.6}
{\it The} $\goth A_{2}(\su(2))^{{\Bbb H}_{     8}}$ {\it orbifold
and its Clifford algebra extension}.  The study of level 2
$\SU(2)$-orbifolds is simplified by the observation that $\goth
A_{2} \equiv \goth A_{2}(\su(2))$ is the even part of the Clifford
algebra $Cl_{3}$ of $3$ anticommuting Major\-ana-Weyl spinor fields
$\psi_{j}(z)$, $j = 1,2,3$.  Indeed, the $\Lambda = 2$ $\goth
A_{2}$-module is generated by an ``isotopic triplet'' of primary
fields of dimension $\Delta_{\Lambda} = \frac{1}{4h} \Lambda
(\Lambda + 2)$ (for $\Lambda = k = 2, h = k+2 = 4$), the Virasoro
central charge being $c = 3 \frac{k}{h} = \frac{3}{2}$.  The
fields $\psi_{j}(z)$ are single-valued in the vacuum
(Neveu-Schwarz) sector and satisfy the canonical anticommutation
relations (and hermiticity)
$$
[\psi_{i}(z), \psi_{j}(w)]_{+} = \delta_{ij} \delta(z-w) \; ,
\quad
\psi^{*}_{j} = \psi_{j} \; , 
\quad  
i,j = 1,2,3 \; .
$$
The $\Bbb Z_{2}$ graded algebra $Cl_{3}$ (with odd generators
$\psi_{j}(z)$) provides a superconformal extension of $\goth
A_{2}$ whose $\SU(2)$ invariant subalgebra is generated by the
$\Delta = \frac{3}{2}$ partner
$$
G(z) = i \psi_{1}(z) \psi_{2}(z) \psi_{3}(z) (= G^{*}(z))
\tag{6.34a}
$$
of the stress energy tensor
$$
T(z) = T_{1}(z) + T_{2}(z) + T_{3}(z) \; ,
\quad 
T_{j}(z) = \frac{1}{4}:[\partial \psi_{j}(z), \psi_{j}(z)] 
\tag{6.34b}
$$
which can be viewed a as composite of two $G$-fields.  The generator
$G(z)$ of the super-Virasoro algebra is a primary field with
respect to $T$ but not with respect to $\goth A_{2}$; its
commutator with a Cartan current is
$$    
[J(z),G(w)] = \delta^{\prime}(z-w)\psi_{3}(w)\ \text{for}\  J(z)
=  -i \psi_{1}(z)\psi_{2}(z) \; .
$$
It intertwines the $\Lambda = 0 $ and $\Lambda = 2$ Neveu-Schwarz
modules mapping the $\Lambda = 1$ Ramond sector into itself.
\endremark

Each subgroup $\Gamma$ of $\SU(2)$ acts on $Cl_{3}$ by
automorphisms which form the adjoint group
$$
Ad_{\Gamma} = \Gamma/\Bbb Z_{2} \subset \SO(3) \; ; 
\text{for }
\Gamma = \Bbb H_{8} \, , 
\;
Ad_{\Gamma} = \Bbb Z_{2} \times \Bbb Z_{2}
\; .
\tag{6.35}
$$

In the (orthonormal $\SO(3)$) basis $\{\psi_{j}\}$ the non-trivial
elements $E_{j} = \al_{2}(q_{j})$ of $\Bbb Z_{2} \times \Bbb
Z_{2}$ act as diagonal matrices:
$$
E_{1} = \pmatrix 
1 & 0 & 0 \\
0 & -1 & 0 \\
0 & 0 & -1 
\endpmatrix \; , 
\quad
E_{2} = \pmatrix
-1 & 0 & 0 \\
0 & 1 & 0 \\
0 & 0 & -1 
\endpmatrix \; , 
\quad
E_{3} = E_{1} E_{2} \; .
\tag{6.36}
$$
The $Ad_{\Gamma}(= \Bbb Z_{2} \times \Bbb Z_{2})$ invariant
subalgebra $Cl^{\Gamma}_{3}$ $(\Gamma = \Bbb H_{8})$ of the
$Cl_{3}$ superalgebra is generated by $G$ and by the individual
stress-tensors $T_{j}$ of the $3$ ``Ising models'' (associated
with each $\psi_{j}$) --- see \thetag{6.34b}.  The $3$ commuting
($\Delta = 2$) field operators $T_{j}(z)$ give rise to the even
part $\goth A^{\Gamma}_{2}$ of this superalgebra.  Its positive
energy representations are tensor products of irreducible
representations of the $3$ (minimal) Ising models.  There are, as
expected, $3^{3} = 27$ such $\goth A^{\Gamma}_{2}$ orbifold
modules.  In particular, the characters of the fixed point
modules split into a sum of two irreducible characters:
$$
\align
\chi_{1,\bar{q}_{j},1}(\tau) &= 
    b^{1}_{1} (\tau) K_{0}(\tau ,4) 
    = b^{1}_{1} (\tau) 
      \left\{ [b^{0}_{0}(\tau)]^{2} + [b^{2}_{0}(\tau)]^{2}
      \right\} \; , \\
\chi_{1,\bar{q}_{j},-1}(\tau) &= b^{1}_{1} (\tau) K_{2}(\tau ,4)
    = 2b^{1}_{1}(\tau) b^{2}_{0}(\tau)b^{0}_{0}(\tau) \; . 
\tag{6.37}
\endalign
$$
The asymptotic dimensions of $b^{1}_{1}(b^{\Lambda}_{0})^{2}$ for
$\Lambda = 0,2$ indeed coincide, (the quantum dimension of the $(c
= \frac{1}{2}, \Delta = \frac{1}{2})$ module being $1$.  Here we
have used the expression \thetag{6.30} of the Ising model
characters in terms of the branching coefficients.  The remaining
orbifold modules are identified in the tensor product
$(\Delta_{1},\Delta_{2},\Delta_{3})(\Delta_{i} = 0, \frac{1}{16},
\frac{1}{2})$ of three Ising modules as follows:
$$
\alignat2 
(0,\bar{1},\alpha_{0}) &= (0,0,0) & \qquad  
    (2,1,\alpha_{0}) &=
    \left( \frac{1}{2},\frac{1}{2},\frac{1}{2} \right) \\
(0,\bar{1}, \alpha_{1}) &= \left( 0,\frac{1}{2},\frac{1}{2} \right) & \qquad 
    (2,1,\alpha_{1}) &= \left( \frac{1}{2},0,0 \right) \\
(0,\bar{1},\alpha_{3}) &= \left( \frac{1}{2},0,\frac{1}{2} \right) & \qquad
    (2,1,\alpha_{3}) &= \left( 0,\frac{1}{2},0 \right) \\
(0,\bar{1}, \alpha_{4}) &= \left( \frac{1}{2},\frac{1}{2},0 \right) & \qquad
    (2,1,\alpha_{4}) &= \left( 0,0,\frac{1}{2} \right) \\
(1,\bar{1},\alpha_{2}) &= 
    \left( \frac{1}{16},\frac{1}{16},\frac{1}{16} \right)
    & \qquad  & \\
(0,\bar{q}_{1},0) &= \left( 0,\frac{1}{16},\frac{1}{16} \right) & \qquad
    (2,\bar{q}_{1},0) 
    &= \left( \frac{1}{2},\frac{1}{16},\frac{1}{16} \right) \\
(0,\bar{q}_{2},0) &= \left( \frac{1}{16},0,\frac{1}{16} \right) & \qquad 
    (2,\bar{q}_{2},0) 
    &= \left( \frac{1}{16},\frac{1}{2},\frac{1}{16} \right) \\
(0,\bar{q}_{3},0) &= \left( \frac{1}{16},\frac{1}{16},0 \right) & \qquad
    (2,\bar{q}_{3},0) &= 
    \left( \frac{1}{16},\frac{1}{16},\frac{1}{2} \right)
\; ;
\tag{6.38a}
\endalignat
$$
the reducible (fixed point) modules with characters \thetag{6.37}
split according to the law
$$
\align
(1,\bar{q}_{1},1) 
    &= \left( \frac{1}{16},0,0 \right) 
    + \left( \frac{1}{16}, \frac{1}{2},\frac{1}{2} \right)  \\
(1,\bar{q}_{1},-1)
    &= \left( \frac{1}{16},\frac{1}{2},0 \right) 
    + \left( \frac{1}{16},0,\frac{1}{2} \right)
    \; , \text{ etc.} 
\tag{6.38b}
\endalign
$$
The $\goth A^{{\Bbb H}_{8}}_{2}$ $S$-matrix is the tensor product
of $3$ Ising model $S$-matrices of the form
$$
S_{\text{Ising}} = \frac{1}{2} \pmatrix
1 & \sqrt{2} & 1 \\
\sqrt{2} & 0 & -\sqrt{2} \\
1 & -\sqrt{2} & 1
\endpmatrix 
\; . 
\tag{6.39}
$$
We note that while $S_{1 \bar{q}_{1} \sigma
,1\bar{q}_{2}\sigma^{\prime}} = 0$ according to \thetag{4.25}
(since the conjugacy classes $\bar{q}_{1}$ and $\bar{q}_{2}$ do
not contain commuting elments) the corresponding split $S$-matrix
elements do not vanish:
$$
\split
\left( S_{\frac{1}{16} \Delta_{2}\Delta_{3},\Delta^{\prime}_{1}
    \frac{1}{16}\Delta^{\prime}_{3}}\right) 
    &= \frac{1}{4} 
\pmatrix \format\r&\r&\r&\r\\
   1 & -1 & -1 & 1 \\
   -1 & 1 & 1 & -1 \\
   -1 & 1 & 1 & -1 \\
   1 & -1 & -1 & 1
\endpmatrix 
\; , \\
(\Delta_{i},\Delta_{j}) &= (0,0), 
    \left( \frac{1}{2}, \frac{1}{2} \right),
    \left( \frac{1}{2},0 \right) \left( 0,\frac{1}{2} \right)
    \; .  
\endsplit
\tag{6.40}
$$
Note that the sum of $\goth A^{\Gamma}_{2}$-modules in each line
of equation \thetag{6.38} is irreducible with respect to the
conformal superalgebra $Cl^{\Gamma}_{3}$.  The characters of the
subset of Neveu-Schwarz modules spanned by the direct sum of
$\Lambda = 0$ and $\Lambda = 2$ representations give rise to a
$7$-dimensional representation of the subgroup $\Gamma^{0}(2)$ of
$SL_2 ({\Bbb Z})$ generated by $T^{2}$ and $S$.  In particular,
the Neveu-Schwarz $S$-matrix is
$$
S_{NS} = \frac{1}{4} 
\pmatrix \format\r&\r&\r&\r&\r&\r&\r\\
1 & 1 & 1 & 1 & 2 & 2 & 2 \\
1 & 1 & 1 & 1 & 2 & -2 & -2 \\
1 & 1 & 1 & 1 & -2 & 2 & -2 \\
1 & 1 & 1 & 1 & -2 & -2 & 2 \\
2 & 2 & -2 & -2 & 0 & 0 & 0 \\
2 & -2 & 2 & -2 & 0 & 0 & 0 \\
2 & -2 & -2 & 2 & 0 & 0 & 0 
\endpmatrix 
\; . 
\tag{6.41}
$$
The importance of this example stems from the fact that it has a
bearing on other $\SU(2)$ orbifold models.  The three conjugacy
classes of imaginary quaternion units $\{\pm q_{j}, j=1,2,3\}$ of
$\Bbb H_{8}$ combine in a single $6$-element conjugacy class in
the binary tetrahedral group  $\tilde{A}_{4}$ which in turn is a
part of a $12$-element conjugacy class of the binary octahedral
group $\tilde{S}_{4}$ and of a $30$ element class of the binary
icosahedral group $\tilde{A}_{5}$.  Here $S_{n}$ is the
permutation group of $n$ letters, $A_{n}$ is its alternating
invariant subgroup, $\tilde{G} \subset \SU(2)$ denotes, in
general, the double cover of a subgroup $G$ of $\SO(3)$.  In all
three cases the centralizer $\Gamma_{q_{j}}$ of an element
$q_{j}$ of this conjugacy class is $\Bbb Z_{4}$.  Hence, the
reducible character $\chi_{1\bar{q}_{j}\sigma}$ is the same for
all three orbifold modules and splits in the same way --- according
to \thetag{6.37} --- for all three binary polyhedral groups.  There
are no other conjugacy classes $\bar{b}$ in either
$\tilde{A}_{4}$ or $\tilde{A}_{5}$ such that both $b$ and
$\varepsilon b$ belong to $\bar{b}$.  Futhermore, for all finite
$\SU(2)$ subgroups $\Gamma$ the Neveu-Schwarz module of
$Cl^{\Gamma}_{3}$ contain no fixed points and give rise to a
$\Gamma^{0}(2)$-invariant subset of characters.  Furthermore, a
similar argument extends to a level $n$ representation of $\SU(n)$
which also involves fixed points of the action of the centre.
Indeed, there is a conformal embedding
$$
\goth A_{n} \equiv \goth A_{n} (\su(n)) \subset \goth
A_{1}(spin(n^{2}-1))  
\left( c= \frac{1}{2} (n^{2}-1) \right)
$$
allowing to extend an $\goth A^{\Gamma}_{n}$ orbifold to a
$Cl^{\Gamma}_{n^{2}-1}$-orbifold.

The $k=1$ tetrahedral $(\tilde{A}_{4} \subset \SU(2))$ orbifold
and its fusion rules are displayed in~\cite{DV$^{3}$}.  The octahedral
$(\tilde{S}_{4} \subset \SU(2))$ and the icosahedral
$(\tilde{A}_{5} \subset \SU(2))$ orbifolds can be studied with
equal ease.  We shall reproduce in Table~2 for a later reference the
character table for the $120$ element binary icosahedral group
$\tilde{A}_{5}$ (associated with $E^{(1)}_{8}$ under the McKay
correspondence).

\pageinsert 
\topcaption{Table 2} 
        Characters of $A_{5} = \tilde{A}_{5}/\Bbb Z_{2}$ 
        and of its double cover $\Gamma = \tilde{A}_{5}$.
\endcaption
\vskip-2ex
$$
\text{\boxit{$\matrix
 \format
    \mmsquad\c\mmsquad    &&
    \vrule height5ex depth3ex \mmsquad\c\mmsquad \\
\lowerthis{cc} \raisethisby{0.5ex}{\Bigg \backslash} \raisethis{IR} 
    & \omit\span 1 & \omit\span \overline{\{ p, p^4 \}} 
    & \omit\span \overline{\{ p^2, p^3 \}} 
    & \omit\span \overline{\{ t, t^2 \}} 
    & \overline{E} = \overline{\alpha_2 (q)} \\
\noalign{\hrule height 1pt}
\alpha_0 & \omit\span 1 & \omit\span 1 & \omit\span 1 
    & \omit\span 1 & 1 \\
\noalign{\vskip-0.5ex}
\alpha_2 & \omit\span 3 & \omit\span x_{+} & \omit\span x_{-} 
    & \omit\span 0 & -1 \\
\noalign{\vskip-0.5ex}
\alpha_4 & \omit\span 5 & \omit\span 0 & \omit\span 0 
    & \omit\span -1 & 1 \\
\noalign{\vskip-0.5ex}
\alpha_6 & \omit\span 4 & \omit\span -1 & \omit\span -1 
    & \omit\span 1 & 0 \\
\noalign{\vskip-0.5ex}
\alpha_8 & \omit\span 3 & \omit\span x_{-} & \omit\span x_{+} 
    & \omit\span 0 & -1 \\
\noalign{\vskip-0.5ex}
(A_5)_g & \omit\span A_5 & \omit\span \Z_5 & \omit\span \Z_5 
    & \omit\span \Z_3 & \Z_2 \times \Z_2 \\
\noalign{\hrule height 1pt}
\lowerthis{cc} \raisethisby{0.5ex}{\Bigg \backslash} \raisethis{IR} 
    & 1 & \varepsilon & \overline{p} & \overline{p^4} &
    \overline{p^2} & \overline{p^3} &
    \overline{t} & \overline{t^2} &
    \overline{q} \\
\noalign{\hrule height 1pt} 
\alpha_1 & 2 & -2 & x_{+} & -x_{+} & -x_{-} & x_{-} & 1 & -1 & 0 \\
\noalign{\vskip-0.5ex}
\alpha_3 & 4 & -4 & 1 & -1 & -1 & 1 & -1 & 1 & 0 \\
\noalign{\vskip-0.5ex}
\alpha_5 & 6 & -6 & -1 & 1 & 1 & -1 & 0 & 0 & 0 \\
\noalign{\vskip-0.5ex}
\alpha_7 & 2 & -2 & x_{-} & -x_{-} & -x_{+} & x_{+} & 1 & -1 & 0 \\
\noalign{\hrule height 1pt} 
\Gamma_g & \omit\span \Gamma = \tilde{A}_5 & \omit\span \Z_{10} 
    & \omit\span \Z_{10} & \omit\span \Z_6 & \Z_4 \\
%%%\noalign{\hrule height 1pt} \crcr
\endmatrix$}} 
%%%
$$
\bigskip
$A_{5} = \alpha_{2}(\tilde{A}_{5})$,
$x_{\pm} = \frac{1\pm \sqrt{5}}{2}$,
$\Gamma = \alpha_{1}(\tilde{A}_{5}) \simeq \tilde{A}_{5}$,
$p^{5} = t^{3} = q^{2} = \varepsilon $,\break
$\theta^{E_{8}} = 2(\alpha_{1}+\alpha_{7}) + 3(\alpha_{2} 
    + \alpha_{8} ) + 4(\alpha_{3}+\alpha_{6}) + 5\alpha_{4} 
    + 6\alpha_{5}$.
\bigskip
\endinsert

Equation~\thetag{6.28} implies: $N(\tilde{A}_{5} \subset \SU(2); k = 1)
= \frac{1}{2} (9 \times 2 + 10 \times 4 + 6 \times 2 + 4 ) = 37$.
It is a straightforward exercise to write down, using Table~2,
the characters of $\goth A_{1}(\su(2))^{\Tilde{A}_{5}}$.

\goodbreak
\head\totoc\nmb0{6C.} A level $1$ $\SU(3)$ orbifold.  Charge
conjugation associated with a non-abelian centralizer.
\endhead

We shall consider the subgroup $\Gamma$ of $\SU(3)$ of order
$|\Gamma| = 1080$ which is a non-trivial central extension of the
simple alternating group $A_{6}: 1 \rightarrow {\Bbb Z}_{3}
\rightarrow \Gamma \rightarrow A_{6} \rightarrow 1$.  It is
generated by the ($60$ element) isoahedral group $A_{5} \subset
\SO(3)$ and by one more element of order $2$.  In a basis in which
a selected ${\Bbb Z}_{2} \times {\Bbb Z}_{2}$ subgroup of $A_{5}$
(see Table~2) is generated by any two of the matrices $E_{i}$ 
($=\alpha_{2}(q_{i})$, $i = 1,2,3$) given by \thetag{6.36} while the
generators of its ${\Bbb Z}_{3}$ and ${\Bbb Z}_{5}$ subgroups are
chosen as
$$
t = \frac{1}{2} \pmatrix
x_{-} & 1 & -x_{+} \\
1 & x_{+} & -x_{-} \\
x_{+} & x_{-} & -1 
\endpmatrix 
\; , 
\quad
p = \frac{1}{2} \pmatrix
-x_{-} & 1 & -x_{+} \\
-1 & x_{+} & -x_{-} \\
x_{+} & -x_{-} & 1 
\endpmatrix 
\; , 
\tag{6.42}
$$ 
where $t^{3} = p^{5} = (tp)^{2} = 1$, $tp = E_{2}$, and the
additional involutive generator $E_{4}$ of $\Gamma$ is given by
$(\omega^{2} + \omega + 1 = 0)$: 
$$
E_{4} = - \pmatrix
0 & \omega & 0 \\
\bar{\omega} & 0 & 0 \\
0 & 0 & 1
\endpmatrix 
\; , 
\quad 
E^{2}_{4} = 1= (E_{3}E_{4})^{2} 
\; . 
\tag{6.43}
$$ 
It is the $360$ element factor group $A_{6} = \Gamma/{\Bbb
Z}_{3}$ that acts by non-trivial automorphisms on the $\su(3)$
current algebra.  There are $17$ conjugacy classes of $\Gamma$
versus $7$ of $A_{6}$.  Both are listed in the combined character
table below (see Table~3).

We observe that to each of the first 5 conjugacy
classes in $A_{6}$ correspond 3 such classes (of the same size)
in $\Gamma$ while the last two are mapped into classes of triple
size: $|\bar{t}_{\Gamma}| = 3|\bar{t}_{A_{6}}| = 3 \times 40 (=
|\bar{t}^{\prime}_{\Gamma}|)$.  The essential difference between
$A_{6} =\Gamma / \Bbb Z_{3}$ and the subgroups $\Gamma / \Bbb
Z_{2}$ of $\SO(3)$ is the presence of elements $E(\in \bar{E})$
with a non-abelian centralizer $\Gamma_{8}$.  Table~4 is its
character table ($E_{5} = E_{3}E_{4}$, $q = E_{4}E_{2}$, $q^{3} =
E_{2}E_{4}$).

We note that the centralizer $\Bbb Z_{4}$ of $q$ in $A_{6}$ is
a normal subgroup of~$\Gamma_{8}$.

There are (according to \thetag{6.28}) altogether
$\frac{1}{3} \sum_{\bar{g} \subset \Gamma} |\hat{\Gamma}_{g}| = 17
+ 3.15 + 12 + 3 +3 = 80$ level $1$ $\Gamma \subset \SU(3)$
orbifold modules.  Although it is not practical to write down the
$80 \times 80$ $S$-matrix, one can extract the relevant
information about $E$-twisted orbifolds.

The multipliers $\mu(h|\sum \beta_{i})$ give rise to a new notion
of conjugation whenever the class $\bar{E}$ of involutions labels
a sector.  To display this fact we first observe that the set of
$(45)^{2}$ pairs $(E,E')$ splits into $9$ different orbits
displayed in Table~5.

The stabilizer $\Gamma_{E,E^{\prime}E}$ of the pair
$E,E^{\prime}E$ in $\Gamma$ is the direct product of the central
subgroup $\Bbb Z_{3}$ with the above
$\Gamma^{(0)}_{E,E^{\prime}E} \subset A_{6}$.  To verify the data
of Table~5 one needs to construct a representative pair in each
orbit.  The number of elements of such an orbit is $|A_{6}| =
360$ devided by $|\Gamma^{(0)}_{E,E^{\prime}E}|$.  For instance,
the orbit $O_{\bar{p}}$ is obtained by conjugation of the pair
$(E_{p},E_{1})$ where
$$
\align
E_{p} &= p^{-1}E_{3}p = \frac{1}{2} 
    \pmatrix
       -x_{-} & 1 & x_{+} \\
       1 & -x_{+} & -x_{-} \\
       x_{+} & -x_{-} & -1 
    \endpmatrix \; , \\
E_{1}E_{p} &= E_{2}p^{-1} E_{2} \in \bar{p}
    \quad
    \left( x_{\pm} 
    = \frac{1 \pm \sqrt{5}}{2} \right) \; . 
\endalign
$$
We shall now prove that the oppositely ordered pairs
$(E_{2},E_{4})$ and $(E_{4},E_{2})$ belong to different orbits
$O_{\bar{q}}$ although they belong to the same $\SU(3)$ orbit.  To
this end we construct the most general $u \in \SU(3)$ such that
$$
uE_{2}u^{*} = E_{4} \; , \quad uE_{4}u^{*} = E_{2} \; ; 
\tag{6.44a}
$$
it is given by a 2(real) parameter family.
$$
u = \pmatrix
   u_{1} & u_{2} & 0 \\
   -\zeta\bar{u}_{2} & \zeta\bar{u}_{1} & 0 \\
   0 & 0 & \bar{\zeta}
\endpmatrix
\text{ with } |\zeta|^{2} = 2|u_{1}|^{2} =
2|u_{2}|^{2} = 1 \; , 
\quad
2\bar{\zeta}u_{1}u_{2} = -\omega \; . 
\tag{6.44b}
$$
\midinsert 
\topcaption{Table 3} 
        $\hat{A}_{6} \subset \hat{\Gamma}$: Zero versus non-zero
        triality representations
\endcaption
\topcaption{Table 3a} 
        $\hat{A}_6$.
\endcaption
\vskip-1.5ex
$$
\text{\boxit{$\matrix
 \format
    \smallquad\c\smallquad    &&
    \strut\vrule height5ex depth3ex \smallquad\c\smallquad \\
%%%\noalign{\hrule height 1pt}
\lowerthis{IR} \raisethisby{0.5ex}{\Bigg \backslash} \raisethis{cc} 
    & \spancol 1
    & \spancol \overline{E} (E^2 = 1) 
    & \spancol \overline{q} (q^2 \in \overline{E})
    & \spancol \overline{p} (p^5 = 1) 
    & \spancol \overline{p^2} 
    & \overline{t} (t^3 = 1 = t^{\prime 3}) 
    & \overline{t'}  \\
\noalign{\hrule height 1pt}
1 & \spancol 1 & \spancol 1 & \spancol 1 
    & \spancol 1 & \spancol 1 & 1 & 1 \\
\noalign{\vskip-0.5ex}
5 & \spancol 5 & \spancol 1 & \spancol -1 
    & \spancol 0 & \spancol 0 & 2 & -1 \\
\noalign{\vskip-0.5ex}
5' & \spancol 5' & \spancol 1 & \spancol -1 
    & \spancol 0 & \spancol 0 & -1 & 2 \\
\noalign{\vskip-0.5ex}
8 & \spancol 8 & \spancol 0 & \spancol 0 
    & \spancol x_{+} & \spancol x_{-} & -1 & -1 \\
\noalign{\vskip-0.5ex}
8' & \spancol 8' & \spancol 0 & \spancol 0 
    & \spancol x_{-} & \spancol x_{+} & -1 & -1 \\
\noalign{\vskip-0.5ex}
9 & \spancol 9 & \spancol 1 & \spancol 1 
    & \spancol -1 & \spancol -1 & 0 & 0 \\
\noalign{\vskip-0.5ex}
10 & \spancol 10 & \spancol -2 & \spancol 0 
    & \spancol 0 & \spancol 0 & 1 & 1 \\
\noalign{\vskip-0.5ex}
(A_6)_g & \spancol A_6 & \spancol \Gamma_8 & \spancol \Z_4 
    & \spancol \Z_5 & \spancol \Z_5 
    & \Z^2_3 & \Z^2_3 \\
\endmatrix$}}
$$
\endinsert
\midinsert 
\topcaption{Table 3b} 
        $\hat{\Gamma}$.
\endcaption
\vskip-1.5ex
$$
{\eightpoint
\text 
{\boxiteight{$\matrix
 \format
    \tinyquad \c \tinyquad    &&
    \strut\vrule height5ex depth3ex 
    \tinyquad \c \tinyquad \\
%%%\noalign{\hrule height 1pt}
 \! \lowerthis{IR} \!\! \raisethisby{0.5ex}{\Bigg \backslash}
     \!\!\! \raisethis{cc} \!
    & 1
    & \omega
    & \omega^2
    & \overline{E}
    & \overline{\omega E}
    & \overline{\omega^2 E}
    & \overline{q}
    & \omega \overline{q}
    & \omega^2 \overline{q}
    & \overline{p}
    & \omega \overline{p}
    & \omega^2 \overline{p}
    & \overline{p^2}
    & \omega \overline{p^2}
    & \omega^2 \overline{p^2}
    & \overline{t}
    & \overline{t'}
\\
\noalign{\hrule height 1pt}
3_\omega & 3 & 3 \omega & 3 \omega^2 & -1 & -\omega & -\omega^2 
    & 1 & \omega & \omega^2 & x_{+} & \omega x_{+} & \omega^2 x_{+} 
    & x_{-} & \omega x_{-} & \omega^2 x_{-} & 0 & 0 \\
\noalign{\vskip-0.5ex}
3^*_\omega & 3 & 3 \omega^2 & 3 \omega & -1 & -\omega^2 & -\omega 
    & 1 & \omega^2 & \omega & x_{+} & \omega^2 x_{+} & \omega x_{+} 
    & x_{-} & \omega^2 x_{-} & \omega x_{-} & 0 & 0 \\
\noalign{\vskip-0.5ex}
3'_\omega & 3 & 3 \omega & 3 \omega^2 & -1 & -\omega & -\omega^2 
    & 1 & \omega & \omega^2 & x_{-} & \omega x_{-} & \omega^2 x_{-} 
    & x_{+} & \omega x_{+} & \omega^2 x_{+} & 0 & 0 \\
\noalign{\vskip-0.5ex}
3^{\prime *}_\omega & 3 & 3 \omega^2 & 3 \omega & -1 & -\omega^2 & -\omega 
    & 1 & \omega^2 & \omega & x_{-} & \omega^2 x_{-} & \omega x_{-} 
    & x_{+} & \omega^2 x_{+} & \omega x_{+} & 0 & 0 \\
\noalign{\vskip-0.5ex}
6_\omega & 6 & 6 \omega & 6 \omega^2 & 2 & 2\omega & 2\omega^2 
    & 0 & 0 & 0 & 1 & \omega & \omega^2
    & 1 & \omega & \omega^2 & 0 & 0 \\
\noalign{\vskip-0.5ex}
6^*_\omega & 6 & 6 \omega^2 & 6 \omega & 2 & 2\omega^2 & 2\omega 
    & 0 & 0 & 0 & 1 & \omega^2 & \omega
    & 1 & \omega^2 & \omega & 0 & 0 \\
\noalign{\vskip-0.5ex}
9_\omega & 9 & 9 \omega & 9 \omega^2 & 1 & \omega & \omega^2 
    & 1 & \omega & \omega^2 & -1 & -\omega & -\omega^2
    & -1 & -\omega & -\omega^2 & 0 & 0 \\
\noalign{\vskip-0.5ex}
9^*_\omega & 9 & 9 \omega^2 & 9 \omega & 1 & \omega^2 & \omega 
    & 1 & \omega^2 & \omega & -1 & -\omega^2 & -\omega
    & -1 & -\omega & -\omega^2 & 0 & 0 \\
\noalign{\vskip-0.5ex}
15_\omega & 15 & 15 \omega & 15 \omega^2 & -1 & -\omega & -\omega^2 
    & -1 & -\omega & -\omega^2 & 0 & 0 & 0
    & 0 & 0 & 0 & 0 & 0 \\
\noalign{\vskip-0.5ex}
15^*_\omega & 15 & 15 \omega^2 & 15 \omega & -1 & -\omega^2 & -\omega 
    & -1 & -\omega^2 & -\omega & 0 & 0 & 0
    & 0 & 0 & 0 & 0 & 0 \\
\noalign{\hrule height 1pt} 
\Gamma_g & \spancol \Gamma & \spancol \Z_3 \times \Gamma_8 
    & \spancol \Z_{12} & \spancol \Z_{15} & \spancol \Z_{15} 
    & \Z^2_3 & \Z^2_3 \\
\noalign{\hrule height 1pt} 
| \overline{g} | & \spancol 1 & \spancol 45
    & \spancol 90 & \spancol 72 & \spancol 72 
    & 120 & 120 \\
%%%\noalign{\hrule height 1pt} \crcr
\endmatrix$}}}
$$
\endinsert
It remains to prove that this family of $3 \times 3$ matrices
does not intersect our group~$\Gamma$.  To this end we note that
$|\tr(u+uE_{1})| = |2u_{1}| = \sqrt{2}$; a glance at Table~3 tells
us that this cannot be the case for $u \in \Gamma$.  It turns out
that the same $2$-parameter family of $u$'s is the most general
subset of $\SU(3)$ elements that transforms the two $O_{\bar{E}}$
orbits among themselves:
$$
uE_{3}u^{*} = E_{3} \Rightarrow uE_{1}u^{*} = uE_{2}E_{3}u^{*} =
E_{4}E_{3} = E_{5} \; . 
\tag{6.45}
$$
This completes the proof that each of the two pairs of
representatives in the last column of Table~5 belongs to a
different $\Gamma$-orbit.  We finally note that the sum of all
$|O_{\bar{g}}| (4.360 + 2.180 + 2.90 + 45)$ adds up, as it
should, to $(45)^{2} = 2025$.

\midinsert 
\topcaption{Table 4} 
        Characters of $\Gamma_{8} = \Gamma_{E_{3}} \subset A_{6}$.
\endcaption
\vskip-1.5ex
$$
\text{\boxit{$\matrix
 \format
    \medquad\c\medquad    &&
    \strut\vrule height5ex depth2.5ex \medquad\c\medquad \\
\lowerthis{IR} \raisethisby{0.5ex}{\Bigg \backslash} \raisethis{cc} 
& 1 & E_{3} & E_{1},E_{2} & E_{4},E_{5} & q,q^{3} \\
\noalign{\hrule height 1pt} 
1_{0} & 1 & 1 & 1 & 1 & 1 \\
\noalign{\vskip-0.5ex}
1_{1} & 1 & 1 & 1 & -1 & -1 \\
\noalign{\vskip-0.5ex}
1_{2} & 1 & 1 & -1 & 1 &  -1 \\
\noalign{\vskip-0.5ex}
1_{3} & 1 & 1 & -1 & -1 & 1 \\
\noalign{\vskip-0.5ex}
2 & 2 & -2 & 0 & 0 & 0 \\
\noalign{\hrule height 1pt} 
\Gamma_{E_{3},g} & \Gamma_{8} & \Gamma_{8} & \Bbb Z_{2} \times \Bbb Z_{2} 
        & \Bbb Z_{2} \times \Bbb Z_{2} & \Bbb Z_{4}
\endmatrix$}}
$$
\endinsert

\proclaim{Proposition 6.1} The charge conjugation matrix
\thetag{4.27} for the $ \goth A_{1}(\su(3))^{\Gamma}$ orbifold
involves a non-trivial involution $\sigma \rightarrow \sigma^{c}$
for $b \in E$, $\sigma \in \hat{\Gamma}_{E}$:
$$
C_{\Lambda_0 E_{3}\sigma, \Lambda_0 E_{3}\sigma^{\prime}} 
   = \delta_{\sigma^{\prime}\sigma^{c}} \; ,
\quad 
\sigma_{c} = \sigma^{*} \otimes \sigma^{E} \; ,
\quad 
\sigma^{E} = e^{2\pi i(2\beta_{3}|\beta(h))} \tag{6.46}
$$
where $\Gamma_{E} = \Bbb Z_{3} \times \Gamma_{8}$, $\sigma^{*} =
\sigma$ (i.e., $\sigma (\omega h) = \sigma(h)$ for $h \in
\Gamma_{E})$,
$$
\split
\beta_{i} &= \beta (E_{i}): \beta_{3} = \frac{1}{2} \pmatrix
   1 & 0 & 0 \\
   0 & 1 & 0 \\
   0 & 0 & -2
\endpmatrix 
\Rightarrow \\
\beta_{1} &= \frac{1}{2} \pmatrix
   -2 & 0 & 0 \\
   0 & 1 & 0 \\
   0 & 0 & 1
\endpmatrix 
\; , \quad
\beta_{2} = \frac{1}{2} \pmatrix
   1 & 0 & 0 \\
   0 & -2 & 0 \\
   0 & 0 & 1
\endpmatrix 
\; , \\
\beta_{\underset 5\to4} &= = \frac{1}{4} \pmatrix
   -1 & \pm 3\omega & 0 \\
   \pm 3\bar{\omega} & -1 & 0 \\
   0 & 0 & 2
\endpmatrix 
\; . 
\endsplit
\tag{6.47}
$$
\endproclaim  

\demo{Proof} The statement is a straightforward consequence of
\thetag{4.27} 
(Theorem 4.3e) and of the observation that
$\beta_{3} = \beta(E_{3}) = \beta(E^{-1}_{3})$.  The
representation $\sigma$ is trivial on $\Bbb Z_{3}$ (and hence,
selfconjugate;  see Table~4), since it has to agree with the
representation $\Lambda_0 = 0$ of $\SU(3)$ on the small center.
\quad $\square$
\enddemo

\midinsert
\topcaption{Table 5} 
        Orbits $O^{(i)}_{\overline{E^{\prime}E}}$ of pairs
        $(E,E') \subset \bar{E}$ and their stabilizers $(i = 1,2)$:
\endcaption
%%$$\matrix
\vskip-1.5ex
$$
\text{\boxit{$\matrix
 \format
    \smallquad\c\smallquad    &&
    \vrule height5ex depth3ex \smallquad\c\smallquad \\
%%%\c & \quad \c & \quad \c & \quad \c \\
%%
CC \text{~of~} E^{\prime}E 
   & \Gamma^{(0)}_{E,E^{\prime}E} \subset A_{6} 
   & \left| O^{(i)}_{\overline{E^{\prime}E}} \right| 
   & \text{Representative pairs} \\ 
\noalign{\hrule height 1pt} 
E^{\prime}E = 1 (E=E^{\prime}) & \Gamma_{8} & 45 & \\
\noalign{\vskip0ex}
E^{\prime}E = EE^{\prime} \in \bar{E} 
   & \Bbb Z_{2} \times \Bbb Z_{2} \\
\noalign{\vskip-10ex} 
&  & \left| O^{(i)}_{\bar{E}} \right| = 90 \; , 
   &  O^{(1)}_{\bar{E}} = O(E_{1},E_{2}) \; , \\
\noalign{\vskip-3ex} 
&  &  i = 1, 2
   &  O^{(2)}_{\bar{E}} = O(E_{5},E_{4})  \\
\noalign{\vskip0ex}
E^{\prime}E \in \bar{q} 
   & \Bbb Z_{2} \\
\noalign{\vskip-10ex} 
&  & \left| O^{(i)}_{\bar{q}} \right| = 180 \; , 
   & O^{(1)}_{\bar{q}} = O(E_{2},E_{4}) \; , \\ 
\noalign{\vskip-3ex} 
&  & i = 1, 2 
   &  O^{(2)}_{\bar{q}} = O (E_{4},E_{2}) \\
\noalign{\vskip-0.5ex}
E'E \in \bar{p}^{n} \; , \quad  n=1,2 & 
    \{ 1\} & 360 & \\
\noalign{\vskip-0.5ex}
E'E \in \bar{t} \text{~or~}  \bar{t}' & \{ 1 \} & 360 & \\
\endmatrix$}}
$$
\endinsert

%%%\goodbreak
%%%\subhead{\nmb.{}Remark 6.2}\endsubhead
\remark {Remark 6.2}  
The appearance of a non-trivial conjugation depends
on the choice of a representative in a class of equivalent
quadruples.  Had we chosen instead of the involution element
$E_{3} \in \bar{E}$ a representative of a minimal phase like
$\bar{\omega}E_{3} \in \omega^{2}\bar{E}$ for which
$$
\tilde{\beta}_{3} := \beta(\bar{\omega}E_{3}) = \frac{1}{6} 
\pmatrix
   -1 & 0 & 0 \\
   0 & -1 & 0 \\
   0 & 0 & 2 
\endpmatrix  
\text{ so that } 
|\tilde{\beta}_{3}|^{2} = \frac{1}{6} 
\left( = \frac{1}{9} |\beta_{3}|^{2}\right) \; , 
\tag{6.48}
$$
then we would have dealt with complex representations since
$$
\chi^{\beta_{3}}_{\Lambda_0,\bar{E},\sigma}(\tau) =
\chi^{\tilde{\beta}_{3}}_{\Lambda_{2},\omega^{2}\bar{E},\sigma_{2}}
(\tau) 
\text{ with } \sigma_{2}(h) = \sigma(h)e^{2\pi i(\Lambda_{2}|\alpha)} 
\tag{6.49a}
$$
where $\Lambda_{2}$ is the fundamental weight of the
``antiquark'' representation $3^{*}$, 
$$
\Lambda_{2} = \frac{2}{3} \pmatrix
   -1 & 0 & 0 \\
   0 & -1 & 0 \\
   0 & 0 & 2 
\endpmatrix 
= \tilde{\beta}_{3} - \beta_{3} \; ,
\quad 
h = e^{2\pi i \alpha} \; ,
\quad 
[\alpha, \Lambda_{2}] = 0(= [\alpha , \beta_{3}])
\; . 
\tag{6.49b}
$$
The charge conjugation matrix in these new labels would assume
its usual form with non-zero entry
$$
C_{\Lambda_{2}b\sigma_{2}, \Lambda^{*}_{2}b^{-1}\sigma^{*}_{2}} 
   = 1 \; ,
\quad 
b \in \omega^{2}\bar{E} \; ,
\quad 
b^{-1} \in \omega\bar{E}
\; . 
\tag{6.50}
$$
\endremark

\bigskip
{\eightpoint
I.T. acknowledges the support of a Fulbright grant 19684 and the
hospitality of the Department of Mathematics at M.I.T. during the
course of this work.  Both authors acknowledge the hospitality of
the Erwin Schr\"odinger International Institute for Mathematical
Physics where this paper was completed.  The authors thank Bojko
Bakalov who took part in the computations of the $S$ matrix and
the associated fusion rules presented in Sect.~6B.
}

\vskip1ex
\head\totoc\nmb0{Appendix A.} Action of the center of a simply  
connected simple Lie group on the coroots and fundamental
weights\endhead
We shall display the action of $w_j$ for the classical Lie algebras as  
well as for $E_6$ and $E_7$ (the simply connected groups with Lie  
algebras $G_2,F_4$ and $E_8$ have a trivial center). We let  
$\tilde J=J \cup \{0\}$, $a_0=a^{\spcheck}_{0}=1$. 
 
\subhead\nmb.{A1}. Simply laced algebras $(\al_i^{\spcheck}=\al_i,  
a_i^{\spcheck}=a_i)$ \endsubhead  

The center $\Bbb Z_{l+1}$ of  
$\SU(l+1)$ acts on both the (co) roots and weights of $A_l^{(1)}$ via  
cyclic permutations: 
$$ 
\alignat2
w_1(\al_0,\al_1,\dots,\al_l) 
   & = (\al_1,\al_2,\dots,\al_l.\al_0) \; , 
       & \qquad
       w_j &= w_1^j \\
\tilde w_1(\tilde \La_0,\tilde\La_1,\dots,\tilde\La_l)  
   &= (\tilde\La_1,\tilde\La_2,\dots,\tilde\La_l,\tilde\La_0)
      \; , & \qquad
      w_1^{l+1} &= 1
\; .
\tag {A.1} 
\endalignat 
$$ 
Here $\tilde\La_{\nu}$ are the extended fundamental weights  
$$ 
\tilde\La_{\nu}=d+\La_{\nu}+\kappa_{\nu}K\tag {A.2} 
$$ 
chosen to have equal norm squares: 
$$ 
|\tilde\La_{\nu}|^2=2\kappa_{\nu}+ 
\frac{\nu(l+1-\nu)}{l+1}= 2\kappa_{0} \; . 
\tag {A.3} 
$$ 
The set $\tilde J$ consists of all indices $0,1,\dots,l$. The element  
$w_1$ is a Coxeter element of the finite Weyl group $W(A_l)=S_{l+1}$.  
In terms of the elementary Weyl reflections $s_i$ it is written as: 
$$ 
w_1=s_1\dots s_l \Rightarrow w_1 \Lambda_{j}=\Lambda_{j}-\al_1 -  
\dots -\al_j  \; . 
\tag {A.4} 
$$ 
 
The center of the simply connected group Spin $(2l)$ with Lie algebra  
$D_l$ is $\Bbb Z_2\times \Bbb Z_2$ for $l$ even and $\Bbb Z_{4}$
for $l$ odd.  To exhibit its action on roots and weights of
$D_l^{(1)}$ it is convenient to use an orthonormal basis  
$\{e_i\}$ in the $l$ dimensional root space of $D_l$ setting 
$$ 
\al_i=e_i-e_{i+1} \; ,
\quad 
i=1,\dots, l-1 \; , 
\quad
\al_l=e_{l-1}+e_l \; , 
\quad
\al_0=K-e_1-e_2 
\; , 
\tag {A.5} 
$$ 
$$ 
\La_i=\sum_{s=1}^i e_s \; ,
\quad 
\La_{l-1}= \Lambda_l - e_l \; , 
\quad
\La_l=\frac{1}{2}\sum_{i=1}^l e_i
\; .
\tag {A.6} 
$$ 
The set $\tilde J$ of indices $\mu$ for which $a_{\mu}=1$ consists of  
4 elements: $0,1,l-1,l$. Writing again 
$$ 
\tilde\La_{\nu}=a_{\nu}d+\La_{\nu}+\ka_{\nu}K\tag {A.7a}  
$$ 
we restrict $\ka_{\nu}$ demanding that the norm squares of  
$\tilde\La_{\mu} (\mu\in\tilde J)$ coincide: 
$$ 
|\tilde\La_0|^2=2\ka_0=|\tilde\La_j|^2=1+2\ka_1= 
\frac{l}{4} +2\ka_{l-1}=\frac{l}{4}+2\ka_l \; .
\tag {A.7b} 
$$ 
We shall first determine the finite part $w_l$ of $\tilde{w}_l$ defined  
by $w_l\al_0=w_l(-\th)=\al_l$ and hence (being a permutation of  
$\al_{\mu},\mu\in\tilde J$), $w_l\al_1=\al_{l-1}$. As a consequence  
of invariance of inner products we further deduce  
$w_l\al_i=\al_{l-i},i=1,\dots,l-2$; hence, in view of \thetag {A.5}, 
$$ 
w_le_i=-e_{l+1-i} \; ,
\quad 
i=1,\dots,l-1 \; ;
\tag {A.8a} 
$$ 
$w_le_l$ is then determined from the condition that an element of  
$W(D_l)$ should involve an even number of reflections: 
$$ 
w_le_l=-(-1)^le_1
\tag {A.8b} 
$$ 
As a result, we have $w_l^2=w_1$ for $l$ odd, $w_l^2=1$ for $l$ even;  
in both cases $\bar w_1^2=1$;  
$$ 
w_1(e_1,e_2,\dots,e_{l-1},e_l)=(-e_1,e_2,\dots,e_{l-1},-e_l)
\tag {A.9} 
$$ 
The corresponding permutations of fundamental weights are  
$$ 
\split
\tilde w_l\tilde\La_0 & = \tilde\La_l \; ,
\quad 
\tilde w_{l-1}\tilde\La_1 =\tilde\La_{l-1} \; , 
\quad
\tilde w_l\tilde\La_{l-\nu}= 
\cases 
   \tilde\La_{\nu} & \text{ for } l \text{ even } \\ 
   \tilde\La_{1-\nu} & \text{ for } l \text{ odd } 
\endcases 
\nu=0,1 \; ; \\  
\tilde w_1\tilde\La_0 & =\tilde\La_1 \; ,
\quad 
\tilde w_1\tilde\La_l=\tilde\La_{l-1} \; ,
\quad 
\tilde w_1^2=1 \; ,
\quad 
\tilde w_{l-1}=\tilde w_1\tilde w_l \; .
\endsplit
\tag {A.10} 
$$ 
 
The center of the group $E_6$ is $Z_3$.  Choosing a basis of
simple roots of $E_{6}$ in such a way that the highest root is
$\theta = \alpha_{2} + \alpha_{4} +
2(\alpha_{1}+\alpha_{3}+\alpha_{5}) + 3\alpha_{6}$ we have
$\tilde{J} = \{ 0,2,4\}$.  The center acts on an arbitrary weight
$\Lambda$ according to the law $\tilde{w}_{j}\Lambda =
k\Lambda_{j} +w_{j}\Lambda$, $j = 2,4$, where
$$
w_{2}(-\theta,\alpha_{1},\alpha_{2},\alpha_{3},\alpha_{4},\alpha_{5},
   \alpha_{6})=
   (\alpha_{2},\alpha_{3},\alpha_{4},\alpha_{5},-\theta,\alpha_{1},\alpha_{6})
   \; ,
   \quad
   w^{2}_{2} = w_{4} 
\tag{A.11a}
$$
$$ 
w_{2}\Lambda_{2} = \Lambda_{4}-\Lambda_{2} =
\frac{1}{3}(\alpha_{5}-\alpha_{3} + 2\alpha_{4}-2\alpha_{2})
\Rightarrow \tilde{w}_{2} \tilde{\Lambda}_{2} =
\tilde{\Lambda}_{4}
\; ,
\quad 
w^{3}_{2} = 1
\; . 
\tag{A.11b}
$$
Here we have
used the expressions for the fundamental weights in terms of simple
roots:
$$ 
\align \Lambda_{2} &= \alpha_{1} +
\frac{1}{3}(4\alpha_{2}+5\alpha_{3}+2\alpha_{4}+4\alpha_{5}+6\alpha_{6})
        \; ,
        \\ 
\Lambda_{4} &= \alpha_{1} +
         \frac{1}{3}(2\alpha_{2}+4\alpha_{3}+4\alpha_{4}
         +5\alpha_{5}+6\alpha_{6}) \; ,
         \quad
         \left( |\Lambda_{2}|^{2} = |\Lambda_{4}|^{2} =
           \frac{4}{3} \right)
\endalign
$$ 
as well as the relations $\tilde{\Lambda}_{\nu} =
d+\Lambda_{\nu} + \kappa_{\nu}K$ with
$$ 
|\tilde{\Lambda}_{0}|^{2} = |\tilde{\Lambda}_{2}|^{2} =
|\tilde{\Lambda}_{4}|^{2} \Rightarrow 2\kappa_{0} = \frac{4}{3} +
2\kappa_{2} = \frac{4}{3} + 2\kappa_{4}\ \text{or}\ \kappa_{2} =
\kappa_{4} = \kappa_{0} - \frac{2}{3} \; .
$$

The center of $E_{7}$ is $\Bbb Z_{2}$.  Choosing a
basis of simple roots of $E_{7}$ such that the highest root is
$\theta = \alpha_{6} +
2(\alpha_{1}+\alpha_{5}+\alpha_{7})+3(\alpha_{2}+\alpha_{4})+4\alpha_{3}$,
we have $\tilde{J} = \{ 0,6\}$.  The non-trivial element of the
center is $\tilde{w}_{6} = t_{6}w_{6}$ where
$$
w_{6}(-\theta, \alpha_{1}, \alpha_{2}, \alpha_{3}, \alpha_{4},
     \alpha_{5},\alpha_{6},\alpha_{7})
= (\alpha_{6}, \alpha_{5}, \alpha_{4}, \alpha_{3}, \alpha_{2},
     \alpha_{1}, -\theta, \alpha_{7}) 
\tag{A.12a}
$$
$$
w_{6}\Lambda_{6} = -\Lambda_{6} \Rightarrow
\tilde{w}_{6}\tilde{\Lambda}_{6} = \tilde{\Lambda}_{0}\
\text{for}\ \Lambda_{6} =
\alpha_{1}+2\alpha_{2}+3\alpha_{3}+2\alpha_{5}+\frac{1}{2}(5\alpha_{4}
+3\alpha_{6}+3\alpha_{7})
\; . \tag{A.12b}
$$ 
Here again $\tilde{\Lambda}_{\nu} = d +
\Lambda_{\nu} + \kappa_{\nu}K$ where $|\tilde{\Lambda}_{6}|^{2} =
2\kappa_{6} + \frac{3}{2} = 2 \kappa_0$.

\subhead\nmb.{A2} $\Bbb Z_2$ action on $B_l$ and $C_l$
\endsubhead 

The simple roots, the highest root and the fundamental weights of  
$B_l$ can be written in an orthonormal basis $\{e_i\}$ as 
$$ 
\split
\al_i &= e_i-e_{i+1} \; ,
   \quad 
   i=1,\dots,l-1 \; , \\
\al_l &= e_l \; ,  
   \quad
   \th=\al_1+2(\al_2+\dots+\al_l)=e_1+e_2 \\ 
\La_i &= \sum_{s=1}^i e_s \; , 
   \quad
   i=1,\dots,l \; . 
\endsplit 
\tag {A.13}
$$ 
The center $\Bbb Z_2$ of the simply connected group Spin $(2l+1)$  
acts on $(\al_0 = K - \th,\al_i)$ and on $(\La_0,\La_i)$ as  
$\tilde{w}_1=t_1w_1$ where  
$$ 
w_1(e_1,e_2,\dots,e_l)=(-e_1,e_2,\dots,e_l)
\tag {A.14a} 
$$ 
$$ 
t_1\al^{\spcheck}=\al^{\spcheck}-(\al^{\spcheck}|\La_1)K  
\text{ for } \al^{\spcheck} \in M \; ,
\quad 
t_1\tilde{\La}_{\nu}=\tilde{\La}_{\nu}+(\tilde{\La}_{\nu}|K)\tilde{\La}_1  
\text{ for } \tilde{\La}_{\nu} \in M^*
\; ; 
\tag {A.14b} 
$$ 
thus 
$$ 
\align 
t_1w_1\al_0 &= t_1(K+\al_1)=\al_1 \; , 
   \quad  
   w_1\al_i=\al_i=t_{1} w_1\al_i \text{ for } i=2,\dots,l \; , \\ 
t_1w_1\al_0 &= t_1(-\th)=-\th+K=\al_0 \; ; \\ 
t_1w_1\tilde{\La}_0 &= t_1\tilde{\La}_0 
   = \tilde{\La}_0+\tilde{\La}_1 \; , \\
t_1w_1(\tilde{\La}_0+\tilde{\La}_{1}) 
   &= t_1(\tilde{\La}_0-\tilde{\La}_1)
   =\La_0-\La_1+\La_1=\La_0 
\endalign 
$$ 
 
The simple roots, the highest root and the fundamental weights for  
$C_l$ are expressed as  
$$ 
\align 
\al_i &= \frac{1}{\sqrt 2}(e_i-e_{i+1}) \; , 
   \quad
   i=1,\dots l-1 \; , \\
\al_l &= \sqrt 2 e_l \; ,
   \quad  
   \th=2\sum_{i=1}^{l-1}\al_i+\al_l=\sqrt 2 e_1 \\ 
\La_i &=\sqrt 2\sum_{s=1}^i e_s \; , 
   \quad
   i=1,\dots, l-1 \; ,
   \quad   
   \La_l=\frac{1}{\sqrt 2}\sum^{l}_{s=1} e_s \; . 
\tag {A.14} 
\endalign 
$$ 
The non-trivial element $\tilde{w}_{l} = t_{l}w_{l}$ of the
center $\Bbb Z_2$ of $Sp(2l)$ acts on these orthonormal basis
$e_{i}$ as
$$ 
w_l(e_1,e_2,\dots,e_{l-1},e_l)= (-e_l,-e_{l-1},\dots, -e_2,-e_1) \; ; 
\tag{A.16}
$$ 
hence
$$
w_{l}(-\theta ,\alpha_{1},\ldots ,\alpha_{l}) =
(\alpha_{l},\alpha_{l-1},\ldots, \alpha_{1},-\theta) \; ,
\tag{A.17a}
$$
$$
\split
\tilde{w}_{l}\tilde{\Lambda}_{0}
   &= \tilde{\Lambda}_{l}(=d+\Lambda_{l}+\kappa_{l}K)
   \; ,
   \quad
   w_{l}\Lambda_{l}
    = -\Lambda_{l}\Rightarrow \tilde{w}_{l}\tilde{\Lambda}_{l}
    = \tilde{\Lambda}_{0}\\
\left( |\tilde{\Lambda}_{0}|^{2} \right.
   &= \left. 2\kappa_{0} = |\tilde{\Lambda}_{l}|^{2} 
    = 2\kappa_{l}+\frac{l}{2} \right)
    \; . 
\endsplit
\tag{A.17b}
$$

\vskip1ex
\head\totoc\nmb0{} Appendix B.  Exceptional elements of a compact
Lie group.\endhead

Let $G$ be a connected compact Lie group with a simple Lie algebra  
$\goth g$ of rank $l$, and let $Ad _G$ denote the adjoint group.
An element $g\in G$ is called $ad${\it -exceptional} if it
cannot be written in the form $g = \exp 2 \pi i \beta$, where
$\beta \in i {\goth g}$ is such that $Ad_g x = x$ iff $[\beta, x]
= 0$ for all $x \in {\goth g}$. 
Note that an element $g \in G$ is $Ad$-exceptional iff it is
$ad$-exceptional or its centralizer in $G$ is not connected.
(Recall that in a simply connected $G$ the centralizer of any
element is connected.)  In this Appendix we classify
$ad$-exceptional elements of finite order of the group $Ad_G$.

The finite order inner  
automorphisms of the simple Lie algebra $\goth g$ belong to $Ad_G$  
and can be described as follows (see Theorem~8.6 and Proposition~8.6b  
of~\cite{K1}). 
 
\proclaim{\nmb.{} Proposition B.1} 
Each order $N$ inner automorphism  
of $\goth g$ is conjugate to 
$$ 
Ad_{b(s)} \; , 
\quad
b(s) =\exp 2\pi i \be(s) \; , 
\tag {B.1} 
$$ 
where
$$ 
\be(s)=\frac{1}{N}\sum_{j=1}^l s_j \La_j^{\spcheck}
\tag {B.2} 
$$ 
and $s_0, s_j, j=1,\dots,l$ are relatively prime non-negative  
integers such that: 
$$ 
s_0+\sum_{j=1}^l a_j s_j=N \; .
\tag {B.3} 
$$ 
Here $\La_j^{\spcheck}$ are the fundamental co-weights: 
$$ 
(\al_i|\La_j^{\spcheck})=(\al_i^{\spcheck}|\La_j)=\de_{ij} \; ,
\quad 
i,j=1,\dots,l \; .
\tag {B.4} 
$$ 
\endproclaim  
 
\proclaim{\nmb.{} Proposition B.2} 
The centralizer of $Ad_{b(s)}$
in $\goth g$ is generated by the $E^{\pm\al_{\nu}}$, $\nu = 0, 1,
\ldots, l$, for which
$s_{\nu}=0$ and by the Cartan subalgebra.
\endproclaim 

According to Definition 4.1 an element $b\in\Ga$ is exceptional if  
there is no $\be\in\goth g$ such that  
$$ 
b=e^{2\pi i\be} \text{ and } \Ga_b=\Ga_{\be} \; .
\tag {B.5} 
$$ 
As noted, $G=U(l)$ has no exceptional elements. By contrast, for each  
partition of the positive integer $n\geq 2$ of the type 
$$ 
n=k_1+\dots+k_{\rh} \; , 
\quad
k_{\min}=\min (k_1,\dots,k_{\rh})=2
\tag {B.6} 
$$ 
there are exceptional elements of $\SU(n)$ conjugate to
diagonal matrices with $k_j$ eigenvalues $\exp(2\pi
i\frac{\nu_{j}}{N})$, $j=1,\dots,\rh$, where the $\nu_j$ are
subject to the conditions: (i) $(\nu_1,\dots,\nu_{\rh},N)=1$
(i.e.~these $\rh+1$ integers have no common factor) and
$\sum_{j} k_j\nu_j=kN$ with $1\leq k< k_{\min}$. For $n=2,3$ all
such elements belong to the center $Z_n$ of $\SU(n)$.  More
generally, for any $n$, one can find an element $\ze\in Z_n$ such
that $g=b\ze$ is non-exceptional. (In the above example it
suffices to choose $\ze=\exp(-2\pi i \frac{k}{n})$.) This agrees
with the remark (of Sect.~4B) that $\SU(n)$ contains no
exceptional subgroups. 

Recall that an element $b\in G$ is $Ad$-{\it
exceptional} if $b\ze$ is exceptional for any choice of $\ze \in
Z(G)$. The following theorem describes all 
finite order 
$ad$-excep\-tional elements of $Ad_G$ (for a simple $\goth g$),
and hence all finite order $Ad$-exceptional elements of a simply
connected $G$.
 
\proclaim{\nmb.{} Proposition B.3} 
The finite order automorphism  
$Ad_{b(s)}$ is $ad$-exceptional iff the marks $a_{\nu}$ with  
$s_{\nu}>0$ have a non-trivial common factor. 
\endproclaim 
 
\demo{Proof} It follows from Proposition B.2 that it suffices to  
study the commutator of $\be(s)$ with $E^{\al_{\nu}}$ for those  
$\nu(=0,\dots,l)$ for which $s_{\nu}=0$. 
\enddemo 
 
This commutator is trivial for $j=1,\dots,l$ and $s_j=0$ since
Eqs.~(B. 1-4) imply 
$$ 
[\be(s), E^{\al_j}] = 
(\al_j|\be(s)) E^{\al_j} \; ,
\quad 
(\al_j|\be(s))=s_j=0 \; .
\tag {B.7} 
$$ 
Thus $Ad_b$ can only be $ad$-exceptional if $s_0=0$; in this case 
$$ 
[\be(s), E^{\al_0}] = [\be(s), E^{-\th}] 
= \left( \frac{s_0}{N}-1 \right) E^{\alpha_0} = - E^{\alpha_0} \; .
\tag {B.8} 
$$ 
This is still not sufficient to assert that $Ad_b$ is $ad$-exceptional  
since $\be(s)$ is not unique: we can add to it $\sum_{i=1}^l  
m_i\La_i^{\spcheck}$ for $m_i\in \Bbb Z$ without changing the  
automorphism. That would give 
$$ 
[\be(s)+\sum_i m_i\La_i^{\spcheck}, E^{-\th}] 
= \left( -1-\sum\Sb i \\ s_i\neq 0 \endSb a_i m_i \right) E^{- \theta} 
$$ 
which can be made zero iff the $a_i$ in the sum have no common  
factor. \quad $\square$
 
Proposition~B.3 shows that $\SU(l)$ has no $Ad$-exceptional elements,  
whereas all other simple simply connected compact groups do. Examples of  
$Ad$-exceptional $b$ are provided by the {\it special elements} with  
$\be(s)=\frac{1}{a_j}\La_j^{\spcheck}$ for $a_j>1$, corresponding to  
$s_{\nu}=\de_{\nu j}$. Such is, for instance, the diagonal symplectic  
matrix 
$$ 
\split 
b_1  =e^{2\pi i\La_1}& = 
\pmatrix 
-1 & 0 & 0 & 0 \\ 
0 & 1 & 0 & 0 \\ 
0 & 0 & 1 & 0 \\ 
0 & 0 & 0 & -1  
\endpmatrix 
\in Sp (4) =\{ g \in \SU(4)|^t g C g=C\} \; , \\ 
C & =\pmatrix 
0 & 0 & 0 & 1 \\ 
0 & 0 & 1 & 0 \\ 
0 & -1 & 0 & 0 \\ 
-1 & 0 & 0 & 0 
\endpmatrix     \; ,
\quad
\La_1=\frac{1}{2}\La_1^{\spcheck}=\frac{1}{2} 
\pmatrix 
1 & \\ 
& 0 \\ 
& & 0 \\ 
& & & -1  
\endpmatrix \\ 
& (=\al_1+\frac{1}{2}\al_2) \; .
\endsplit
\tag {B.9} 
$$
($\La_1$ is only stabilized by $U(2)$ while the centralizer of $b_1$  
in $Sp(4)$ is $\SU(2)\times \SU(2)$).

If $\Ga_1\subset \SU_2$ is the  
binary icosahedral group, then  
$\Ga=<b_1,-1>\times \Ga_1\times\Ga_1\subset Sp(4)$ is clearly an  
exceptional subgroup containing the center of $Sp(4)$. 

The simplest example of a non-special $Ad$-exceptional element is  
provided by the simply laced Lie algebra $D_5$ (corresponding to the  
simply connected group Spin (10)). If we label the nodes of the  
affine diagram $D^{(1)}_5$ so that $a_2=a_3=2$ (while  
$a_0=a_1=a_4=a_5=1$) then the non-special $Ad$-exceptional element of  
Spin (10) correspond to  
$\be=\frac{1}{4}(\La_2^{\spcheck}+\La_3^{\spcheck})$.

An example of an element of $\SO (3) = Ad_{\SU (2)}$ with a
disconnected centralizer is provided by either of the diagonal
matrices $E_i$, $i = 1, 2, 3$ of Eq.~\thetag{6.36}.  Indeed,
there is no Cartan subalgebra of $\SO(3)$ containing the
infinitesimal generators of both $E_1$ and $E_2$.  Note that the
preimages of $E_i$ in the simply connected double cover $\SU(2)$
of $\SO(3)$ do not commute (in fact, they anticommute).  This
example extends to the $n^3$ element Heisenberg subgroup $H_n$ of
$\SU(n)$ generated by the $n \times n$ matrices $a$ and $b$
satisfying 
$$
a^n = b^n = 1 \; ,
\quad
a b = e^{2\pi i/n} b a \; .
\tag{B.10}
$$
Clearly, $Ad_a$ and $Ad_b$ commute but their infinitesimal
generators do not.  This happens since $Ad_{\SU(n)}$ (unlike $\SU
(n)$) is not simply connected and the centralizer of either
$Ad_a$ or $Ad_b$ is disconnected.

%%%\newpage
\goodbreak
 

\vskip1ex
\Refs 
\widestnumber\key{AFMO} 
 
\ref 
\key AFMO 
\by J.H. Awata, M. Fukuma, Y. Matsuo, S. Odake  
\book  Representation theory of the $W_{1+\infty}$ 
\publ Prog. Theor. Phys. 
\publaddr Proc. Suppl. 
\yr  1995
\vol 118
\pages 343--373
\endref 
 
\ref 
\key BGT 
\by B.N. Bakalov, L.S. Georgiev, I.T. Todorov 
\paper A QFT approach to $W_{1+\infty}$ 
\paperinfo New Trends in Quantum Field Theory 
\inbook Proc. of the 1995 Razlog (Bulgaria) Workshop, A. Ganchev et al. (eds.) 
\publ Heron Press, Sofia
\yr 1996
\pages  147--158
\endref 
 
\ref 
\key BPZ
\by A.A. Belavin, A.M. Polyakov, A.B. Zamolodchikov
\paper Infinite conformal symmetry in two-dimensional quantum
field theory
\jour Nucl. Phys. 
\vol B241
\yr 1984
\pages 333--381
\endref

\ref 
\key Bor 
\by R. Borcherds 
\paper Vertex algebras, Kac-Moody algebras and the Monster 
\jour Proc. Natl. Acad. Sci. USA  
\vol 83 
\yr 1986 
\pages 3068--3071 
\moreref  
\paper Monstrous moonshine and monstrous Lie Superalgebras 
\jour Invent. Math. 
\vol 109 
\pages 405--444 
\yr 1992 
\endref 
 
\ref 
\key BMP  
\by  P. Bouwknegt, J. McCarthy, K. Pilch 
\paper Semi-infinite cohomology and $w$-gravity 
\jour J. Geom. Phys. 
\vol 11 
\yr 1993 
\pages 225--249 
\endref 
 
\ref 
\key BMT 
\by J.D. Buchholz, G. Mack, I.T. Todorov 
\paper The current algebra on the circle as a germ of local field  
 theory 
\jour Nucl. Phys. B (Proc. Suppl.) 
\vol 5B 
\yr 1988 
\pages 20--56 
\endref 
 
\ref 
\key CTZ 
\by A. Cappelli, C.A. Trugenberger, G.R. Zemba 
\paper Stable hierarchical quantum Hall fluids on $W_{1+\infty}$  
minimal models 
\jour Nucl. Phys. 
\vol B448 [FS] 
\yr 1995 
\pages 470--504 
\moreref 
\paper $W_{1+\infty}$ dynamics of edge excitations, 
\jour cond-mat 
\vol 9407095 
\endref 
 
\ref CZ 
\key CZ 
\by A. Cappelli, G. Zemba 
\paper Modular
invariant partition functions in the quantum Hall effect,
hep-th/9605127
\endref

\ref 
\key DFSZ 
\by P. Di\ Francesco, H. Saleur, J.-B. Zuber 
\paper Modular invariance in non-minimal two-dimen\-sion\-al
conformal theories
\jour Nucl. Phys. 
\vol B285 [FS19] 
\yr 1987 
\pages 454--480 
\endref 
 
\ref 
\key DV$^3$ 
\by R.K. Dijkgraaf, C. Vafa, E. Verlinde, H. Verlinde 
\paper The operator algebra of orbifold models 
\jour Commun. Math. Phys. 
\vol 123 
\yr 1989 
\pages 485--526 
\endref 
 
\ref \key DGM 
\by L. Dolan, P. Goddard, P. Montague 
\paper Conformal field theory of twisted vertex operators 
\jour Nucl. Phys. 
\vol B338 
\yr 1990 
\pages 529--601 
\endref 
 
\ref 
\key FFK 
\by W.M. Fairbairn, T. Fulton, W.H. Klink 
\paper Finite and disconnected subgroups of $\SU_3$ and their  
 application to the elementary particle spectrum 
\jour J. Math. Phys. 
\vol 5 
\yr 1964 
\pages 1038--1051 
\endref 
 
\ref 
\key F 
\by M. Flohr 
\paper $W$-algebras, new rational models and completeness of the  
 $c=1$ classification  
\jour Commun. Math. Phys. 
\vol 157 
\yr 1993 
\pages 179--212 
\endref 
 
\ref 
\key FK 
\by I.B. Frenkel, V.G. Kac 
\paper Basic representations of affine Lie algebras and dual  
 resonance models 
\jour Invent. Math. 
\vol 62 
\yr 1980 
\pages 23--66 
\endref 
 
\ref 
\key FKRW 
\by E. Frenkel, V. Kac, A. Radul, W. Wang 
\paper $W_{1+\infty}$ and $W(gl_N)$ with central charge $N$ 
\jour Commun. math. Phys. 
\vol 170 
\yr 1995 
\pages 337--357 
\endref 
 
\ref 
\key FKW 
\by E. Frenkel, V. Kac, M. Wakimoto 
\paper Characters and fusion rules for $W$-algebra via quantized  
Drinfield-Sokolov reduction 
\jour Commun. Math. Phys. 
\vol 147 
\yr 1992 
\pages 295--328 
\endref 
 
\ref 
\key FLM 
\by I.B. Frenkel, J. Lepowsky, A. Meurman 
\book Vertex Operator Algebras and the Monster 
\publ Academic Press 
\publaddr New York 
\yr 1988 
\endref 
 
\ref
\key FZ
\by I.B. Frenkel, Y. Zhu
\paper Vertex operator algebras associated to representations of
affine and Virasoro algebra
\jour Duke Math. J.
\vol 66
\yr 1992
\pages 123--168
\endref

\ref 
\key FT  
\by J. Fr\"ohlich, E. Thiran 
\paper Integral quadratic forms, Kac-Moody algebras, and fractional  
 quantum Hall effect;
An $ADE$-O classification 
\jour J. Stat. Phys. 
\vol 76 
\yr 1994 
\pages 209--283 
\endref 
 
\ref 
\key FST 
\by P. Furlan, G.M. Sotkov, I.T. Todorov 
\paper Two-dimensional conformal quantum field theory 
\jour Riv. Nuovo Gim 
\vol 12:6 
\yr 1989 
\pages 1--202 
\endref 

 
\ref 
\key Gep 
\by D. Gepner 
\paper New conformal field theories associated with Lie algebras and  
 their partition functions 
\jour Nucl. Phys. 
\vol B285 [FS20] 
\yr 1987 
\pages 10--24 
\endref 
 
\ref 
\key G 
\by P. Ginsparg 
\paper Curiosities at $c=1$ 
\jour Nucl. Phys. 
\vol B295 [FS21] 
\yr 1988 
\pages 153--170 
\endref

\ref
\key Go
\by P. Goddard
\paper Meromorphic conformal fields theory
\inbook Infinite-dimensional Lie algebras and groups
\bookinfo Adv. Ser. Math. Phys. 7
\ed V.~Kac
\publ World Sci.
\publaddr Singapore
\yr 1989
\pages 556--587
\endref

\ref
\key Gor 
\by D. Gorenstein 
\book Finite Groups 
\publ Harper \& Row 
\publaddr New York 
\yr 1968 
\endref 
 
\ref 
\key H 
\by G. Harris 
\paper $\SU(2)$ current algebra orbifolds of the Gaussian model 
\jour Nucl. Phys. 
\vol B300 [FS22] 
\yr 1988 
\pages 588--610 
\endref 
 
\ref 
\key K1 
\by V.G. Kac 
\book Infinite Dimensional Lie Algebras 
\bookinfo Third edition 
\publ Cambridge Univ. Press 
\publaddr 1990 
\yr  
\endref 
 
\ref 
\key K2 
\by V.G. Kac 
\book Vertex algebras 
\publ in: New
Trends in Quantum Field Theory, 
Proceedings of the 1995 Razlog (Bulgaria)
Workshop, A. Ganchev et al. editors, Heron Press, Sofia 
\yr 1996
\pages 261--358
\endref 
 
\ref 
\key KP0 
\by  V.G. Kac, D.H. Peterson 
\paper Affine Lie algebras and Hecke modular forms 
\jour  Bull. Amer. Math. Soc.
\vol  3
\yr  1980
\pages  1057--1061
\endref 
 
\ref 
\key KP1 
\by V.G. Kac, D.H. Peterson 
\paper Spin and wedge representations of infinite dimensional Lie  
 algebras and groups 
\jour Proc. Nat. Acad. Sci. USA 
\vol 78 
\yr 1981 
\pages 3308--3312 
\endref 
 
\ref 
\key KP2 
\by V.G. Kac, D.H. Peterson 
\paper Infinite dimensional Lie algebras, theta-functions and modular  
 forms 
\jour Adv. in Math. 
\vol 53 
\yr 1984 
\pages 125--264 
\endref 
  
\ref 
\key KR1 
\by V.G. Kac, A. Radul 
\paper Quasi-finite highest weight modules over the Lie algebra of  
 differential operators on the circle 
\jour Commun. Math. Phys. 
\vol 157 
\yr 1993 
\pages 429--457 
\endref 
 
\ref 
\key KR2 
\by V.G. Kac, A. Radul 
\book Representation theory of the vertex algebra $W_{1+\infty}$ 
\publ Transformation groups 1 
\publaddr  
\yr 1996
\pages 41--70 
\endref 
 
\ref 
\key KW  
\by V.G. Kac, M. Wakimoto 
\paper Modular and conformal invariant constraints in   
 representation theory of affine algebras 
\jour Advances in Math. 
\vol 70 
\yr 1988 
\pages 156--236 
\endref 
 
\ref 
\key Kos 
\by B. Kostant 
\paper The McKay correspondence, the Coxeter element and  
 representation theory 
\paperinfo in: {\it The Mathematical Heritage of Eli Cartan} 
\jour Soc. Math. de France, Asterisque, hors series 
\yr 1985 
\pages 209--255 
\endref 
 
\ref 
\key LZ 
\by B.H. Lian, G. Zuckerman 
\paper Commutative quantum operator algebras 
\jour J. Pure Appl. Alg. 
\vol 100 
\yr 1995 
\pages 117--139 
\endref 
 
\ref 
\key LR 
\by R. Longo, K.-H. Rehren 
\paper Nets of subfactors 
\jour Rev. Math. Phys. 
\vol 7 
\yr 1995 
\pages 567--597 
\endref 
 
\ref 
\key Lus1 
\by G. Lusztig 
\paper Unipotent representations of finite Chevalley groups of type  
 $E_8$ 
\jour Quart. J. Math. Oxford (2) 
\vol 30 
\yr 1979 
\pages 315--338 
\endref 
 
\ref 
\key Lus2 
\by G. Lusztig 
\paper Leading coefficients of character values of Hecke alebras 
\paperinfo in: Arcata Conference of Representations of Finite Groups,  
\inbook Proceedings of Symposium in Pure Math (P. Fong, ed.) 
\vol 47 
\yr 1987 
\pages  
\endref 
 
\ref 
\key MST 
\by L. Michel, Ya.S. Stanev, I.T. Todorov 
\paper $D-E$-classification of the local extensions of the $su_2$  
 current algebras 
\jour Teor Mat. Fiz. 
\vol 92 
\yr 1992 
\pages 507--521 
\moreref 
\jour (American edition: Theor. Math. Phys. 
\vol 92 
\yr 1993 
\pages 1063) 
\endref 
 

 
\ref 
\key R 
\by K.-H. Rehren 
\paper A new view of the Virasoro algebra 
\jour Lett. Math. Phys. 
\vol 30 
\yr 1994 
\pages 125--130 
\endref 
 
\ref 
\key RST 
\by K.-H. Rehren, Ya.S. Stanev, I.T. Todorov 
\paper Characterizing invariants for local extensions of current  
 algebras 
\paperinfo hep-th/9409165 
\jour Commun. Math. Phys. 
\vol 174 
\yr 1996 
\pages 605--633 
\endref 
 
\ref 
\key V 
\by E. Verlinde 
\paper Fusion rules and modular transformations in $2D$ conformal  
 field theory 
\jour Nucl. Phys. 
\vol B300 [FS22] 
\yr 1988 
\pages 360--376 
\endref 
 
\endRefs
\enddocument